\documentclass[11pt,a4paper]{article}
\usepackage[T1]{fontenc}
\usepackage{braket}
\usepackage{amsmath}
\usepackage{slashed}
\usepackage{verbatim}
\usepackage[titletoc]{appendix}
\usepackage{graphicx}
\usepackage{sidecap}
\usepackage{amsthm}
\usepackage{bbm}
\usepackage{color}
\usepackage{subcaption}
\usepackage{cite}
\usepackage{fullpage}
\usepackage{amsthm}
\usepackage{amssymb}
\usepackage{hyperref}
\usepackage{authblk}

\newcommand{\Tr}{\mathrm{Tr}}

 \newcommand{\Srad}{S_\text{rad} } 

\begin {document}

\title{\bf Entanglement wedge reconstruction and the information paradox}
\author[1]{Geoffrey Penington}
\affil[1]{\small \em  Stanford Institute for Theoretical Physics, Stanford University, Stanford CA 94305 USA}
\date{}
\maketitle
\begin{abstract}
When absorbing boundary conditions are used to evaporate a black hole in AdS/CFT, we show that there is a phase transition in the location of the quantum Ryu-Takayanagi surface, at precisely the Page time. The new RT surface lies slightly inside the event horizon, at an infalling time approximately the scrambling time $\beta/2\pi \log S_{BH}$ into the past. We can immediately derive the Page curve, using the Ryu-Takayanagi formula, and the Hayden-Preskill decoding criterion, using entanglement wedge reconstruction. Because part of the interior is now encoded in the early Hawking radiation, the decreasing entanglement entropy of the black hole is exactly consistent with the semiclassical bulk entanglement of the late-time Hawking modes, despite the absence of a firewall.

By studying the entanglement wedge of highly mixed states, we can understand the state dependence of the interior reconstructions. A crucial role is played by the existence of tiny, non-perturbative errors in entanglement wedge reconstruction. Directly after the Page time, interior operators can only be reconstructed from the Hawking radiation if the initial state of the black hole is known. As the black hole continues to evaporate, reconstructions become possible that simultaneously work for a large class of initial states. Using similar techniques, we generalise Hayden-Preskill to show how the amount of Hawking radiation required to reconstruct a large diary, thrown into the black hole, depends on both the energy and the entropy of the diary. Finally we argue that, before the evaporation begins, a single, state-independent interior reconstruction exists for any code space of microstates with entropy strictly less than the Bekenstein-Hawking entropy, and show that this is sufficient state dependence to avoid the AMPSS typical-state firewall paradox.

\noindent
\footnotetext{\hspace{-0.75cm}{\tt geoffp@stanford.edu}}

\end{abstract}

\tableofcontents

\section{Introduction}
By discovering the AdS/CFT correspondence \cite{maldacena1999large, witten1998anti}, Maldacena definitively answered the question of whether information can escape from a black hole. It can. 

While there remains debate about whether information is lost during black hole evaporation in the real universe \cite{unruh2017information}, in AdS/CFT, the bulk quantum gravity theory in $d+1$ spacetime dimensions is dual to an ordinary $d$-dimensional conformal field theory that lives on the asymptotic boundary of the bulk spacetime. The unitarity of the boundary conformal field theory means that information must be preserved. 

However, on its own, boundary unitarity is not sufficient to consider the information paradox `solved', even in the restricted context of AdS/CFT. We also need to understand what is \emph{wrong} with the Hawking calculation \cite{hawking1974black, hawking1975particle}, which apparently suggests that the radiation should be completely thermal until the black hole has almost entirely evaporated,\footnote{This is a slight over-simplification. In reality, part of the Hawking radiation will be reflected back into the black hole, adding `greybody factors' to the radiation that escapes.} or at least why the conclusion of information loss is na\"{i}ve. 

Conventional effective field theory suggests that the bulk evaporation should be semiclassical, in agreement with Hawking's calculation, so long as the black hole is large compared to the string and Planck scales. In this paper, we will assume that this is indeed the case. In particular, we assume that, from a semiclassical bulk perspective, the Hawking radiation continues to be in a thermal state (up to greybody factors) that is purified by interior modes, even late in the evaporation.

However, as we shall show, this does not mean that no information escapes the black hole. By assuming the quantum version of the Ryu-Takayanagi formula and entanglement wedge reconstruction, we will show that, at late times, the interior degrees of freedom are not microscopically independent of the early Hawking radiation. Instead, a large part of the interior is \emph{encoded in} the early Hawking radiation, in exactly the same way that the bulk in AdS/CFT is microscopically encoded in its asymptotic boundary. This is essentially a formal realisation of the notion of black hole complementarity \cite{susskind1993stretched}. We will precisely identify the part of the interior that is encoded in the Hawking radiation, and thereby derive all the expected properties of unitary black hole evaporation.

In particular, we will show that
\begin{itemize}
\item Only a non-perturbatively small amount of information escapes the black hole before the so-called Page time, when the entropy of the Hawking radiation becomes equal to the Bekenstein-Hawking entropy of the black hole.\footnote{The Page time is commonly called the `halfway point' in the black hole evaporation, although, because of the thermodynamic irreversibility of the evaporation and the time dependence of the black hole temperature, it does not occur halfway through the evaporation either by time or by horizon area/entropy \cite{page2013time}.} However, the existence of such non-perturbatively small corrections is crucial in allowing information to later escape.

\item A small diary, thrown into the black hole early in the evaporation, can be reconstructed at the Page time, so long as the state of the black hole is known. A diary thrown into the black hole after the Page time can be reconstructed after waiting for the scrambling time $\beta/2\pi \log S_{BH}$.\footnote{In this formula, $S_{BH}$ is the Bekenstein-Hawking entropy of the black hole, and $\beta$ is the black hole inverse temperature.} These twin results are known as the \emph{Hayden-Preskill decoding criterion} and were conjectured based on toy models \cite{hayden2007black}. We also derive generalisations of the Hayden-Preskill decoding criterion to large diaries and partially unknown black hole states, where we continue to find exact agreement with toy models.

\item The microscopic entanglement entropy of the black hole obeys the so-called \emph{Page curve}, which was similarly conjectured based on toy models of black hole evaporation \cite{page1993information}. Before the Page time, the entanglement entropy is equal to the entropy of the Hawking radiation, while, after the Page time, it is equal to the Bekenstein-Hawking entropy of the black hole. Crucially, the firewall paradox, which we discuss below, is avoided because the black hole interior is partially encoded in the early Hawking radiation.
\end{itemize}
The firewall paradox \cite{almheiri2013black} suggests that the combination of standard quantum field theory in the bulk, together with global unitarity, is inconsistent with the existence of a smooth horizon after the Page time. For a smooth horizon with finite energy density to exist, outgoing modes close to the horizon must be entangled with the interior outgoing modes, just inside the horizon. As we evolve forwards in time, these outgoing modes become late-time Hawking radiation. 

Furthermore, because the black hole is already close to maximally entangled with the early Hawking radiation, the late-time Hawking radiation must be entangled with the early Hawking radiation, to avoid violating unitarity. However, strong sub-additivity means that a single system cannot be strongly entangled with two different systems at once \cite{lieb1973proof}. We therefore have a paradox. To resolve the paradox, the authors of \cite{almheiri2013black}, known by the acronym AMPS, suggested that a `firewall', or region of very high energy density, must form at the horizon at some point at or before the Page time.

The publication of \cite{almheiri2013black} provoked a flood of responses, including \cite{nomura2013complementarity, bousso2013complementarity, susskind2012singularities, verlinde2013black, susskind2012transfer, harlow2013quantum}. Perhaps most compellingly, in the ER=EPR proposal \cite{maldacena2013cool}, it was pointed out that the thermofield double state of a CFT,
\begin{align}
\sum_i e^{- \beta E_i / 2} \ket{\bar{i}} \ket{i},
\end{align}
is also an example of a black hole that is close to maximally entangled with an external system. Rather than being entangled with the early Hawking radiation, it is entangled with the second copy of the CFT. Hence, the AMPS paradox should apply and the thermofield double state should have a firewall at its horizon. However, the thermofield double state is well-known to be dual to a two-sided Schwarzschild black hole, which has a smooth horizon. 

The resolution, in this case, is obvious: the `interior modes' (which are really exterior modes from the perspective of the second asymptotic boundary) are encoded, from a boundary perspective, in the second copy of the CFT. Hence, there is no contradiction in the Hawking radiation being entangled with both the interior modes and the second copy of the CFT; in fact, the first statement directly implies the second. As one of the results of this paper, we will show that the firewall paradox for one-sided black holes is resolved in exactly the same way. 

To show this, and to show all our other results, we will need to use entanglement wedge reconstruction. The entanglement wedge reconstruction conjecture was developed in \cite{czech2012gravity, headrick2014causality, wall2014maximin} and then established with increasing levels of rigour in \cite{jafferis2016relative}, \cite{dong2016reconstruction} and \cite{cotler2017entanglement}. It has long been known that bulk operators in AdS/CFT can have multiple, distinct boundary representations, which are known as `reconstructions'. Moreover, there exist reconstructions of any local bulk operator that act on only part of the boundary. Bulk information  is encoded redundantly on the boundary. This redundancy is best understood in the language of quantum error correction as the statement that bulk operators, acting on the `code space' of states with a given bulk geometry, are protected against the erasure of certain boundary subregions \cite{almheiri2015bulk}. 

Entanglement wedge reconstruction tells us which part of the bulk is encoded in a given part of the boundary. A bulk operator can be reconstructed using a given boundary region $B$, if, and only if, the bulk operator is contained in a region known as the \emph{entanglement wedge} $b$ of the boundary region $B$. 

To define the entanglement wedge of $B$, we first need to define the Ryu-Takayanagi surface $\chi_B$ associated to $B$. This surface was originally defined for static spacetimes as the bulk surface $\chi_B$ of minimal area $A(\chi_B)$, lying within a static timeslice and homologous to the boundary region $B$ \cite{ryu2006holographic, headrick2007holographic}. However, for general dynamic spacetimes \cite{hubeny2007covariant}, and taking into account quantum corrections \cite{engelhardt2015quantum,dong2018entropy}, it is the quantum extremal surface, homologous to $B$, with the smallest generalised entropy $$A(\chi_B)/4G_N + S_\text{bulk}(\chi_B).$$ 
Here, the bulk entropy $S_\text{bulk}(\chi_B)$ is the von Neumann entropy of the bulk fields contained in the entanglement wedge of $B$, as defined using the candidate surface, and a quantum extremal surface is defined as a $(d-1)$-dimensional surface of extremal generalised entropy.\footnote{In this paper, we will always use the term quantum extremal surface to refer to any surface that is an extremum of the generalised entropy. Similarly, classical extremal surface refers to any extremal area surface. We use Ryu-Takayanagi surface, or quantum Ryu-Takayanagi surface, to refer to the quantum extremal surface of minimal generalised entropy and classical Ryu-Takayanagi surface to refer to the minimal area classical extremal surface.} The Ryu-Takayanagi formula, including quantum corrections, states that entanglement entropy of the boundary region $B$ is equal to the generalised entropy of the Ryu-Takayanagi surface. 

The entanglement wedge is now simple to define. It is the bulk region, or, more precisely, the bulk domain of dependence, bounded by the Ryu-Takayanagi surface $\chi_B$ and the boundary region $B$.

An inportant breakthrough was made recently by Almheiri \cite{almheiri2018holographic}, who used entanglement wedge reconstruction to understand how the ER=EPR proposal continues to resolve the firewall paradox for a two-sided black hole, even when the black hole is dynamically evolving in time. He considered a two-dimensional, two-sided black hole, which has an approximate `boundary' description as an entangled state in a pair of SYK models, and imagined extracting Hawking radiation, using absorbing boundary conditions, from one side of the black hole, and then throwing it into the other side. 

When the final state was evolved backwards in time, this time without any interaction between the two sides, he argued that the Ryu-Takayanagi surface was different from the Ryu-Takayanagi surface in the initial state. Degrees of freedom had moved from the entanglement wedge of the `evaporating' side to the entanglement wedge of the `growing' side. Information was `escaping in the Hawking radiation'. Moreover, this change in Ryu-Takayanagi surface meant that the interior modes, with which the Hawking radiation on the `evaporating' side was entangled, were still encoded in the `growing' side. The two-sided black hole therefore continued to evade the firewall paradox, even as one side of the black hole shrank and the other grew.

The basic conceptual story of this paper will be similar to \cite{almheiri2018holographic}, and indeed to the original ER=EPR proposal \cite{maldacena2013cool}. However, rather than relying on the toy model of the thermofield double state, which is well understood but does not actually describe an evaporating black hole, we will work directly with one-sided evaporating black holes. 

More specifically, in this paper, we consider an evaporating black hole, formed from collapse, where the Hawking radiation is extracted into an auxiliary reservoir $\mathcal{H}_\text{rad}$ using absorbing boundary conditions. By doing so, we will be able to make precise quantitative statements about where, and when, information is encoded. 

Unlike in \cite{almheiri2018holographic}, where only classical Ryu-Takayanagi surfaces were considered, it is crucial, when studying an evaporating black hole, that we look at quantum RT surfaces. Since we are extracting the Hawking radiation into an auxiliary reservoir $\mathcal{H}_\text{rad}$, the microscopic entanglement entropy of the black hole is simply the entanglement entropy between the entire boundary Hilbert space $\mathcal{H}_\text{CFT}$ and the reservoir $\mathcal{H}_\text{rad}$. The Ryu-Takayanagi formula states that this entropy is equal to the generalised entropy of the Ryu-Takayanagi surface $\chi$ associated to the \emph{entire boundary}. 

For an evaporating black hole formed from collapse, the only classical extremal surface, homologous to the entire boundary (i.e. trivial homology), is empty. If entanglement wedge reconstruction was based on the classical Ryu-Takayanagi surface, the interior of the black hole would always be encoded in $\mathcal{H}_\text{CFT}$ and no information would ever escape the black hole.\footnote{In this case, we would have to believe either in remnants, or in a complete breakdown of the semiclassical description of the evaporation, as in the firewall proposal. However, remnants are inconsistent with the spectral density of CFTs and there is no reason within the bulk effective field theory to expect the semiclassical bulk description to breakdown until the black hole has almost entirely evaporated.}

The empty surface is also a quantum extremal surface, with generalised entropy equal to the bulk entanglement entropy $\Srad$ between the Hawking radiation and the interior of the black hole. Since we are assuming that the semiclassical Hawking calculation is valid so long as the black hole is large compared to the string/Planck scales, this bulk entanglement entropy will continue to grow, in agreement with semiclassical calculations, even after the Page time.

However, even early in the evaporation of the black hole, it will turn out that there also exists a second, non-empty quantum extremal surface, which lies just inside the event horizon of the black hole. In Eddington-Finkelstein coordinates, the infalling time of this extremal surface is exactly the scrambling time, to leading order, before the `current time', when Hawking radiation was most recently extracted into $\mathcal{H}_\text{rad}$.

Initially, this quantum extremal surface will not be the Ryu-Takayanagi surface. Its generalised entropy will be approximately the Bekenstein-Hawking entropy $S_\text{BH} = A_\text{hor} / 4G_N$ of the black hole, which is much larger than the generalised entropy $\Srad$ of the empty surface. However, at the Page time, there will be a phase transition and the non-empty quantum extremal surface will become the Ryu-Takayanagi surface. 

From this, one can easily use the Ryu-Takayanagi formula to find the entanglement entropy $S$ between the CFT and the reservoir. To leading order, it is given by
\begin{align}
S = \min(\Srad, A_\text{hor}/4G_N).
\end{align}
The entanglement entropy therefore peaks at the Page time (defined by $\Srad = A_\text{hor}/4G_N$) before beginning to decrease. It has long been conjectured that the entanglement entropy of an evaporating black hole is given by this formula, which is known as the Page curve \cite{page1993information}. In particular, a version of the Page curve can be derived if we model the CFT boundary dynamics by a Haar random unitary acting on a large number of qubits.\footnote{This is slightly ahistorical. The Page curve was conjectured well before AdS/CFT was known. However the conjecture was still based on the assumption that the black hole evaporation could be modelled by a Haar random unitary.}
Here we derive it directly from a bulk calculation.\footnote{We do, of course, need to assume the Ryu-Takayanagi formula and entanglement wedge reconstruction, which are both fundamentally holographic ideas. The Page curve cannot be found using the semiclassical bulk description alone, because it results from the build-up of non-perturbatively small effects. See Section \ref{sec:approx} for more details.}

On its own, an explanation of the Page curve using the Ryu-Takayanagi formula is not entirely satisfactory. It does not explain \emph{why} extracting Hawking radiation into the reservoir should decrease the entanglement entropy. 

In particular, it does not resolve the firewall paradox. If the entanglement entropy $S$ is to decrease over time, the Hawking radiation that is transferred over from the CFT to the reservoir must itself be entangled with the reservoir. In the semiclassical bulk picture of the evaporation, however, it is instead entangled with the interior of the black hole.

Fortunately, we also know about entanglement wedge reconstruction. The newly emitted Hawking radiation is indeed entangled with interior modes, but some of these modes are now in the entanglement wedge of, and so encoded in, the reservoir $\mathcal{H}_\text{rad}$. The same resolution of the firewall paradox that worked for the thermofield double state also works for evaporating black holes. 

In the thermofield double state, the Hawking radiation is perfectly thermally entangled with the second CFT. If the late-time Hawking radiation in an evaporating black hole was perfectly thermally entangled with the reservoir, we would find
\begin{align}
\frac{dS}{dt} = -\frac{d\Srad}{dt} < \frac{1}{4 G_N}\frac{d A_\text{hor}}{dt}.
\end{align}
This inequality is strict, even at leading order, because generically black hole evaporation is a strictly thermodynamic-entropy-increasing process. The entanglement structure of the bulk modes would therefore be inconsistent with the Page curve.

However, unlike in the thermofield double state, the Ryu-Takayanagi surface of an evaporating black hole does not lie exactly on the event horizon. Instead it lies an $O(G_N)$ radial distance inside the horizon. Some of the interior outgoing modes are still encoded in the CFT, and so the new radiation is not perfectly entangled with the reservoir $\mathcal{H}_\text{rad}$.

In simple cases, one can explicitly calculate the rate of change in the entanglement entropy $S$ that results from extracting a small amount of new Hawking radiation. It agrees exactly with the rate of change we found using the Ryu-Takayanagi formula. This is not a coincidence. In fact, we shall show that this agreement must always exist; it is a necessary consequence of the Ryu-Takayanagi surface being an extremum of the generalised entropy.

We are also interested in the question of when information about objects thrown into the black hole reappears in the Hawking radiation. If a small diary had been thrown into the black hole more than one scrambling time ago, it would now lie in the entanglement wedge of the reservoir $\mathcal{H}_\text{rad}$. It is therefore in principle possible to recover the state of the diary by looking only at the Hawking radiation in the reservoir. At least from a boundary perspective, the information contained in the diary has escaped the black hole.

By modelling the boundary dynamics of the CFT as a fast scrambling unitary, Hayden and Preskill famously conjectured in \cite{hayden2007black} that the state of a small diary thrown into a black hole early in the evaporation could be decoded from the Hawking radiation at the Page time, while the state of a diary thrown in after the Page time could be decoded after waiting for the scrambling time. Just as for the Page curve, by assuming entanglement wedge reconstruction, we can derive the Hayden-Preskill decoding criterion from a bulk description of the evaporation.

So far, we have avoided any discussion of the crucial issue of state dependence. The idea that there does not exist any single boundary operator that always corresponds to a given interior bulk operator, and instead different boundary operators must be used for different states, goes back to Papadodidimas and Raju \cite{papadodimas2013infalling, papadodimas2014state}.  As with the ER=EPR proposal, it was partially inspired as a response to the AMPS firewall paradox. Since then, there has been considerable work on understanding whether such state dependence exists and, if so, how it works \cite{harlow2014aspects, papadodimas2016remarks, de2018interior}.

In particular, great progress has been made in the context of the SYK model, a toy model of quantum gravity, where it was shown that there exists a complete basis (in fact an overcomplete basis) of pure black hole microstates, whose interior geometries are well understood \cite{kourkoulou2017pure}. Interior operators can be reconstructed on the boundary for each individual microstate, but there is no single reconstruction that works for all the microstates. The idea that the state dependence of interior operators could be interpreted in the language of quantum error correction was suggested in \cite{hayden2018learning} and developed in detail in \cite{almheiri2018holographic}.

In many ways, however, the simplest case of interior state dependence is the Hayden-Preskill decoding criterion. As discussed above, a small diary thrown into a known black hole state can be reconstructed from the Hawking radiation immediately after the Page time. However, to do this, we have to know the state of the black hole. 

If there was a way of extracting information about the diary from the Hawking radiation that worked for \emph{any} initial black hole state, then, by linearity, we could also extract information for highly mixed initial black hole states. But for highly mixed intial states, the Hawking radiation will look completely thermal until long after the Page time. So it is clear that the interior operators describing the state of the diary can only be encoded in the Hawking radiation in a highly state-dependent way.

It was shown in \cite{hayden2018learning} that state dependence can arise as a consequence of entanglement wedge reconstruction. By using the formalism of approximate operator algebra quantum error correction, specifically the results of \cite{beny2007generalization, beny2009conditions, beny2010general}, one can show that there only exists a single state-independent reconstruction on a boundary region $B$ of a given bulk operator and for a given code space, if the bulk operator is contained in the entanglement wedge of $B$ for all states, \emph{both pure and mixed}, with support only in the code space. In contrast, the existence of state-dependent reconstructions is possible so long as the bulk operator is contained in the entanglement wedge of $B$ for all \emph{pure states}.

Suppose, as before, we want to reconstruct a small diary, thrown into the black hole at an early time, from the Hawking radiation reservoir $\mathcal{H}_\text{rad}$. However, rather than knowing the exact initial state of the black hole, we now only know that the black hole was in some large code space of possible initial microstates. As a simple example, we can imagine that we started with a smaller black hole, in a completely unknown state, and then threw in a large amount of additional energy.

For any pure initial microstate in this code space, the Ryu-Takayanagi surface will jump to the non-empty quantum extremal surface near the horizon, at the Page time, and so the entanglement wedge of $\mathcal{H}_\text{rad}$ will contain the diary. If we knew the initial state of the black hole, we could reconstruct the diary. On the other hand, for a highly mixed initial state, the Ryu-Takayanagi surface of the reservoir $\mathcal{H}_\text{rad}$ will remain empty until much later. 

To be able to find a single state-independent reconstruction that works for the entire code space, we need the interior to be in the entanglement wedge of the reservoir, even for such highly mixed initial states. We therefore need the entropy $S_\text{code}$ of the code subspace to satisfy
\begin{align}
S_\text{code} < \Srad - S_{BH}.
\end{align}
Not only does this agree with a conjecture from \cite{hayden2018learning} based on random unitary toy models, it is also provides the mechanism by which information is able to escape the black hole. Regardless of the initial state of the black hole and the state of any diary that was thrown in, the outgoing Hawking radiation is entangled with interior modes in exactly the same way. However, because the interior modes are themselves encoded in the reservoir $\mathcal{H}_\text{rad}$ in a state-dependent way, the new Hawking radiation still provides information about the state of the black hole to an observer with access to $\mathcal{H}_\text{rad}$.

A similar effect happens before the Page time. At this point, the interior is encoded in the boundary $\mathcal{H}_\text{CFT}$ rather than the reservoir $\mathcal{H}_\text{rad}$. However the encoding is still necessarily state dependent; if we allow too large a class of initial black hole microstates, the interior will no longer be contained in the entanglement wedge of $\mathcal{H}_\text{CFT}$ for highly mixed states. To reconstruct interior operators on the CFT, we need the code subspace of allowed initial microstates to satisfy
\begin{align} \label{eq:bbbbbbb}
S_\text{code} < S_\text{BH} - \Srad.
\end{align}

If the black hole has not evaporated at all, the bulk entanglement entropy $\Srad$ is zero. Hence, \eqref{eq:bbbbbbb} suggests that we can reconstruct the interior for code spaces of microstates whose entropy is \emph{almost} as large as, but still strictly less than, the Bekenstein-Hawking entropy of the black hole. We will say that the interior of an unevaporated black hole is encoded in $\mathcal{H}_\text{CFT}$ in a minimally state-dependent way.

Most of the explicit state-dependent interior reconstructions that have appeared in the literature, for example \cite{papadodimas2013infalling, papadodimas2014state, kourkoulou2017pure, de2018interior}, are only intended to work for a single black hole microstates, or a code space with $O(1)$ dimension. However, in an appendix, we show that the Kourkoulou-Maldacena construction for the SYK model \cite{kourkoulou2017pure} can be trivially extended to work for  a set of microstates with entropy almost as large as the Bekenstein-Hawking entropy. We also show that minimal state dependence is sufficient to avoid the AMPSS typical-state firewall paradox.

The structure of the paper is as follows. In Section \ref{sec:evaporation}, we study entanglement wedge reconstruction in an evaporating black hole that was formed by collapse. By restricting our attention to a single initial microstate, we avoid the issue of state dependence. We find the location of the non-empty quantum extremal surface explicitly, in a simplified evaporation process where the Hawking radiation is extracted from close to the horizon, in Section \ref{sec:extremal}, and then use this calculation to explain the Hayden-Preskill decoding criterion and the Page curve in Section \ref{sec:haydenpage}. Finally, we show how one can still derive Hayden-Preskill and the Page curve, even when non-trivial greybody factors are present, in Section \ref{sec:greybody}.

In Section \ref{sec:statedependence}, we consider large code spaces of initial black hole microstates, and show how the state dependence of interior reconstructions depends on time. We also generalise the Hayden-Preskill decoding criterion to large diaries in Section \ref{sec:largediaries}. In Section \ref{sec:minimal}, we argue that the interior of black holes that have not evaporated at all is encoded in the boundary with only minimal state dependence.

Finally, Section \ref{sec:discuss} includes a detailed summary of the results of the paper, as well as discussion on various topics. In particular, we argue in Section \ref{sec:postevaporation} that, from a bulk perspective, information must escape the black hole through a version of the Horowitz-Maldacena final state proposal \cite{horowitz2004black}. In appendices, we generalise the calculations from Section \ref{sec:extremal} to finite temperature infalling modes, and show how the Kourkoulou-Maldacena construction can easily be made minimally state dependent.

After the completion of this manuscript, the author became aware of independent related work by Almheiri, Engelhardt, Marolf and Maxfield \cite{almheiri2019entropy}, which was published simultaneously.

\section{Entanglement Wedge Reconstruction in an Evaporating Black Hole} \label{sec:evaporation}
In this section, we study an evaporating black hole formed from collapse. For simplicity, we assume throughout that the collapsing matter, and hence the entire spacetime, is rotationally symmetric. We show that no information about the black hole escapes in the Hawking radiation, until the Page time, when the bulk entropy $\Srad$ of the Hawking radiation becomes equal to the Bekenstein-Hawking entropy $S_{BH}$ of the black hole. After the Page time, a large part of the interior of the black hole becomes encoded in the early Hawking radiation. In particular, a diary thrown into the black hole becomes encoded in the Hawking radiation after waiting for the scrambling time. The microscopic entanglement entropy of the black hole begins to decrease, in accordance with the Page curve, because the new Hawking radiation is entangled with interior modes that are encoded in the early Hawking radiation.

Our main focus will be on black holes with fixed Schwarzschild radius $r_s$ in AdS units in the limit $G_N \to 0$. All such black holes are microcanonically stable: if we evolve the system with reflecting boundary conditions, the black hole will quickly reach equilibrium with the Hawking radiation and remain constant in size (up to small fluctuations).\footnote{This should not be confused with the fact that black holes that are sufficiently small (below the Hawking-Page transition \cite{hawking1983thermodynamics}) in AdS units are thermodynamically unstable, even if their size is fixed in the semiclassical limit $G_N \to 0$.}

To study the evaporation of microcanonically stable black holes, we instead impose absorbing boundary conditions.\footnote{For similar use of absorbing boundary conditions to evaporate black holes in AdS/CFT see \cite{almheiri2018holographic, rocha2008evaporation, van2014evaporating}.} The outgoing modes are absorbed by the boundary and so the infalling modes are always in the vacuum state. The Hawking radiation never returns to the black hole, which gradually evaporates.

The dynamics of the system are now irreversible. Rather than evolving unitarily, the boundary state $\ket{\psi} \in \mathcal{H}_\text{CFT}$ will obey a Markovian master equation \cite{preskill1998lecture}. However, as usual with any quantum channel, we can make the evolution unitary by adding an auxiliary Hilbert space -- in this case, a large Markovian reservoir $\mathcal{H}_\text{rad}$ that stores the outgoing Hawking radiation once it reaches the boundary. Such a reservoir is sometimes known as an evaporon \cite{rocha2008evaporation}, although we will not use this term. The information paradox will be resolved by simply keeping track of which parts of the bulk are in the entanglement wedge of $\mathcal{H}_\text{CFT}$ and which are in the entanglement wedge of $\mathcal{H}_\text{rad}$.

We assume that the Ryu-Takayanagi surface, associated to a given boundary region, is defined to be the quantum extremal surface, i.e. surface of extremal generalised entropy
\begin{align} \label{eq:genentropy}
\frac{A}{4G_N} + S_\text{bulk},
\end{align}
homologous to the boundary region, with the smallest generalised entropy. Here, the bulk entropy $S_\text{bulk}$ is the von Neumann entropy of the bulk fields in any spacelike surface bounded by the Ryu-Takayanagi surface and the boundary region.\footnote{Bulk causality ensures that this bulk entropy is independent of the choice of spacelike surface. Instead $S_\text{bulk}$ is a function of the bulk domain of dependence of the spacelike surface, which itself depends only on the RT surface and the boundary region.} If the overall bulk state is pure, this bulk von Neumann entropy is simply entanglement entropy between bulk fields inside and outside the entanglement wedge. 

We will also generally assume that this prescription is equivalent to a maximin prescription,
\begin{align}
\max_{\text{Cauchy}} \min_{\chi} \left[\frac{A(\chi)}{4 G_N} + S_\text{bulk}(\chi)\right],
\end{align}
where one first finds the surface of (globally) minimal generalised entropy within fixed Cauchy slices, and then selects the Cauchy slice which (globally) maximises this minimal generalised entropy.\footnote{As just described, the maximin prescription will not generally pick out a unique surface $\chi$, since we can generally find maximising Cauchy slices with more than one minimal surface. However, generically, it should pick out a unique \emph{stable} surface $\chi$, which continues to be close to a minimal surface under any small perturbation of the Cauchy slice. This stable surface should be extremal and the quantum RT surface.} When this paper first appeared on arXiv, the equivalence of the maximin and extremal (HRT) prescriptions had only been formally shown for classical surfaces (assuming the null energy condition) \cite{wall2014maximin}, although it was expected to also be true for quantum surfaces \cite{engelhardt2015quantum}. It has since been shown for quantum surfaces \cite{akers2019quantum} (assuming the quantum focussing conjecture \cite{bousso2016quantum}). We shall therefore only use the maximin prescription to provide intuition about the location of the quantum RT surface; all our actual results will be found using the extremal surface prescription.

If a classical extremal surface is homologous to any boundary component\footnote{By this we mean a boundary region $B$, whose boundary $\partial B$ is empty.} at some time $t$, it will also be homologous to the same boundary component at any other time and, trivially, will still be a classical extremal surface. The classical Ryu-Takayanagi surface therefore cannot change as a function of the boundary time.\footnote{One might worry that the classical Ryu-Takayanagi surface might not be spacelike separated from the boundary at some boundary time. However, this cannot happen, assuming the null energy condition \cite{wall2014maximin}.}

However, this is only true for a quantum extremal surface when we have reflecting boundary conditions. The bulk entropy term in the generalised entropy \eqref{eq:genentropy} depends not only on local data at the Ryu-Takayanagi surface, but on the state of the bulk fields in entire bulk region, bounded by the RT surface and the boundary.  
\begin{figure}[t]
\includegraphics[width = 0.5\linewidth]{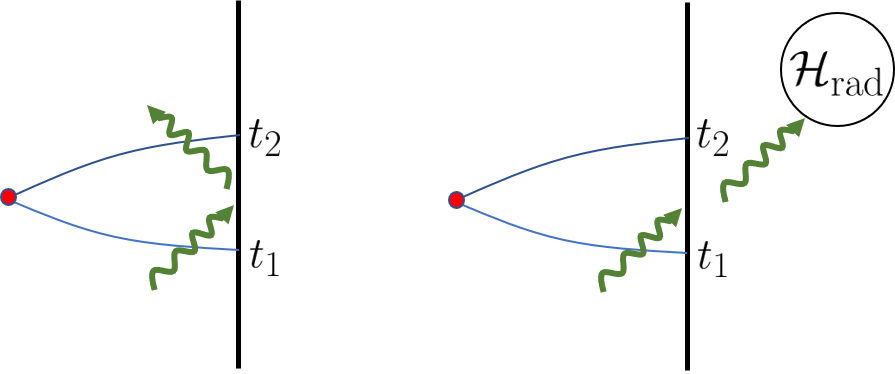}
\centering
\caption{With reflecting boundary conditions, outgoing modes in a surface ending on the boundary at time $t_1$ may become ingoing modes on a surface ending at time $t_2$, but the same degrees of freedom will always be contained in each surface. The bulk entropy cannot depend on the boundary time. In contrast, with absorbing boundary conditions, outgoing modes at time $t_1$ may escape the bulk in the reservoir $\mathcal{H}_\text{rad}$ by time $t_2$, and so no longer be contained in a surface ending at time $t_2$. The bulk entropy, and hence the notion of quantum extremality, depends on the boundary time.}
\label{fig:bcs}
\end{figure}

As shown in Figure \ref{fig:bcs}, with reflecting boundary conditions, the same degrees of freedom are contained in this region, independent of the boundary time. However, with absorbing boundary conditions, outgoing modes, which are contained in a spacelike surface that ends on the boundary at time $t_1$, may escape the boundary and not be contained in a spacelike surface ending at a later time $t_2$. The bulk entropy, and hence, more importantly, the \emph{gradient} of the bulk entropy, may depend on the boundary time. A quantum extremal surface for the entire boundary at time $t_1$ may no longer be a quantum extremal surface for the entire boundary at time $t_2$. 

To understand how the information paradox is resolved in AdS/CFT, we will need not only to find the entanglement wedge of the CFT, but also the entanglement wedge of the reservoir $\mathcal{H}_\text{rad}$. This is not a `boundary region' in the usual sense and we therefore need to be careful about how to define a Ryu-Takayanagi surface and an entanglement wedge for it. 

It is a general fact that, if we divide the boundary of a pure holographic state into two complementary regions, the quantum Ryu-Takayanagi surface will be the same for both regions, and hence the entanglement wedges of the two regions will also be complementary. The same will be true here. The Ryu-Takayanagi surface for $\mathcal{H}_\text{rad}$ will be the same as the RT surface for $\mathcal{H}_\text{CFT}$, and the entanglement wedge of $\mathcal{H}_\text{rad}$ will contain the outgoing modes that were extracted into the reservoir, along with the bulk domain of dependence of any spacelike surface in the black hole interior that is bounded only by the RT surface. All the bulk degrees of freedom will either be in the entanglement wedge of the CFT, or be in the entanglement wedge of the reservoir $\mathcal{H}_\text{rad}$.

However, because our `boundary region' is not actually holographic, let us take a moment to see explicitly why this is true. The simplest way to do so is to imagine throwing the radiation in $\mathcal{H}_\text{rad}$ into the bulk of an auxiliary holographic CFT, with a parametrically smaller Newton's constant $G_N'$ (i.e. a parametrically larger central charge) than the original CFT. The small gravitational coupling $G_N'$ ensures that there won't be any significant backreaction.\footnote{To avoid talking about more than one holographic CFT, we could instead avoid backreaction by taking a large number of copies of the original CFT and throwing a small amount of radiation into each one.} Since the entanglement entropy of this auxiliary CFT must be the same as the entanglement entropy of $\mathcal{H}_\text{rad}$, we will define the RT surface of $\mathcal{H}_\text{rad}$ to be the RT surface of this auxiliary CFT, which is, by definition the smallest generalised entropy quantum extremal surface homologous to the entire boundary of this auxiliary bulk geometry (i.e. a closed surface with trivial homology). Note that, for a one-sided black hole, the entire boundary of the original CFT also has trivial homology; the two homology constraints are the same.

One might worry that the RT surface that we define in this way could depend on the details of the auxiliary spacetime, in which case it would not be well-defined as an RT surface for $\mathcal{H}_\text{rad}$. However, it is easy to see that the RT surface cannot contain a non-empty component in the auxiliary geometry, since in the limit $G_N' \to 0$ the area term will always be dominant and vacuum AdS contains no classical extremal surface.\footnote{The quantum extremal surface prescription for the RT surface can be derived from the replica trick \cite{dong2018entropy}. In the limit $G_N' \to 0$, the auxiliary spacetimes will be identical in the replicated and unreplicated geometries (because of the lack of backreaction). It follows that the derivation from  \cite{dong2018entropy} can be used to directly calculate the entropy of $\mathcal{H}_\text{rad}$, without having to worry about the auxiliary geometry we introduced here. After this paper first appeared on arXiv, these replica trick calculations were done explicitly in \cite{penington2019replica, almheiri2019replica}, where it was shown that the relevant topologies for the second half of the Page curve involve spacetime wormholes connecting different replicas.} It follows that the only place where a nonemtpy component of the RT surface can exist is in the original black hole geometry.

The entanglement wedge of the auxiliary CFT is the domain of dependence of any spacelike surface bounded by the RT surface and the auxiliary boundary. This will therefore include the entire auxiliary geometry, plus, if the RT surface is non-empty, the domain of dependence of any spacelike region bounded by the RT surface alone.

The RT surface, and entanglement wedge, of the \emph{combination} of a boundary region and a nonholographic system was previously considered in \cite{hayden2018learning}. It was argued there that the RT surface should be the minimal generalised entropy quantum extremal surface, homologous to the boundary region, defined with the nonholographic system automatically included as part of the fields in $S_\text{bulk}$ (see for example eqn. 4.14 of \cite{hayden2018learning}). This was justified by similar arguments to those above (see footnote 15 of \cite{hayden2018learning}). In the special case considered here where the boundary region is empty and so we only have a nonholographic system, this rule leads to the same conclusions that we reached above.

We have already argued that the RT surfaces for $\mathcal{H}_{CFT}$ and $\mathcal{H}_\text{rad}$ satisfy the same (trivial) homology constraint. Moreover, for any given candidate RT surface, the resulting entanglement wedges for $\mathcal{H}_\text{CFT}$ and $\mathcal{H}_\text{rad}$ are complementary within a Cauchy slice (of the original black hole geometry plus the auxiliary geometry). Since the overall state of the bulk modes is pure, the bulk entropy associated to each entanglement wedge will always be the same. It follows that a quantum extremal surface with respect to $\mathcal{H}_\text{CFT}$ will also be a quantum extremal surface with respect to $\mathcal{H}_\text{rad}$ and vice versa. The Ryu-Takayanagi surfaces for $\mathcal{H}_\text{CFT}$ and $\mathcal{H}_\text{rad}$ will therefore be the same.

The simplest quantum extremal surface, for both $\mathcal{H}_\text{CFT}$ and $\mathcal{H}_\text{rad}$, is, of course, the empty surface. The generalised entropy of the empty surface will be equal to the bulk entanglement entropy $\Srad$ between the Hawking radiation and the interior of the black hole. Recall that we assume that the bulk evaporation is semiclassical, so long as the black hole is large compared to the string and Planck scales. Hence the bulk entanglement entropy $\Srad$ will continue to grow, in accordance with semiclassical calculations,  until the black hole has almost completely evaporated.

Before the Page time, the empty surface will be the Ryu-Takayanagi surface. This can easily be seen using the maximin prescription.  Since the empty surface lies in every Cauchy slice, $\Srad$ upper bounds the generalised entropy of the Ryu-Takayanagi surface. However, it is also easy to find a Cauchy slice for which the empty surface has minimal generalised entropy. We simply choose a Cauchy slice that stays a small, but fixed, radial distance inside the event horizon of the black hole. Within this Cauchy slice, we cannot choose an RT surface $\chi$ that excludes the interior modes, entangled with the outgoing radiation in $\mathcal{H}_\text{rad}$, without the area $A(\chi)$ of this surface satisfying
\begin{align}
\frac{A(\chi)}{4 G_N} > \Srad.
\end{align}
The surface of minimal generalised entropy within this Cauchy slice is therefore the empty surface, with generalised entropy $\Srad$.\footnote{Assuming that the minimal generalised entropy surface is unique, it is sufficient, because of the rotational symmetry of the system, to only consider rotational symmetric candidate surfaces. However, it is not difficult to verify that the empty surface does indeed have minimal generalised entropy, within this Cauchy slice, even when we consider surfaces that are not rotationally symmetric.}
\begin{figure}[t]
\vspace{2cm}
\begin{subfigure}{.48\textwidth}
  \centering
  \vspace{-2cm}
 \includegraphics[width = 0.8\linewidth]{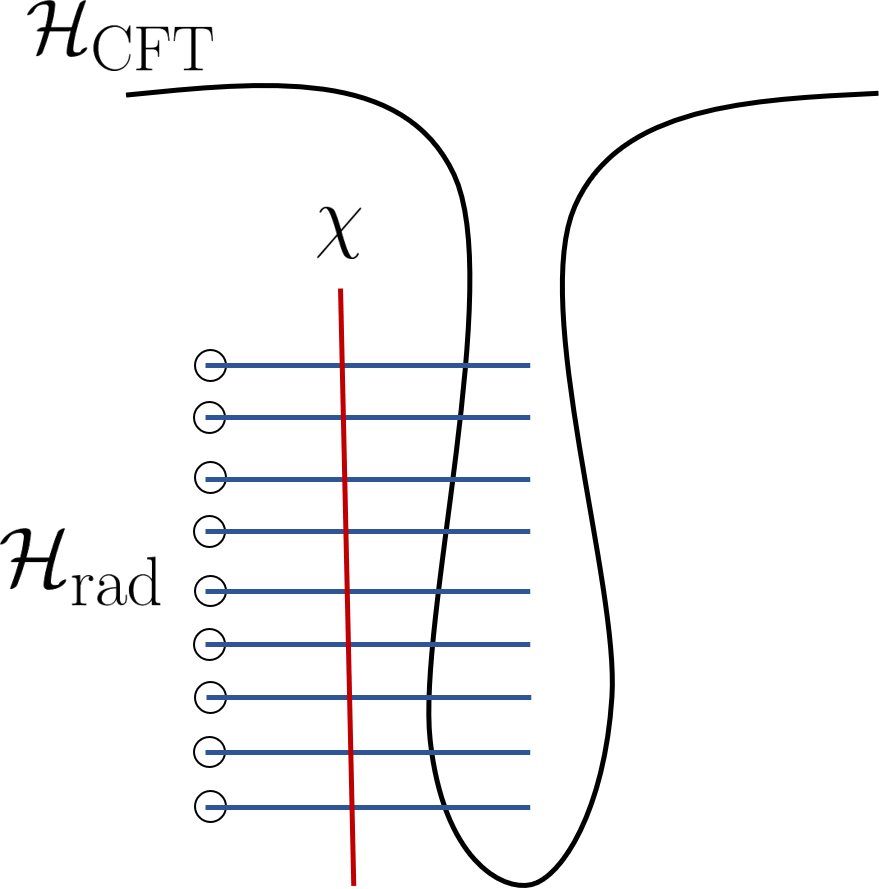}
\end{subfigure}
\begin{subfigure}{.48\textwidth}
  \centering
  \vspace{-2cm}
 \includegraphics[width = 0.7\linewidth]{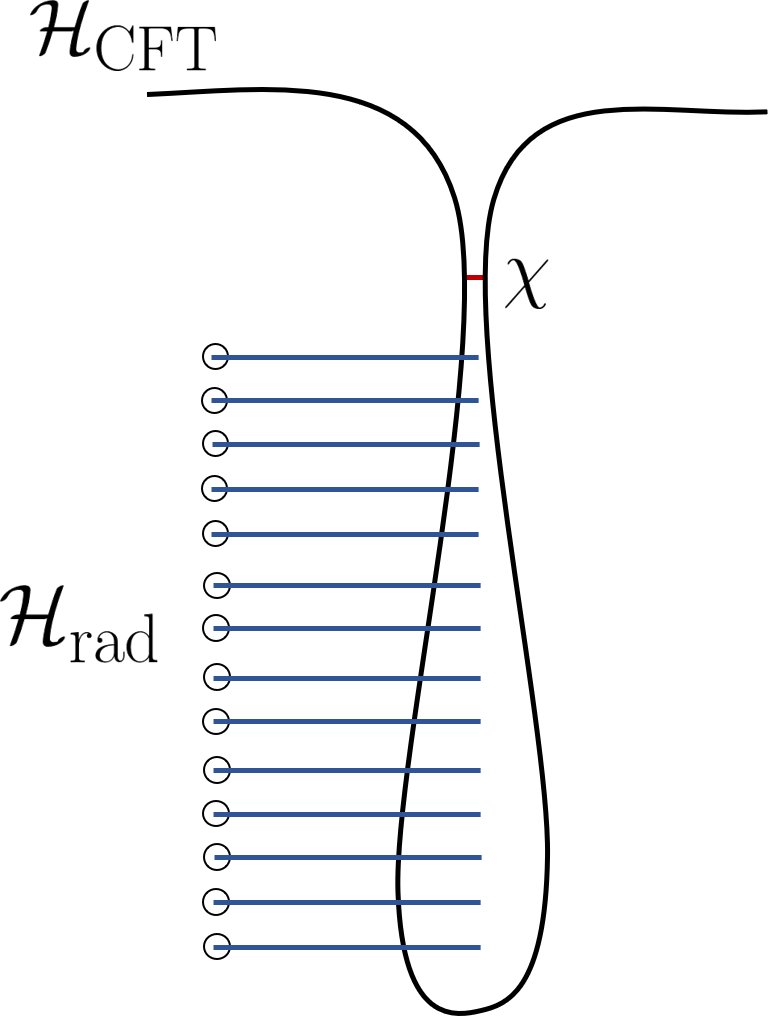}
\end{subfigure}
\centering
\caption{Schematic drawings of Cauchy slices through  the black hole interior, both before the Page time (left) and after the Page time (right). The blue lines indicate entanglement between the interior of the black hole and the reservoir $\mathcal{H}_\text{rad}$. Before the Page time, there exist Cauchy slices where the empty surface is the surface homologous to the boundary with minimal generalised entropy. It is therefore the Ryu-Takayanagi surface $\chi$. For illustrative purposes, we draw this surface cutting the entanglement between the interior and the reservoir. After the Page time, however, no such Cauchy slice exists. Within any Cauchy slice, there will always exist a surface, near the horizon and homologous to the boundary, with smaller generalised entropy. The Ryu-Takayanagi surface must become non-empty at the Page time.}
\label{fig:schematic}
\end{figure}

It follows that the RT surface is empty and the interior of the black hole is in the entanglement wedge of the boundary CFT, and not the entanglement wedge of the Hawking radiation reservoir. No information has escaped the black hole. The Hawking radiation is thermally entangled with the interior, which is encoded in the CFT, so the redused density matrix of the reservoir $\mathcal{H}_\text{rad}$ will be thermal (with appropriate greybody factors). We can also see from the Ryu-Takayanagi formula that the entanglement entropy between the CFT and the reservoir will indeed be the bulk entanglement entropy $\Srad$ between the Hawking radiation and the interior. We have derived the first half of the Page curve.

As an aside, we emphasize that, because the evaporation is thermodynamically irreversible, the bulk entanglement entropy $\Srad$ is strictly greater than $(A_\text{hor}^0 -  A_\text{hor})/ 4G_N$ where $A_\text{hor}^0$ is the initial horizon area of the black hole. This means that the Page time, which we recall is defined by 
\begin{align}
\Srad = \frac{A_\text{hor}} {4G_N},
\end{align}
occurs when the horizon area $A_\text{hor} > A_\text{hor}^0 /2$, despite commonly being called the halfway point of the evaporation \cite{page2013time}.

What happens after the Page time? The empty surface is still extremal, but it is easy to see from the maximin prescription that it \emph{cannot} be the Ryu-Takayanagi surface. In any Cauchy slice, we can construct a surface homologous to the boundary that is (a) outside the event horizon, and (b) has only slightly greater area than the current horizon area.\footnote{In fact the area of the `classical maximin surface' that we find in Section \ref{sec:classicalmaximin} gives an upper bound on this area.} Since the only source of greater-than-$O(1)$ bulk entropy is the interior modes that are entangled with radiation that escaped to $\mathcal{H}_\text{rad}$, the generalised entropy of this surface will be less than the generalised entropy of the empty surface.\footnote{One might worry that late-time Hawking radiation, which has yet to reach the boundary could provide a source of greater-than-$O(1)$ bulk entropy. However, because the Cauchy slice must be achronal, the infalling time of any surface in the Cauchy slice cannot be later than the boundary infalling time. Hence, the bulk entropy of any surface in the exterior will be at most $O(1)$ (once it has been regulated using a cut-off).} The two cases, before and after the Page time, are shown schematically in Figure \ref{fig:schematic}.

We therefore conclude that the quantum Ryu-Takayanagi surface must become non-empty at the Page time. In fact, we will show that there exists a non-empty quantum extremal surface even before the Page time. At the Page time, there is a phase transition, and the non-empty quantum extremal surface becomes the Ryu-Takayanagi surface. The main focus of this section will be on identifying the location of, and the consequences of the location of, this non-empty quantum extremal surface.

\subsection{The Classical `Maximin Surface'} \label{sec:classicalmaximin}

We begin with a warm-up. Since the bulk entropy term $S_\text{bulk}(\chi)$ is subleading compared to the area term $A(\chi)/4G_N$ in the formula for the generalised entropy, we shall initially ignore the local variation in the bulk entropy and instead attempt to simply find a `classical maximin surface' $\chi_c$.\footnote{We specify classical \emph{maximin} surface here, because, unlike the actual quantum extremal surface, this surface will not be extremal, by either the classical or quantum definition.} 

Obviously, for an evaporating black hole formed from collapse, the true classical maximin surface would be empty. We could fix this issue by temporarily assuming that the original black hole was two-sided, with one side allowed to evaporate. However, that would be taking this calculation too seriously. The actual surface that we find will very clearly not make sense as an actual Ryu-Takayanagi surface of any kind (it won't be extremal for example); however, it will turn out to correctly identify the \emph{approximate} location of the quantum extremal surface, which we shall find in the later parts of this section. We shall therefore simply ignore the question of how the black hole formed entirely, and assume that it has been evaporating forever.

Because of the assumed rotational symmetry, the classical maximin surface (and the eventual quantum extremal surface) should be rotationally symmetric. We therefore only need to consider rotationally symmetric Cauchy slices. 

We also know that the area of an infalling lightcone in an evaporating black hole is montonically decreasing. Hence, given any Cauchy slice, we can increase
\begin{align}
 \min_{\chi} \frac{A(\chi)}{4 G_N} 
\end{align}
by pushing the Cauchy slice backwards and outwards along infalling lightrays. The maximising Cauchy slice is therefore simply the past lightcone of the boundary.

In Eddington-Finkelstein coordinates, the metric of a static black hole is
\begin{align} \label{eq:staticmetric}
ds^2 = -f(r) dv^2 + 2 dv dr + r^2 d\Omega^2,
\end{align}
where the co-ordinate $v$ labels the initial Schwarzschild time of an infalling lightray and, for an uncharged AdS-Schwarzschild black hole, 
\begin{align} \label{eq:f(r)exp}
f(r) = 1 + \frac{r^2}{l^2} - \frac{16 \pi G_N M}{(d-1)\Omega_{d-1} r^{d-2}}.
\end{align}
Since our arguments should also be valid for charged black holes (at least when all the particles involved in the Hawking radiation are neutral) and BTZ black holes, we will avoid using \eqref{eq:f(r)exp} directly. 

In the semiclassical limit, the evaporation process is very slow. Over any fixed range of infalling times, the metric will approach the metric of a static black hole of some fixed mass $M$. We can therefore approximate the metric of an evaporating black hole by a static black hole with a mass $M$, and hence Schwarzschild radius $r_s$ (defined by $f(r_s) = 0$), that is slowly varying with infalling time $v$; this is known as an ingoing Vaidya metric.\footnote{An ingoing Vaidya metric is only a simple approximation of the actual semiclassical metric of an evaporating black hole; for example, at large radii compared to the Schwarzschild radius, the metric instead resembles an \emph{outgoing} Vaidya metric. For a detailed calculation of the metric of an evaporating black hole in flat space see \cite{abdolrahimi2016ingoing}. In the limit $G_N \to 0$, the only deviation of this metric from the static metric that will be relevant for our calculations is the infalling-time dependence of the black hole mass.}

The radius $r_{l.c.}(v)$ of an outgoing lightcone satisfies
\begin{align} \label{eq:outgoinglc}
\frac{dr_{l.c.}}{dv} = \frac{f(r)}{2} \approx \frac{2\pi}{\beta} (r - r_s),
\end{align}
where the approximation is valid in the near-horizon region and the inverse temperature $\beta = 4 \pi / f'(r_s)$. If $r' = r_{l.c.} - r_s$, we have
\begin{align} \label{eq:outgoingdiffeq}
\frac{dr'_{l.c.}}{dv} = \frac{dr_{l.c.}}{dv} - \frac{d r_s}{d v} \approx \frac{2\pi}{\beta} r'_{l.c.} - \frac{d r_s}{d v},
\end{align}
At leading order in the semiclassical limit, we can assume that the inverse temperature $\beta$ and the evaporation rate $d r_s / d v$ are constant.

Integrating \eqref{eq:outgoingdiffeq}, we find
\begin{align} \label{eq:arbitraryoutgoinglc}
r_{l.c.} = r_s + C \,e^{\frac{2\pi}{\beta}v} + \frac{\beta}{2 \pi} \frac{d r_s}{d v},
\end{align}
for some arbitrary constant $C$. If $C>0$, the lightcone will eventually escape the black hole, even if its radius is initially decreasing. In contrast, if $C<0$, the outgoing lightcone will eventually fall into the singularity.

The causal, or event, horizon of the black hole is defined as the boundary of the causal past of future asymptotic infinity. In this case, up to subleading corrections from the time dependence of $dr_s/dv$ and $\beta$, it is the outgoing lightcone \eqref{eq:outgoingdiffeq} with $C=0$.\footnote{We are assuming here that nothing to radical happens (such as the black hole becoming a white hole) when the black hole has almost entirely evaporated and the semiclassical description breaks down.} Its radius $r_\text{hor}$ is therefore given by
\begin{align} \label{eq:rhor}
r_\text{hor} = r_s + \frac{\beta}{2 \pi} \frac{d r_s}{d v}.
\end{align}
Since $d r_s/d v < 0$, this is inside the timelike apparent horizon $r_s$.\footnote{For our purposes, the apparent horizon is the radius at which the area of an outgoing lightcone is locally constant.}

If we instead choose $C=r_s$, we have
\begin{align} \label{eq:outgoinglc2}
r_{l.c.} = r_s + r_s \,e^{\frac{2\pi}{\beta}v} + \frac{\beta}{2 \pi} \frac{d r_s}{d v} = r_\text{hor} + r_s\,e^{\frac{2\pi}{\beta}v},
\end{align}
and the lightray escapes the near horizon region at $v = O(\beta)$.

The radius $r_{l.c.}$, and hence the area $\Omega_{d-1}r^{d-1}$, of the past lightcone reaches a minimum and begins increasing when it reaches the apparent horizon $r_s$. This occurs at
\begin{align} \label{eq:vclassical}
v = - \frac{\beta}{2\pi} \log \frac{r_s}{\beta\, |d r_s/d v|} + O(\beta).
\end{align}
For small AdS-Schwarzschild black holes,\footnote{For large AdS black holes, there is no clear distinction between radiation that is inside the so-called `zone' near the horizon and radiation that has escaped to the boundary. As a result, a large AdS black hole does not have a well-defined evaporation rate, even with absorbing boundary conditions; it depends on the details of the evaporation process.} the only relevant lengthscale is the Schwarzschild radius $r_s$.\footnote{\label{foot:nearextremal} In small, near-extremal Reissner-Nordstr\''{o}m black holes, the inverse temperature $\beta$ is parametrically large compared to the Schwarzschild radius $r_s$. The approximation in \eqref{eq:outgoinglc} is therefore only valid when $r - r_s \ll r_s^2 / \beta$. At larger radii, we have $dr/dv = O(1/f(r)) = O(r_s^2/(r - r_s)^2)$ and so an outgoing lightray escapes the black hole in an $O(\beta)$ time. Hence, \eqref{eq:vclassical} becomes $$v = - \frac{\beta}{2\pi} \log \frac{r_s^2}{\beta^2\, |d r_s/d v|} + O(\beta).$$} Hence, it is easy to see using dimensional analysis and the fact that $d r_s/d v = O(G_N)$ that
\begin{align}
v = - \frac{\beta}{2 \pi} \log S_{BH} + O(\beta).
\end{align}
We have therefore found that the classical maximin surface lies on the classical apparent horizon, one scrambling time into the past.\footnote{In addition to the $O(\beta)$ corrections, if the Schwarzschild radius $r_s$ is parametrically small in AdS units, we also need to add the infalling time $\pi\, l_{AdS}$ for the outgoing lightcone to get from the black hole to the boundary, after it has escaped the near horizon region.} 

This is very hopeful: the Hayden-Preskill decoding criterion says that a small diary, thrown into the black hole after the Page time, should be reconstructable from the Hawking radiation after waiting for the scrambling time. Our calculation suggests that this is because the entanglement wedge of the Hawking radiation reservoir $\mathcal{H}_\text{rad}$ now contains the diary.

However, there are two major problems with this classical maximin surface $\chi_c$ as a candidate Ryu-Takayanagi surface. 
Firstly, the surface $\chi_c$ is not a classical extremal surface. It has extremal area with respect to deformations that stay on the past lightcone, but it certainly does not have extremal area if we allow deformations that move the surface away from the lightcone. It therefore cannot be the Ryu-Takayanagi surface according to the HRT extremal surface prescription, even though the HRT and maximin prescriptions are supposed to be equivalent \cite{wall2014maximin}.

Secondly, if the surface $\chi_c$ was actually the Ryu-Takayanagi surface, the entanglement wedge of the boundary CFT would not contain the causal future of the boundary. This would be highly problematic because the forward time evolution of the CFT is deterministic, and so the future boundary, although not the past boundary, is in the boundary domain of dependence. The entanglement wedge of the CFT would not contain its causal wedge.

Both problems have the same cause and will have the same solution. The cause is that the evaporating black hole spacetime violates the null energy condition. The null energy condition is needed to prove that the classical maximin surface is the same as the classical extremal surface \cite{wall2014maximin}. It is also needed to prove that the classical entanglement wedge contains the causal wedge \cite{wall2014maximin}.

\subsection{The Quantum Extremal Surface} \label{sec:extremal}
The problem of quantum effects leading to spacetimes that violate the null energy condition was the original reason for the conjecture that the Ryu-Takayanagi surface should be a quantum extremal surface, rather than a classical extremal surface \cite{engelhardt2015quantum}. So, we should definitely be hopeful that all these problems will go away, once we fully include the effects of the bulk entropy term.

Indeed, heuristically, it is easy to see that the bulk entropy term can push the Ryu-Takayanagi surface away from the past lightcone. If the RT surface was exactly on the past lightcone, no outgoing modes would be included in the entanglement wedge of the CFT. Since the entropy of the outgoing modes is divergent, moving the RT surface a small distance inside the lightcone should increase the bulk entropy by a formally infinite amount. This strongly suggests that the quantum RT surface, in the maximin prescription, will be stabilised a small radial distance away from the past lightcone, creating an actual quantum extremal surface.

Unfortunately, actually calculating the bulk entropy is complicated by the presence of non-trivial greybody factors. Because the black hole spacetime is curved, outgoing modes close to the black hole do not necessarily escape to infinity. Instead, there is a non-trivial scattering process. The curved spacetime wave equation can be rewritten as a flat space wave equation with a potential barrier. This potential barrier lies an $O(r_s)$ distance away from the black hole horizon and is higher for modes with large angular momentum, causing the Hawking radiation to be dominated by modes with $O(1)$ angular momentum. The region inside the potential barrier is known as the zone.

Within the zone, the Hawking radiation is truly black body radiation at the black hole temperature.\footnote{A more precise statement is that each angular momentum mode looks like two-dimensional black body radiation. For higher-dimensional black bodies, only angular momentum modes with $|J| \lesssim T r$ are significantly excited, and only modes with $|J| \ll T r$ look like two-dimensional thermal radiation. In contrast, sufficiently close to the horizon, Rindler modes with $|J| \gg T r_s = O(1)$ will be excited.} However, the probability of a Hawking mode escaping depends on its angular momentum and, importantly, on its Schwarzschild energy. This probability is known as a greybody factor. 

Outgoing modes in the entanglement wedge of the CFT are entangled both with modes further in towards the black hole, and with outgoing modes further out -- outside of the past lightcone. The non-trivial greybody factors mean that the outgoing modes outside the past lightcone are related to later \emph{ingoing} modes, which are also in the entanglement wedge of the CFT. This dramatically complicates any explicit calculation of the bulk entropy.

As a simple solution to this problem, we shall therefore temporarily assume that the Hawking radiation is extracted from deep inside the zone, close to the horizon, before the mixing of ingoing and outgoing modes occurs. (We will reintroduce the greybody factors in Section \ref{sec:greybody}.) By doing so, we are effectively reducing the problem to a calculation in the two-dimensional effective theory that governs the near horizon region \cite{strominger1995houches}. For this reason, the results we find in this section will be essentially identical (once certain constants are fixed appropriately) to those found independently in an explicitly two-dimensional model in \cite{almheiri2019entropy}.

Extracting the radiation from close to the horizon obviously involves changing the dynamics of the system compared to the original procedure, where the radiation was extracted near the boundary. In particular, the boundary Master equation, and its purification using the reservoir $\mathcal{H}_\text{rad}$, will now involve reconstructions of operators deep in the bulk, which are highly non-local from a boundary perspective. 

We emphasize, however, that there is nothing fundamentally unphysical about this. The outgoing modes will still be extracted at a distance from the horizon (and hence an energy scale) that is fixed in AdS units as $G_N \to 0$; the distance simply needs to be small compared to the Schwarzschild radius $r_s$. We are therefore still well within the domain of validity of the bulk effective field theory.

It is important to note that the closer to the horizon we extract the Hawking radiation, the larger the number of angular momentum modes that are excited.\footnote{Another way of saying this is that modes with larger angular momentum are reflected back into the black hole at a smaller, but still finite, distance from the horizon.} Similarly, increasingly massive fields will be excited very close to the horizon. This can be seen from the explicit form of the potential barrier in tortoise coordinates \cite{harlow2016jerusalem}.

We will assume that we extract some fixed finite number of angular momentum modes for each field, and that these fields are extracted sufficiently close to the horizon that their mass and angular momentum can be safely ignored. In effect, we are changing the dynamics of the theory such that all greybody factors are either zero or one, depending on the angular momentum. We will also assume that all the light fields are free.

Close to the horizon, the spacetime can be approximated by $\mathbbm{R}^{1,1} \times S^{d-1}$ where the radius of the sphere $S^{d-1}$ is the Schwarschild radius $r_s$. Each angular momentum mode acts as an independent free field in an effective two-dimensional theory, with a Kaluza-Klein mass $m_{KK}^2 = L^2 / r_s^2$, which can be ignored at the lengthscales of interest.

Let the number of two-dimensional bosonic modes $N_b$ and fermionic modes $N_f$. The $(1+1)$-dimensional Stefan-Boltzman law \cite{landsberg1989stefan} states that the rate of energy loss from the black hole is given by
\begin{align} \label{eq:energyflux}
\frac{d M}{d v} =  \frac{c_\text{evap}\, \pi}{12 \,\beta^2},
\end{align}
where $c_\text{evap} = N_b + N_f/2$.\footnote{Of course, since Hawking radiation is stochastic, this is only the average rate of energy loss. However, so long as we consider timescales that are large compared to the thermal time $\beta$, the average energy change should be large compared to the fluctuations in the energy change, which can therefore be safely ignored. In our case, the relevent timescale is the scrambling time, which is indeed very large compared to the thermal time $\beta$ in the semiclassical limit. We can also suppress the fluctuations by taking the limit where $c_\text{evap}$ is large.} The first law of black hole thermodynamics says that $\beta dM = dA_\text{hor}/4G_N$. Hence
\begin{align} \label{eq:dAhor/dv}
\frac{d A_\text{hor}}{d v} = \frac{c_\text{evap}\, \pi \,G_N}{3\, \beta},
\end{align}
and
\begin{align} \label{eq:drs/dv}
\frac{d r_s}{d v} = \frac{c_\text{evap} \,\pi\, G_N}{3 \,\beta \,(d-1)\, r_s^{d-2}\, \Omega_{d-1}}.
\end{align}
Substituting \eqref{eq:drs/dv} into \eqref{eq:vclassical} and dropping $O(\beta)$ terms,  we find that the classical maximin surface for this spacetime occurs at\footnote{As noted in Footnote \ref{foot:nearextremal}, for near-extremal black holes, \eqref{eq:vclassical} becomes $$v_c = - \frac{\beta}{2\pi} \log \frac{r_s^2}{\beta^2\, |d r_s/d v|} + O(\beta).$$ Hence, substituting \eqref{eq:drs/dv}, we find that $$v_c = - \frac{\beta}{2\pi} \log \Delta S_{BH} + O(\beta),$$ where $\Delta S_{BH} = S_{BH} - S_{BH}^0 =  O(r_s S_{BH} / \beta)$ with $S_{BH}^0$ the entropy of an extremal black hole with the same charge. The location we find for the non-empty extremal surface is therefore consistent the similar calculations, for two-sided black holes in JT gravity, done in \cite{almheiri2019entropy}. The $O(\beta)$ corrections are also the same in both calculations \cite{brown2019python}.}
\begin{align} \label{eq:vc}
v_c = -\frac{\beta}{2 \pi} \log \frac{S_{BH}}{c_\text{evap}} + O(\beta).
\end{align}

To calculate the quantum extremal surface, we also need to calculate how the bulk entanglement entropy depends on the location of the extremal surface.\footnote{Since the entanglement entropy of gravitons is not well understood, we shall assume here that no graviton modes are extracted into the reservoir $\mathcal{H}_\text{rad}$. One would hope that, if we did understand the entanglement entropy of gravitons, we would find that they would contribute to the location of the quantum extremal surface in a similar way to other bulk modes.} Since we are assuming that all the relevant fields are free and effectively massless at the lengthscales of interest, the ingoing and outgoing modes are decoupled. The total bulk entropy is therefore simply the sum of the bulk entropies of the ingoing and outgoing modes.

\begin{figure}[t]
\includegraphics[width = 0.5\linewidth]{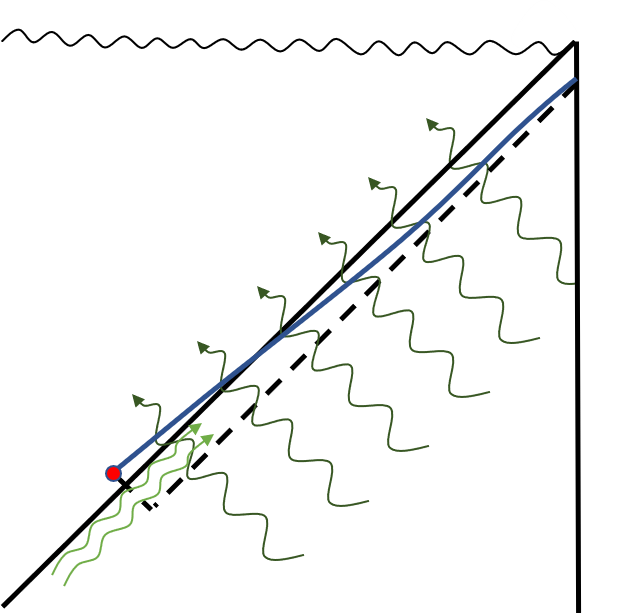}
\centering
\caption{The bulk entropy of the region, shown in blue, between the Ryu-Takayanagi surface and the boundary can be decomposed into the entropy of the ingoing and outgoing modes. The ingoing modes are in the infalling vacuum, and the bulk region includes ingoing modes spread over approximately the scrambling time, which diverges in the semiclassical limit. This means that the gradient, in units of infalling time, of the entropy of the infalling modes tends to zero. In contrast, as the RT surface approaches the past lightcone of the boundary, there will be a negative logarithmic divergence in the (renormalised) entropy of the outgoing modes. This divergence should stabilise the location of the quantum extremal surface a small distance away from the outgoing lightcone.}
\label{fig:cauchy}
\end{figure}

The infalling modes are in the vacuum state with respect to the infalling time $v$. Moreover, if we assume (correctly) that the quantum extremal surface is close to the classical maximin surface, the entanglement wedge of the boundary CFT will contain infalling modes spread over approximately the scrambling time, which diverges as $G_N \to 0$. Since the entanglement entropy of the vacuum state grows only logarithmically with system size, we can treat the entanglement entropy of the infalling modes as approximately independent of the location of the extremal surface, so long as  the cut-off at the extremal surface is fixed in units of infalling time $v$. By this, we really mean that the cut-off is equal to $\varepsilon \partial/ \partial v$ for some constant $\varepsilon$.

What about the outgoing modes? The only relevant modes are the modes that are extracted into the reservoir $\mathcal{H}_\text{rad}$. At sufficiently short lengthscales, the entropy of these modes will be given by the Minkowski vacuum formula \cite{calabrese2004entanglement, calabrese2009entanglement}
\begin{align} \label{eq:bulkentropyoutgoing}
S = \frac{c_\text{evap}}{6}\log\left(\frac{r_{lc}(v) - r}{\sqrt{\varepsilon_1 \varepsilon_2}}\right),
\end{align}
where $r_{lc}$ is the radius of the outgoing lightcone and $\varepsilon_1$ and $\varepsilon_2$ are the cut-offs at the quantum extremal surface and the outgoing lightcone respectively, in units of the radius $r$.\footnote{Here we are implicitly using the fact that the radius $r$, like any smooth, non-singular coordinate, is an approximate affine coordinate along ingoing lightrays at sufficiently small distance scales.}

The cut-offs $\varepsilon_1$ and $\varepsilon_2$ are crucial to the calculation and so we take some time to discuss them in detail. The cut-off $\varepsilon_1$ at the quantum extremal surface is unphysical. Since the bulk entropy would otherwise be formally divergent, we need to cut-off the bulk degrees of freedom at some fixed proper lengthscale. The lengthscale chosen is arbitrary; in the calculation of physical quantities, such as entanglement entropies in the boundary CFT, the cut-off dependence should be cancelled by the scale dependence of the couplings in the effective gravitational theory due to renormalisation.\footnote{Of course, even boundary entanglement entropies are not actually well defined because of UV-divergences in the boundary theory (which correspond to IR-divergences in the bulk theory). However the mutual information between two boundary regions, for example, is a well defined regulator-independent finite quantity.}

Of course, the cut-off $\varepsilon_1$ is not itself a proper lengthscale. Since the radius $r$ is a null coordinate, the proper length of the cut-off $\varepsilon_1$ is zero. The proper cut-off $\varepsilon_\text{prop}$ is instead determined by the \emph{inner product} of the cut-off $\varepsilon_1$ with the cut-off at the extremal surface on the \emph{ingoing} modes. Specifically
\begin{align}
\varepsilon_\text{prop} = \sqrt{g\left(\varepsilon_1 \frac{\partial}{\partial r}, \varepsilon_\text{in} \frac{\partial}{\partial v}\right)},
\end{align}
where $g$ is the metric and $\varepsilon_\text{in}$ is the cut-off on the ingoing modes in units of $v$.

When we claimed that the entropy of the ingoing modes was approximately constant, we implicitly assumed that the cut-off $\varepsilon_\text{in}$ was constant. Since, to leading order, the metric is given by
\begin{align}
ds^2 = 2 dv dr,
\end{align}
everywhere in the near horizon region, this means that the cut-off $\varepsilon_1$ should also be constant, so that the proper cut-off $\varepsilon_\text{prop}$ is constant in AdS units. 

The physical status of the lightcone cut-off $\varepsilon_2$ is very different. The lightcone cut-off $\varepsilon_2$ is related to the cut-off $\varepsilon_0$ on the Schwarzschild frequency of the modes that we extract into the reservoir $\mathcal{H}_\text{rad}$. Unlike the cut-off at the extremal surface, this is a physical cut-off that depends on the dynamics that we use to extract outgoing modes into $\mathcal{H}_\text{rad}$. 

However the cut-off $\varepsilon_2$ is blueshifted as the outgoing modes evolve back in time. If we parallel transport the cut-off $\varepsilon_2 \partial/ \partial r$ backwards along the past lightcone, we find that
\begin{align}
0 &= \nabla_v \left(\varepsilon_2 \frac{\partial}{\partial r}\right),
\\&= \left[\partial_v \varepsilon_2 + \Gamma^r{}_{vr} \,\varepsilon_2 \right] \frac{\partial}{\partial r},
\\& = \left[\partial_v \varepsilon_2 - \frac{2 \pi}{\beta} \varepsilon_2 \right] \frac{\partial}{\partial r},
\end{align}
where we approximated the metric by its leading order (static) approximation \eqref{eq:staticmetric} and used the fact that, in the near horizon region, $f'(r) \approx f'(r_s) = 4 \pi/ \beta$. We therefore find that the cut-off $\varepsilon_2$ is related to the (constant) cut-off $\varepsilon_0$ on the Schwarzschild energy of the extracted modes by
\begin{align} \label{eq:cutoffscaling}
\varepsilon_2 \propto e^{\frac{2\pi}{\beta}v} \varepsilon_0.
\end{align}

We emphasize that the infalling-time dependence of this cut-off is purely a product of the coordinate system that we are using. In Section \ref{sec:greybody}, we do a more general calculation, which includes greybody factors, in Kruskal-Szekeres-like coordinates, where there is no blueshifting, and so the cut-off at the lightcone would be constant.\footnote{In many ways, the nicest coordinates to use for the problem are the outgoing Kruskal coordinate $U$ together with the infalling time $v$. However, the author is too lazy to rewrite all the calculations in these coordinates. See \cite{youtube}.} The calculations done here are rederived as a special case, without any reference to exponentially small cut-offs. For the moment, however, we shall continue to use Eddington-Finkelstein coordinates, which have a more natural physical interpretation. For pedagogical purposes, in Appendix \ref{app:rindler}, we also give an example of a simple Rindler space calculation that illustrates the importance of taking into account the coordinate dependence of cut-offs.

\begin{figure}[t]
\begin{subfigure}{.48\textwidth}
  \centering
 \includegraphics[width = 0.58\linewidth]{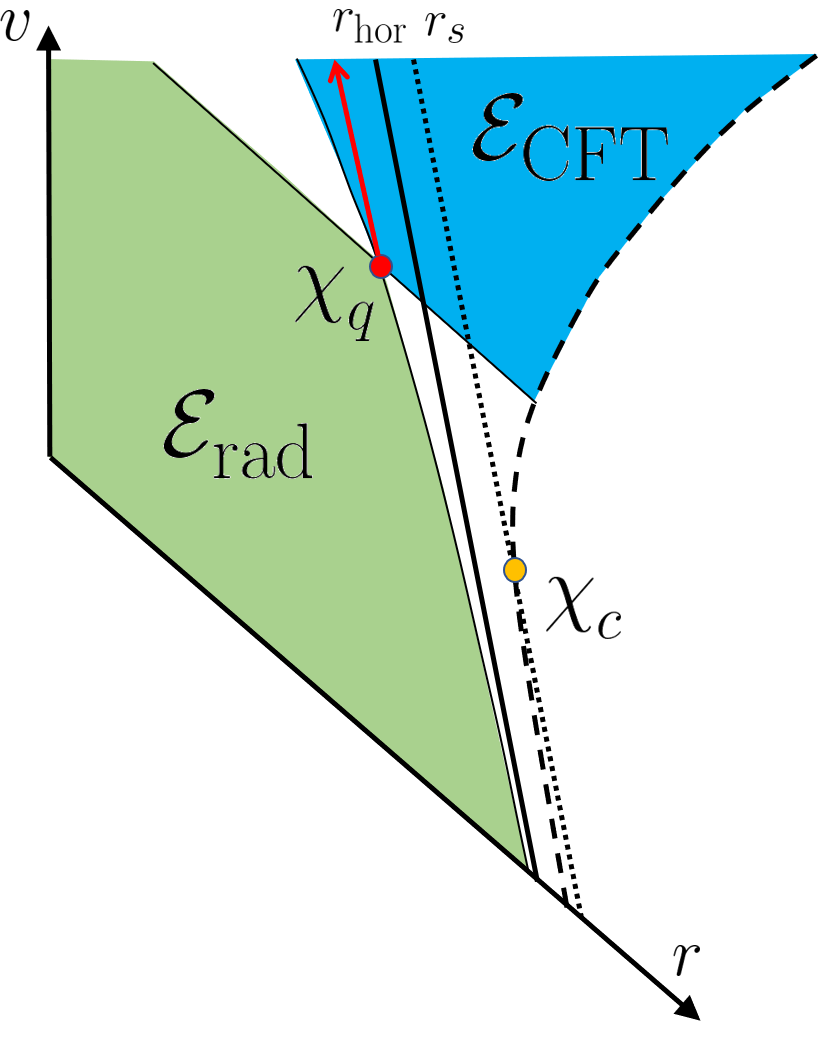}
\end{subfigure}
\begin{subfigure}{.48\textwidth}
  \centering
  \vspace{-2cm}
 \includegraphics[width = 0.9\linewidth]{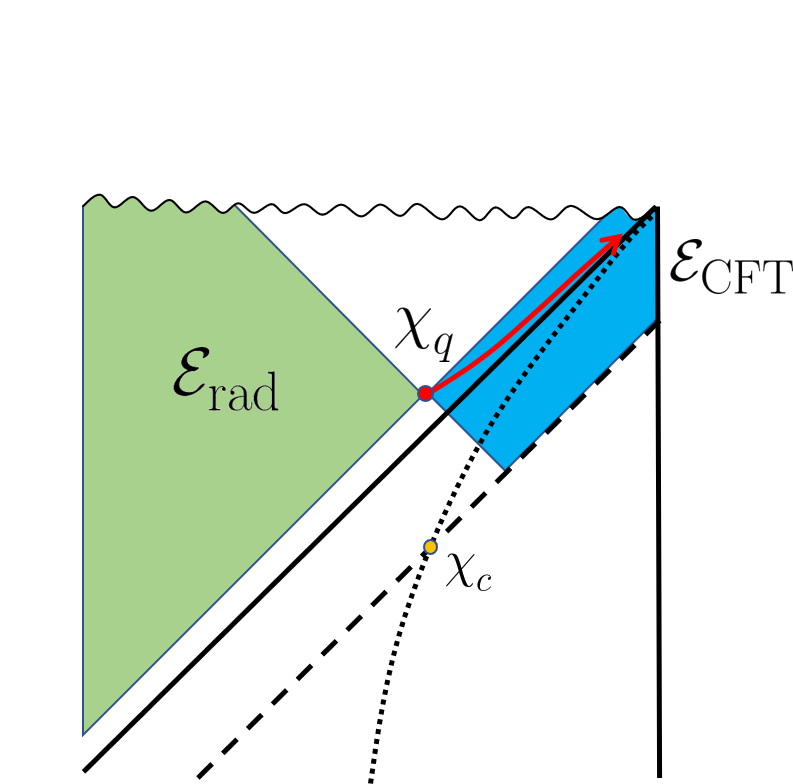}
\end{subfigure}
\centering
\caption{The quantum Ryu-Takayanagi surface $\chi_q$, the classical maximin surface $\chi_c$, and the entanglement wedges $\mathcal{E}_\text{rad}$ and $\mathcal{E}_\text{CFT}$ of the reservoir and CFT, in Eddington-Finkelstein coordinates (left) and in a Penrose diagram (right). In the interests of simplicity, the Penrose diagram does not include the post-evaporation region, which would be in the top right. The classical maximin surfaces lies at the intersection of the past lightcone (dashed) with the apparent horizon $r_s$ (dotted), which is outside the event horizon. The quantum RT surface, in contrast, lies slightly inside the event horizon. Much of the interior is in the entanglement wedge $\mathcal{E}_\text{rad}$ of the reservoir (green), although part of the interior still lies in the entanglement wedge $\mathcal{E}_\text{CFT}$ of the CFT (blue). As the black hole continues to evaporate, the RT surface moves forward in infalling time along a spacelike trajectory, following the red arrow. On timescales that are small compared to the evaporation time, it remains a fixed radial distance inside the event horizon.}
\label{fig:bh_eddington}
\end{figure}
If we drop terms that are independent of position, it follows from \eqref{eq:bulkentropyoutgoing} that the bulk entropy is given by
\begin{align} \label{eq:outgoingentropy}
S_\text{bulk} =  \frac{c_\text{evap}}{6}\log\left(r_{lc}(v) - r\right) - \frac{c_\text{evap} \pi v}{6 \beta} + \dots
\end{align}
Because of the rotational symmetry, it is sufficient to show that the surface is extremal under perturbations that preserve the symmetry. Varying the infalling time $v$, while holding the radius $r$ fixed, we find,
\begin{align}
0 &= \left. \frac{\partial S_\text{bulk}}{\partial v} \right|_r + \frac{1}{4G_N} \left. \frac{\partial A}{\partial v}\right|_r,
\\& = \frac{\partial S_\text{bulk}}{\partial v},
\\& = \frac{d r_{lc}/d v}{6(r_{lc} - r)} - \frac{\pi}{6 \beta},
\\r_{lc} - r & = \frac{\beta}{\pi}\frac{d r_{lc}}{d v}
\\& = 2(r_{lc} - r_s).  \label{eq:dSdv}
\end{align}
In the last equality, we used \eqref{eq:outgoinglc}. Varying the radius $r$, at fixed infalling time $v$, we find,
\begin{align}
0 &= \left. \frac{\partial S_\text{bulk} }{\partial r}\right|_v + \frac{1}{4G_N} \left.\frac{\partial A}{\partial r}\right|_v,
\\&= -\frac{c_\text{evap}}{6(r_{lc} - r)} + \frac{(d-1)\Omega_{d-1} r_s^{d-2}}{4G_N},
\\r_{lc} - r &=\frac{2G_N c_\text{evap}}{3(d-1) \Omega_{d-1} r_s^{d-2}}, \label{eq:dSdr}
\\&= \frac{2\beta}{\pi} \frac{d r_s}{d v} = 4(r_s - r_\text{hor}), \label{eq:rlcr=4rsrhor}
\end{align}
where in the last equality we have used \eqref{eq:drs/dv}. The quantum extremal surface therefore lies at 
\begin{align} \label{eq:vq}
v = v_c + \frac{\beta}{2\pi} \log 3 = - \frac{\beta}{2 \pi} \log \frac{S_{BH}}{c_\text{evap}} + O(\beta),
\end{align}
where $v_c$ is the infalling time of the classical maximin surface found in \eqref{eq:vc}, and
\begin{align} \label{eq:rq}
r = r_s - \frac{\beta}{\pi} \frac{\partial r_s}{\partial v} = r_\text{hor} - (r_s - r_\text{hor}).
\end{align}
The extremal surface is twice as far inside the apparent horizon as the event horizon because the entropy of the Hawking radiation produced by the black hole is twice the Bekenstein-Hawking entropy lost by the black hole, which can be easily seen by noting that the energy $E$ and entropy $S$ of black body radiation in two spacetime dimensions are related by 
\begin{align}
E = \frac{1}{2} T S.
\end{align}
The location of the quantum extremal surface is shown in Figure \ref{fig:bh_eddington}, together with the classical maximin surface and the two entanglement wedges. 

It is important to note that the entanglement wedge of $\mathcal{H}_\text{CFT}$ is bounded by the past lightcone, rather than by the boundary of anti-de Sitter space. This is because the boundary conditions are not deterministic when evolving backwards into the past, without access to $\mathcal{H}_\text{rad}$. The region outside the past lightcone is therefore not in the bulk domain of dependence of a spacelike surface connecting the RT surface to the boundary.

The causal wedge is the intersection of the causal past and future of the boundary domain of dependence. Since the boundary time evolution is irreversible, only the future of the boundary is in its domain of dependence. The causal wedge is therefore the intersection of the exterior of the black hole with the future of an infalling lightcone from the boundary. The entanglement wedge contains the causal wedge, as expected.

\begin{figure} [t]
\vspace{-1.2cm}
\centering
\begin{subfigure}{.48\textwidth}
\vspace{-0.6cm}
  \centering
 \includegraphics[width = 0.8\linewidth]{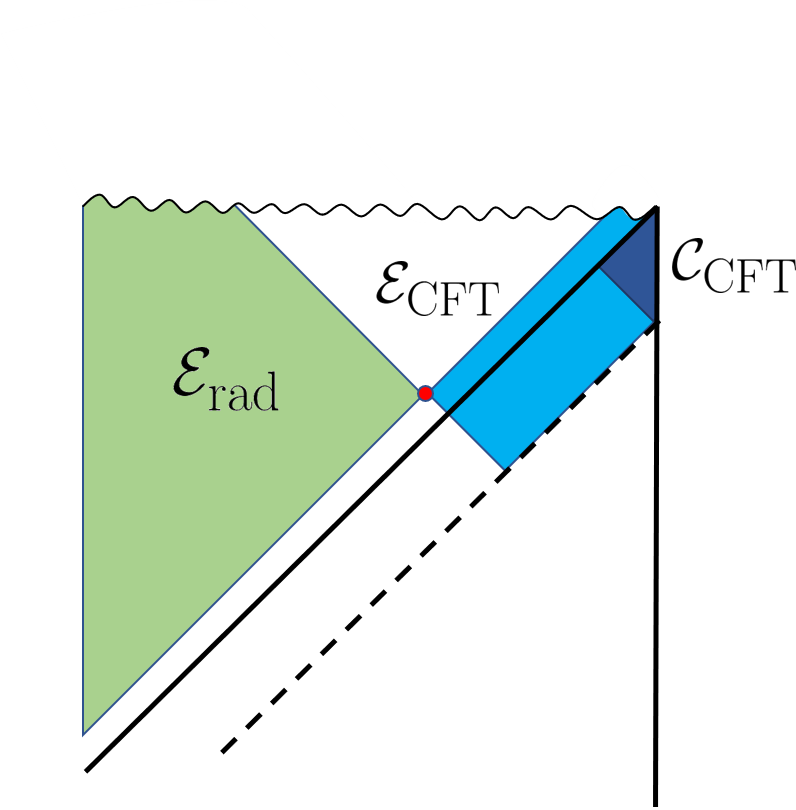}
\end{subfigure}
\begin{subfigure}{.48\textwidth}
 \includegraphics[width = 0.9\linewidth]{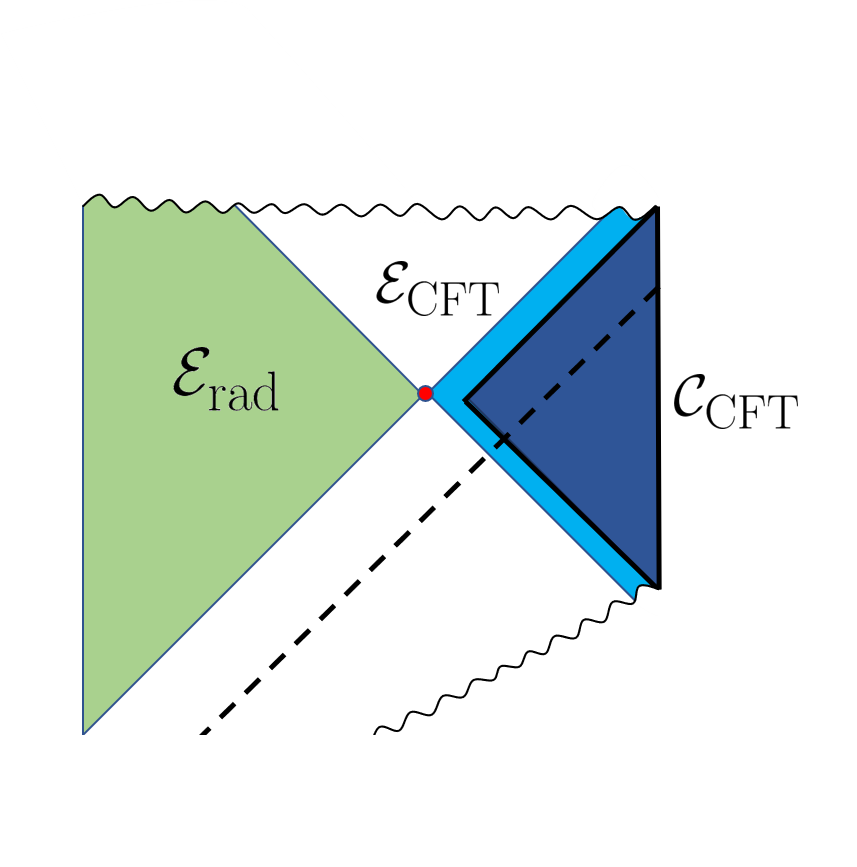}
 \centering
\end{subfigure}
\caption{Because the absorbing boundary conditions are only deterministic when evolving forwards in time, the domain of dependence of the boundary is the future boundary. The entanglement wedge $\mathcal{E}_\text{CFT}$ of the CFT (light blue) contains the causal wedge $\mathcal{C}_\text{CFT}$ (dark blue). When time is reversed using standard reflective boundary conditions, the entanglement wedge still contains the causal wedge because the backreaction on the geometry creates a white hole.}
\label{fig:bh_penrose}
\end{figure}
Of course, we can simply evolve the system back in time using standard reflective boundary conditions; on the boundary, this corresponds to using the ordinary, uncoupled Hamiltonian for the CFT. For this spacetime, both the future and past of the boundary are in its domain of dependence and so the entire exterior of the black hole will be in the causal wedge. One might worry that the entanglement wedge will not contain the causal wedge.

Specifically, in our original spacetime, a signal could easily travel backwards in time from the entanglement wedge of $\mathcal{H}_\text{rad}$ to the boundary. If this was still possible under an evolution where the two systems $\mathcal{H}_\text{rad}$ and $\mathcal{H}_\text{CFT}$ were uncoupled, then we would have a serious problem.

However this fails to take into account the backreaction on the spacetime geometry that happens when we change the dynamics.\footnote{See \cite{almheiri2018holographic} for discussion of essentially the same effect in terms of the dynamics of the boundary particle in $(1+1)$-dimensional gravity.} It is easy to see that the geometry of the spacetime must change when we change the boundary conditions. Without the Hawking radiation from $\mathcal{H}_\text{rad}$, the black hole cannot grow indefinitely as we evolve the state backwards into the past; it does not have the energy to do so. 

Instead, the discontinuity in the outgoing modes will create a shell of high energy density at the past lightcone; the energy of this shell will be proportional to the number of modes $c_\text{evap}$ that were extracted. As the shell evolves back into the past, it will be blueshifted, creating large backreaction on the geometry once it is a distance $O(c_\text{evap}\, G_N)$ from the black hole. This will create a white hole with Schwarzschild radius $O(c_\text{evap}\,G_N)$ larger than the original black hole. The Ryu-Takayanagi surface will now lie slightly inside the bifurcation horizon of the new spacetime and the entanglement wedge of the CFT will continue to contain the causal wedge.

\subsection{Hayden-Preskill and the Page Curve} \label{sec:haydenpage}
In this subsection, we show how the Ryu-Takayanagi surface, calculated in Section \ref{sec:extremal}, explains properties of black hole evaporation, such as the Hayden-Preskill decoding criterion and the Page curve, that have been conjectured based on simple toy models of black hole evaporation.
\begin{figure}[t]
\vspace{-1cm}
\begin{subfigure}{.48\textwidth}
\includegraphics[width = 0.8\linewidth]{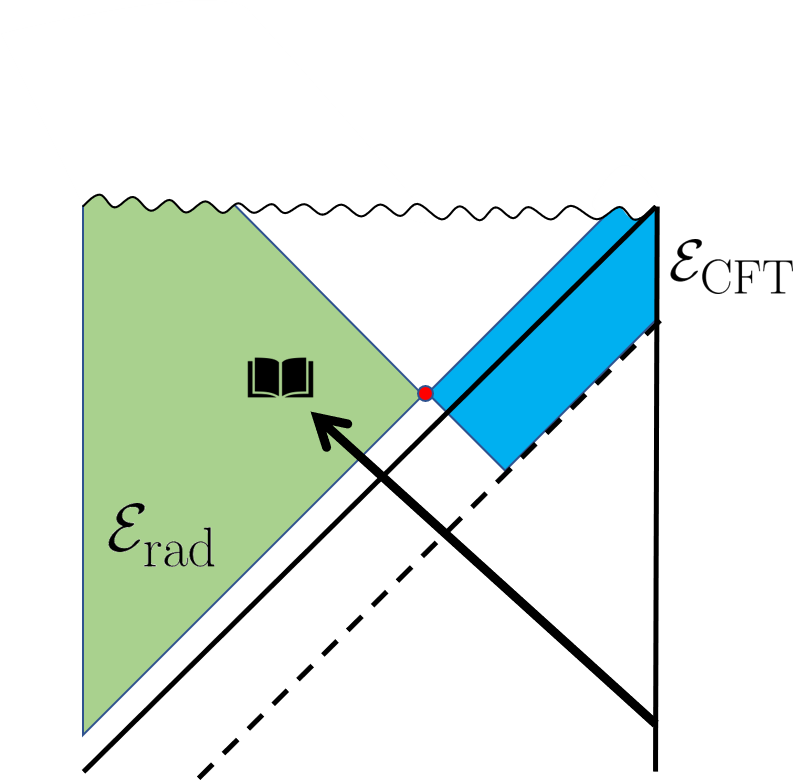}
\centering
\end{subfigure}
\begin{subfigure}{.48\textwidth}
\includegraphics[width = 0.8\linewidth]{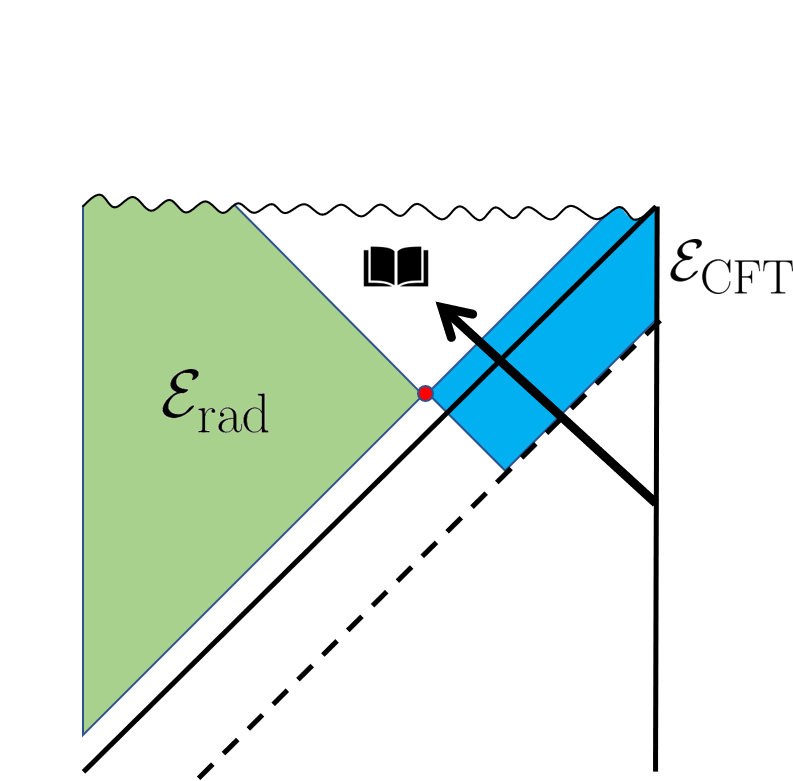}
\centering
\end{subfigure}

\caption{If a diary was thrown into the black hole more than the scrambling time into the past (left), it will now lie in the entanglement wedge of the reservoir $\mathcal{H}_\text{rad}$ and can in principle be decoded using only the Hawking radiation. A diary thrown into the black hole more recently (right) remains in the entanglement wedge of, and encoded in, the CFT.}
\label{fig:haydenpreskill}
\end{figure}

We start with the Hayden-Preskill decoding criterion \cite{hayden2007black}. This says that, if an unknown, small, light diary is thrown into a black hole, whose state is known, at an early stage in its evaporation, the diary can be decoded from the Hawking radiation almost immediately after the Page time. If the small diary is instead thrown into the black hole \emph{after} the Page time, it can be decoded from the Hawking radiation after waiting for the scrambling time.

This indeed exactly what we see from entanglement wedge reconstruction. After the Page time, the quantum extremal surface lies near the black hole horizon at an infalling time
\begin{align} \label{eq:logcevap}
v = - \frac{\beta}{2 \pi} \log \frac{S_{BH}}{c_\text{evap}} + O(\beta).
\end{align}
Assuming $c_\text{evap} = O(1)$, this is exactly one scrambling time, plus subleading corrections, before the current time. A diary thrown into the black hole before this time lies in the entanglement wedge of, and can be decoded from, the reservoir $\mathcal{H}_\text{rad}$ containing the Hawking radiation. Anything thrown in after this time lies in the entanglement wedge of the CFT. This is shown in Figure \ref{fig:haydenpreskill}.

Of course, if we actually throw a diary into the black hole, it will have non-zero energy and will therefore backreact on the geometry. So long as the energy of the diary is $O(1)$, the backreaction will only change the horizon area by an $O(G_N)$ amount. However, the evaporation of the black hole changes the horizon area by an $O(G_N)$ amount over one thermal time $\beta$. Hence it is reasonable to expect that the backreaction will only affect the delay until the diary can be reconstruction from the Hawking radiation by a subleading $O(\beta)$ amount.
When we study the reconstruction of large diaries in Section \ref{sec:largediaries}, we will see that this is indeed the case.

In addition to the well-known scrambling time delay in recovering information thrown into a black hole, \eqref{eq:logcevap} has a small logarithmic correction based on the rate $c_\text{evap}$ at which Hawking radiation is extracted from the black hole. 
While we postpone any formal calculation to future work, it is easy to see heuristically that this is consistent with the boundary dynamics of the theory. 

In a fast scrambling system, the number of degrees of freedom that a `simple' initial perturbation influences grows exponentially with time. This is sometimes described as an `epidemic' where each `infected' qubit infects an $O(1)$ number of other qubits in each timestep (which in our case corresponds to an $O(\beta)$ timescale). After the time given in \eqref{eq:logcevap}, a $O(1/c_\text{evap})$ fraction of the degrees of freedom will be infected. However, the number of degrees of freedom extracted per timestep is $O(c_\text{evap})$. So it is reasonable to expect that an observer with access to the extracted Hawking radiation should be able to detect the perturbation, and hence decode the diary, after the time given in \eqref{eq:logcevap}.

We emphasize that the state of the diary being encoded in the early Hawking radiation does not mean that the diary has been magically extracted out of the interior of the black hole and into the reservoir $\mathcal{H}_\text{rad}$. It is `still' in the interior. There exists a single spacetime that describes the evaporating black hole (which is semiclassical everywhere except for regions of high curvature). In this spacetime, the diary falls into the black hole and keeps falling until it approaches the singularity and the semiclassical spacetime breaks down. It cannot `no longer' have this worldline -- that's not how spacetime works. Spacetime does not change over time; it describes changes over time.

Instead, the encoding of the diary in the Hawking radiation should be understood in terms of the usual story of holography. An object sitting in the middle of the bulk is not `actually' at the boundary; it is in the middle of the bulk. In the effective field theory that describes the bulk, it is an independent degree of freedom from all the fields at asymptotic infinity. 

Nonetheless, by manipulating the fields at asymptotic infinity in a sufficiently complicated way, we can make the bulk effective field theory breakdown and thereby manipulate the object in the middle. Microscopically, the fields at asymptotic infinity contain all the degrees of freedom of the theory. 

We should therefore not be too surprised that, at a microscopic level, the diary in the interior is not an independent degree of freedom from the radiation in the reservoir, and hence, with sufficiently complicated manipulations of the reservoir, one can, in principle, manipulate the diary.\footnote{This is not to say that there aren't serious conceptual questions that remain to be understood about the relationship between the microscopic boundary theory and the effective bulk theory; it is just that they are fundamentally they same conceptual problems that always exist in holography, even without any black holes.}

We have understood the Hayden-Preskill decoding criterion by analysing the entanglement wedge, $\mathcal{H}_\text{CFT}$ or $\mathcal{H}_\text{rad}$, that particular \emph{ingoing} modes are in. The Page curve, and a resolution of the firewall paradox, will follow from analysing the entanglement wedge of \emph{outgoing} modes.

We first note that, since we know the location of the Ryu-Takayanagi surface, it is easy to find the entanglement entropy between the Hawking radiation reservoir $\mathcal{H}_\text{rad}$ and the conformal field theory $\mathcal{H}_\text{CFT}$ (and hence the black hole) using the Ryu-Takayanagi formula. After the Page time, the entanglement entropy is given to leading order by the Bekenstein-Hawking entropy $A_\text{hor}/4G_N$ of the black hole, plus a subleading correction from the bulk entropy term. Since we have already calculated the entanglement entropy before the Page time, at the start of this section, we have therefore successfully derived the entire Page curve.
\begin{figure} [t]
\centering
\vspace{0.5cm}
\begin{subfigure}{.48\textwidth}
  \centering
 \includegraphics[width = 0.58\linewidth]{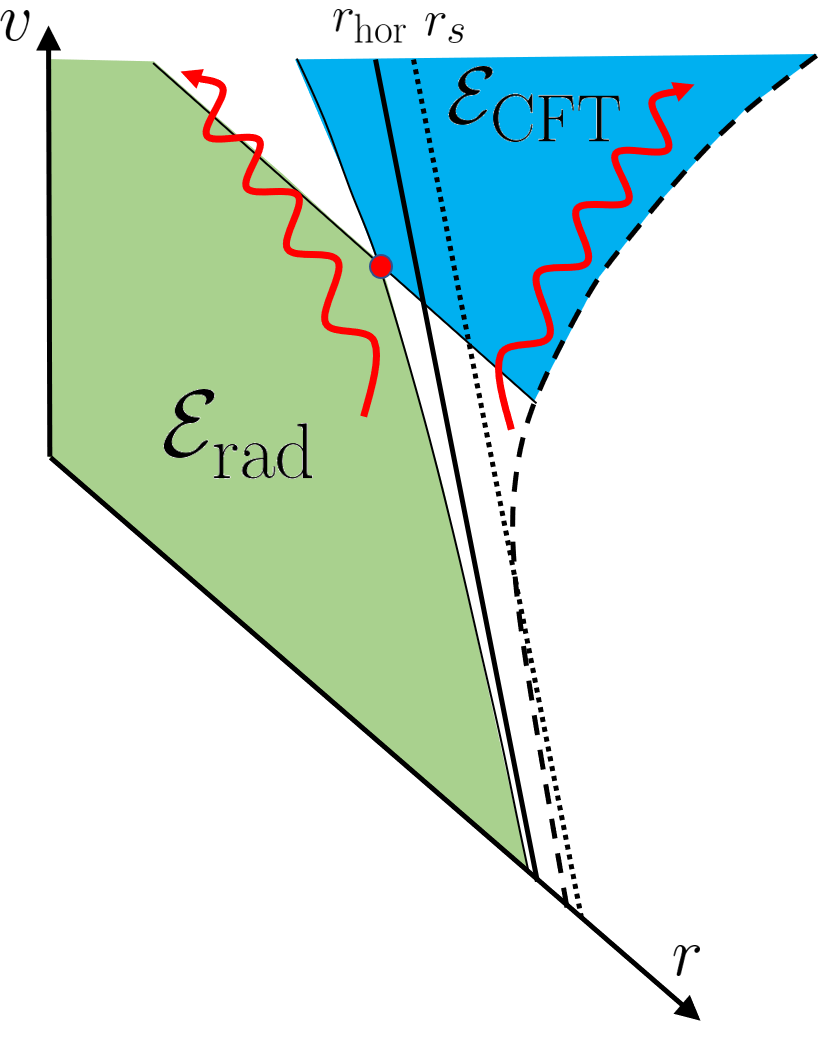}
\end{subfigure}
\begin{subfigure}{.48\textwidth}
\vspace{-2cm}
 \includegraphics[width = 0.8\linewidth]{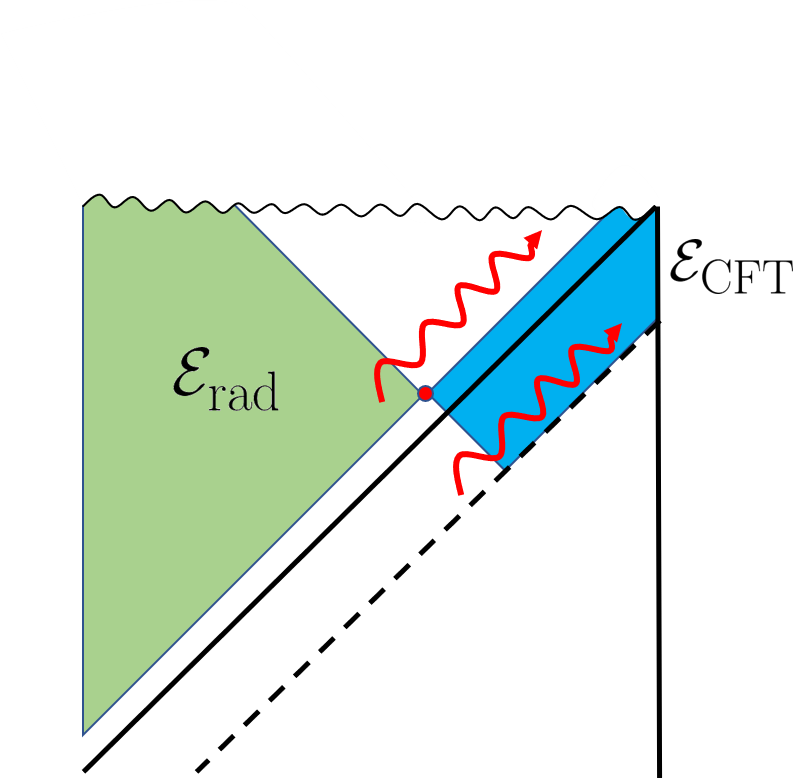}
 \centering

\end{subfigure}
\caption{The next Hawking modes to escape the black hole and be extracted into $\mathcal{H}_\text{rad}$ will be entangled with interior modes that are mostly in the entanglement wedge of the reservoir $\mathcal{H}_\text{rad}$. This causes the entanglement between the black hole and the reservoir to decrease as the new radiation is extracted, in accordance with the Page curve. (Left: Eddington-Finkelstein coordinates; right: a Penrose diagram.)}
\label{fig:hawkingmodes}
\end{figure}

However, on its own, this is somewhat unsatisfying. It does not explain \emph{why} moving outgoing Hawking modes, which we naively thought were unentangled with the earlier radiation, from the CFT to the reservoir $\mathcal{H}_\text{rad}$ should somehow decrease the entanglement between the two. It does not explain how the AMPS firewall paradox \cite{almheiri2013black} is avoided.

Fortunately, in addition to the Ryu-Takayanagi formula, we also know about entanglement wedge reconstruction. Again, we start with heuristic arguments and then progress to more precise statements. Consider a Hawking mode that escapes the black hole slightly into the future. As shown in Figure \ref{fig:hawkingmodes}, we can heuristically think of this mode as being entangled with a partner mode behind the horizon. The partner modes will be in the entanglement wedge of $\mathcal{H}_\text{rad}$. Hence moving this Hawking quanta from $\mathcal{H}_\text{CFT}$ to $\mathcal{H}_\text{rad}$ will \emph{decrease} the entanglement between the two. The same ER=EPR resolution \cite{maldacena2013cool} to the firewall paradox that worked for the two-sided black hole also works for a one-sided evaporating black hole.

Of course, this is only an approximate heuristic picture. For free fields, Rindler modes, with a given Rindler frequency, outside and inside the horizon are indeed perfectly entangled with one another. However such modes are completely delocalised within the exterior and interior respectively. Localised modes outside the horizon will not be perfectly entangled with their reflection inside the horizon, unless the modes have support in only a very narrow range of Rindler frequencies, and hence are delocalised across a large region in Rindler units.

In this case, since part of the interior is in the entanglement wedge of the CFT, we cannot find modes, with support in only a narrow range of Rindler frequencies, whose reflection inside the horizon will be entirely in the entanglement wedge of $\mathcal{H}_\text{rad}$. We should therefore expect that the thermal outgoing radiation will be mostly entangled with the reservoir $\mathcal{H}_\text{rad}$, but also be somewhat entangled with the CFT. Moving the Hawking quanta from $\mathcal{H}_\text{CFT}$ to $\mathcal{H}_\text{rad}$ will decrease the entanglement entropy, but by less than the entropy of the Hawking quanta themselves. This agrees with the Page curve, since the total thermodynamic entropy of the CFT and reservoir is increasing over time and hence
\begin{align}
\frac{1}{4G_N}\frac{d A_\text{hor}}{dv} > - \frac{d \Srad}{dv},
\end{align}
even at leading order. We will do a formal calculation that finds perfect agreement between the bulk entanglement structure and the Page curve below.
\begin{figure} [t]
\centering
\vspace{0.5cm}
\begin{subfigure}{.48\textwidth}
  \centering
 \includegraphics[width = 0.58\linewidth]{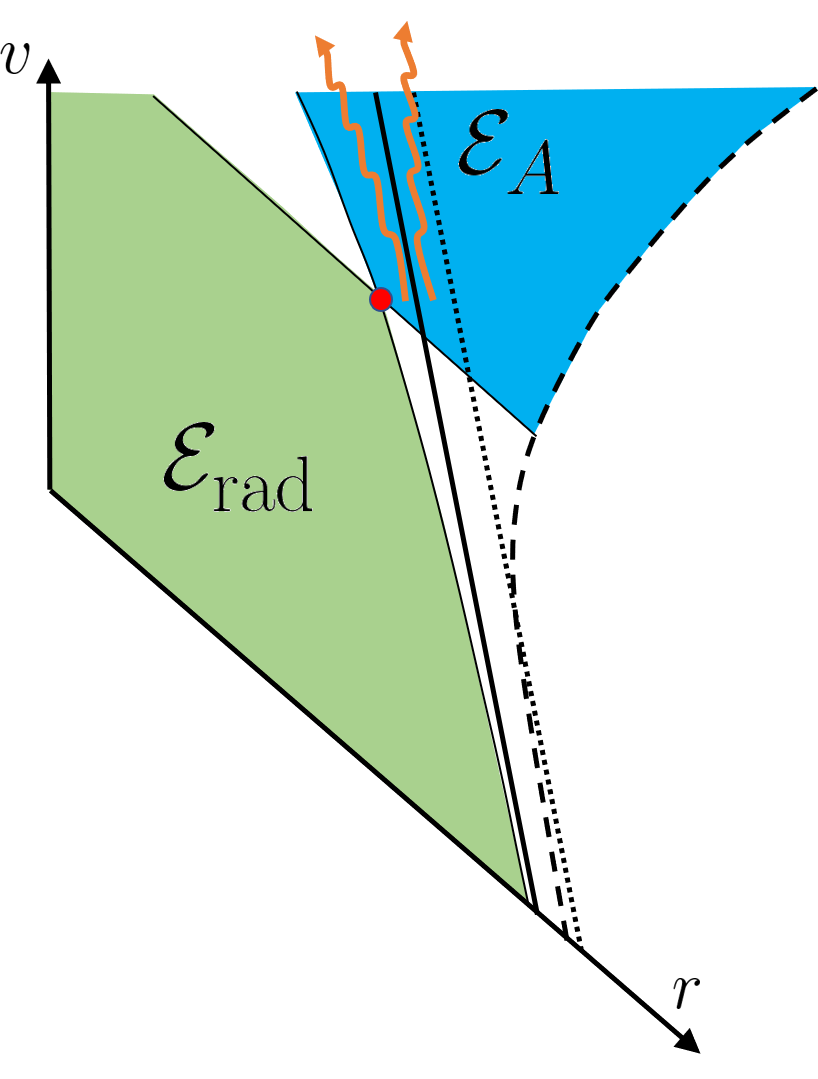}
\end{subfigure}
\begin{subfigure}{.48\textwidth}
\vspace{-2cm}
 \includegraphics[width = 0.8\linewidth]{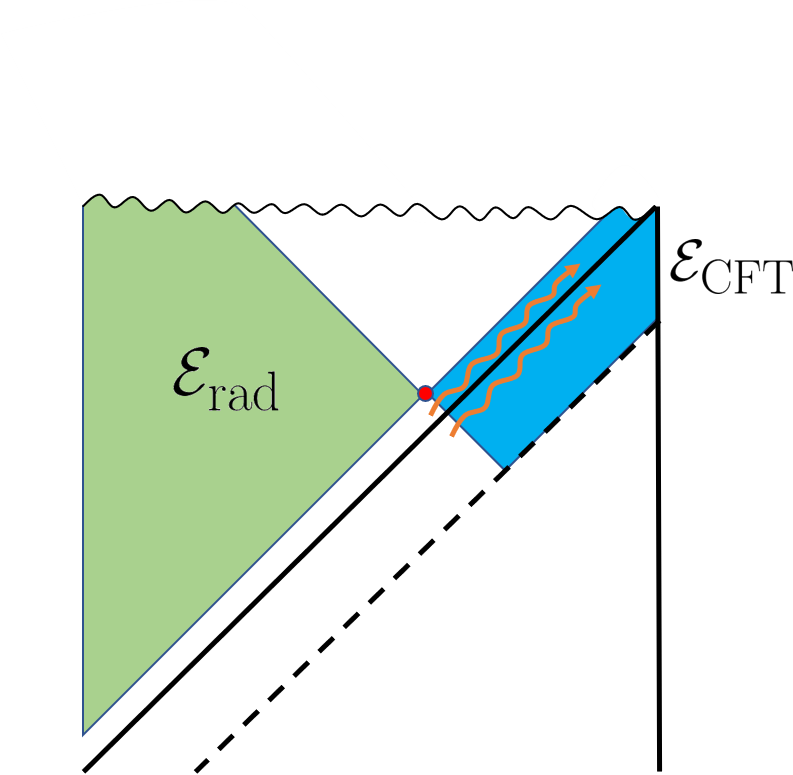}
 \centering

\end{subfigure}
\caption{Hawking modes that will escape the black hole only an $O(\beta)$ time into the future are entangled with interior modes that are almost entirely in the entanglement wedge of the CFT. This is very different from a simple random unitary toy model, but is consistent with toy models where the thermodynamic entropy (i.e. number of qubits) increases over time. (Left: Eddington-Finkelstein coordinates; right: Penrose diagram.)}
\label{fig:hawkingmodes_late}
\end{figure}

Interestingly, and somewhat counterintuitively, Hawking quanta that escape at a time only $O(\beta)$ into the future will be almost perfectly entangled with interior modes that lie almost entirely in the entanglement wedge of $\mathcal{H}_\text{CFT}$, as shown in Figure \ref{fig:hawkingmodes_late}. (In this case, by making the outgoing Hawking mode escape sufficiently far, although still an $O(\beta)$ time, into the future, we can indeed consider outgoing wavepackets with support in only a narrow range of Rindler frequencies, but with `mirror' interior modes that are entirely within the entanglement wedge of $\mathcal{H}_\text{CFT}$.) They will therefore be almost completely \emph{unentangled} with the reservoir $\mathcal{H}_\text{rad}$. 

This is in sharp contrast with the most na\"{i}ve random unitary toy models of black hole evaporation. If our model consists of a single random unitary acting on the initial black hole state and then qubits being released one by one as Hawking radiation, we find that any single qubit of Hawking radiation is almost perfectly entangled with \emph{any} set of more than half the qubits. Hence, if we collect Hawking radiation until after the Page time, throw away qubits of Hawking radiation for a while, and then finally collect one more qubit of Hawking radiation, we still find that the additional qubit of Hawking radiation is almost perfectly entangled with the early Hawking radiation that we collected.

\begin{figure}[t]
\includegraphics[width = 0.95\linewidth]{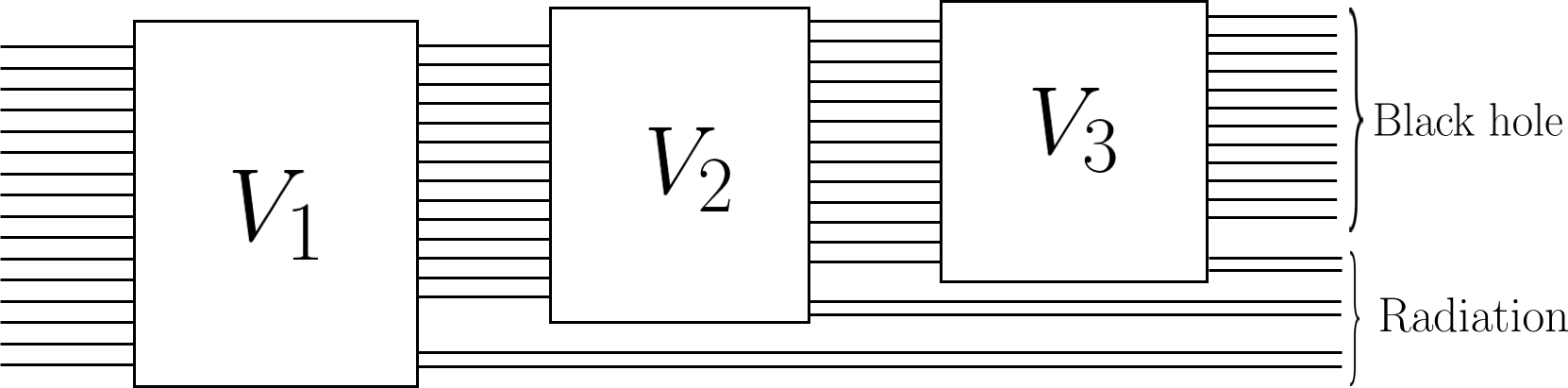}
\centering
\caption{A simple toy model of black hole evaporation that takes into account the increase in thermodynamic entropy as the black holes evaporates. At each timestep, two qubits escape the black hole as Hawking radiation, but then a random isometry is applied to the black hole that increases the number of qubits by one. The total number of qubits (corresponding to the total thermodynamic entropy) therefore increases by one qubit at each timestep. Even if we collect the radiation qubits until long after the Page time, qubits that are radiated only a few timesteps into the future will be completely unentangled with the radiation that we have collected.}
\label{fig:irreversible}
\end{figure}

However such a model does not take into account the fact that the total combined thermodynamic entropy of the black hole and Hawking radiation, which corresponds in the toy model to the total number of qubits, is strictly increasing over time as the black hole evaporates. A more sophisticated toy model, which does take into account this thermodynamic entropy increase, involves a series of nested random isometries, as shown in Figure \ref{fig:irreversible} \cite{hayden2018learning}. 

At each step, qubits are extracted into the Hawking radiation, but then a random isometry is applied to the black hole so that the number of black hole qubits decreases by less than the number of Hawking radiation qubits increases. It can easily be seen using Page's theorem \cite{page1993average} that an additional qubit of Hawking radiation will be almost totally uncorrelated with the early Hawking radiation, so long as a large, but $O(1)$, number of qubits are thrown away in between.\footnote{We call the early radiation Hilbert space $\mathcal{H}_E$, and the black hole Hilbert space, when we stop collecting radiation, $\mathcal{H}_{BH}$. There is then a random isometry $V: \mathcal{H}_{BH} \to \mathcal{H}_T \otimes \mathcal{H}_Q \otimes \mathcal{H}_Z$, where $\mathcal{H}_T$ contains the thrown away radiation, $\mathcal{H}_Q$ is the final collected qubit and $\mathcal{H}_Z$ contains the remaining black hole qubits.  When we stop collecting the Hawking radiation, the state $\ket{\psi} \in \mathcal{H}_{BH} \otimes \mathcal{H}_E$ will be close to maximally entangled. Since we are after the Page time, $|\mathcal{H}_E| \gg |\mathcal{H}_{BH}|$ and so $\ket{\psi}$ only has support in a subspace $\tilde{\mathcal{H}}_E \subseteq \mathcal{H}_E$ with $|\tilde{\mathcal{H}}_E| = |\mathcal{H}_{BH}|$.  However, since $|\tilde{\mathcal{H}}_E \otimes \mathcal{H}_Q| \ll |\mathcal{H}_T \otimes \mathcal{H}_Z |$, the reduced density matrix of the state $V \ket{\psi}$ on $\mathcal{H}_E \otimes \mathcal{H}_Q$ will be very close to maximally mixed. There will be essentially no entanglement, or even correlation, between the early radiation and the final collected qubit.}

If the infalling modes are in a thermal state at a temperature equal to the temperature of the black hole, there will be no increase in thermodynamic entropy. We study finite-temperature infalling modes in Appendix \ref{app:finitetemp}, where we indeed find that, if the temperature of the infalling modes is equal to the black hole temperature, the Hawking radiation will be completely unentangled with any CFT degrees of freedom, even when it escapes far into the future. We also find, in several separate cases such as thermal infalling modes and pure infalling modes with constant energy density, that information stops escaping the black hole and the Hawking radiation becomes completely thermal at exactly the moment when this becomes consistent with unitarity.

Having described heuristically how entanglement wedge reconstruction allows us to avoid the firewall paradox, let us now do a more formal calculation of the change in entanglement entropy from extracting bulk Hawking modes. In fact, we shall prove that this change in entanglement entropy will necessarily always agree with the Ryu-Takayanagi formula. However, we first start by calculating it explicitly.

We want to calculate the change in entanglement entropy from outgoing modes, over some small time range $\delta v$, being transferred from $\mathcal{H}_\text{CFT}$ to $\mathcal{H}_\text{rad}$. We need the time $\delta v$ to be small, because otherwise the Ryu-Takayanagi surface, and hence the entanglement wedges, of $\mathcal{H}_\text{CFT}$ and $\mathcal{H}_\text{rad}$ will depend on whether the transferred modes are included in $\mathcal{H}_\text{CFT}$ or $\mathcal{H}_\text{rad}$. By making $\delta v$ very small, we ensure that all bulk modes (other than the transferred ones) are encoded in one of the two Hilbert spaces, even if the transferred modes themselves are not counted as part of either Hilbert space. One can then calculate the change in entanglement over longer time periods by integrating these infinitesimal changes.

Outgoing modes that are between the Ryu-Takayanagi surface and the past lightcone of the boundary are encoded in the CFT, while all other outgoing modes are encoded in the reservoir $\mathcal{H}_\text{rad}$.\footnote{We are ignoring the fact that entanglement wedge reconstruction is only approximate here. Since the separation between the outgoing lightcone and the extremal surface grows linearly  in the limit of large $c_\text{evap}$, the effect of the reconstruction errors should become small in this limit. Furthermore, it is reasonable to hope that the effect of errors in the reconstruction of bulk modes on each side will cancel and so we will still find the correct answer even at small $c_\text{evap}$. As we shall see, it seems that this is indeed the case.}  As discussed at the beginning of this section, since the overall state of the outgoing modes is pure, we will find the same change in entanglement entropy if we look at the change in the entropy of the outgoing modes in $\mathcal{H}_\text{CFT}$, or the entropy of the outgoing modes in $\mathcal{H}_\text{rad}$. However, since the modes in the CFT consist of a single interval, the change in their entropy is more natural to calculate.

From \eqref{eq:outgoinglc}, it is easy to see that extracting the Hawking radiation for an additional time $\delta v$ will move the radius $r_{lc}$ of the outgoing lightcone by
\begin{align} \label{eq:deltarlcv}
\delta r_{lc}(v) = - \frac{2 \pi \delta v}{\beta} (r_{lc}(v) - r_\text{hor}(v)).
\end{align}
Assuming we keep the cut-off on the extracted outgoing modes constant, extracting the additional Hawking radiation will change the cut-off $\varepsilon_2$ in units of $r$ by
\begin{align} \label{eq:changecutoff}
\delta \varepsilon_2 = -\frac{2 \pi \delta v}{\beta} \varepsilon_2.
\end{align}
To derive this equation, we note that the cut-off $\varepsilon_2$ only depends on the difference between the infalling time at which the radiation is extracted and the infalling time of the quantum extremal surface. Hence, extracting radiation for an additional time $\delta v$ has the same effect on $\varepsilon_2$ as moving the extremal surface backward in infalling time by $\delta v$; \eqref{eq:changecutoff} is therefore an immediate consequence of \eqref{eq:cutoffscaling}.

Using \eqref{eq:bulkentropyoutgoing}, we find that
\begin{align} \label{eq:deltaSbulk}
\delta S &= \frac{c_\text{evap}}{6} \frac{\delta r_{lc}}{r_{lc} - r} - \frac{c_\text{evap}}{12}\frac{\delta \varepsilon_2}{\varepsilon_2}
\\& = -\frac{c_\text{evap}\, \pi\, \delta v\, (r_{lc}(v) - r_\text{hor}(v))}{3\,(r_{lc} - r)} + \frac{c_\text{evap}\, \pi\, \delta v}{6}
\\& = -\frac{c_\text{evap} \pi \delta v}{12 \beta} = \frac{1}{4 G_N}\frac{d A_\text{hor}}{dv} \delta v,
\end{align}
where in the second equality we used \eqref{eq:deltarlcv} and \eqref{eq:changecutoff}, in the third equality we used \eqref{eq:dSdv} and  \eqref{eq:rlcr=4rsrhor} and in the last equality we used \eqref{eq:dAhor/dv}. The change in entanglement entropy exactly agrees with the change in entanglement entropy that one finds using the Ryu-Takayanagi formula. 

At first glance, these two methods of calculating the change in entanglement entropy appear very different, despite the perfect quantitative agreement between them. In the Ryu-Takayanagi formula, the change comes from the change in the horizon area of the black hole, with its somewhat mysterious association with entropy, while, in \eqref{eq:deltaSbulk}, the change occurs because outgoing bulk modes are entanglemed with other outgoing bulk modes in the entanglement wedge of $\mathcal{H}_\text{rad}$. 

However, it is not a coincidence that they give the same answer. The Ryu-Takayanagi formula is really calculating the change in the generalised entropy $A/4G_N + S_\text{bulk}$ of the Ryu-Takayanagi surface, not just the change in the area; it is just that, in this case, the bulk entropy happens to stay approximately constant (because of the approximate time translation invariance of the evaporation process). The bulk entropy calculation can also be thought of as a change in generalised entropy; just one in which the RT surface for which the generalised entropy is evaluated, and therefore the area term, stay fixed. 

In the Ryu-Takayanagi formula calculation, the change in entropy is given by the difference between the new generalised entropy for the new Ryu-Takayanagi surface and the old generalised entropy for the old Ryu-Takayanagi surface. In the bulk entanglement calculation, the change in entropy is the difference between the new generalised entropy and the old generalised entropy, when both are evaluated using the old Ryu-Takayanagi surface. However, by definition, the generalised entropy is constant at leading order if we perturb the Ryu-Takayanagi surface. Since we need $\delta v$ to be infinitessimally small to do the bulk entanglement calculation, the two calculations will always give the same answer. Because the Ryu-Takayanagi surface is a quantum extremal surface, there can never be a firewall paradox. The bulk entanglement structure will always be consistent with the Ryu-Takayanagi formula.

\subsection{Greybody Factors} \label{sec:greybody}
As we discussed in Section \ref{sec:extremal}, there is nothing genuinely unphysical about evaporating a black hole in AdS/CFT by extracting black-body Hawking radiation from well inside the zone. However,  if we eventually want to understand four dimensional black holes in flat space that evaporate naturally (without an external super-observer extracting Hawking radiation from near the horizon), it is important to understand what happens when there are non-trivial greybody factors. 

Although we will not be able to explicitly calculate the location of the quantum extremal surface when greybody factors are present, it turns out that we will still be able to derive both the Hayden-Preskill decoding criterion and the Page curve.\footnote{Since the entanglement entropy of gravitons is not understood, we shall still assume that no graviton modes are extracted using the absorbing boundary conditions and hence that their entanglement entropy can be ignored. Of course, in flat space, gravitons will always contribute to the Hawking radiation, so understanding their entanglement entropy precisely is an important task for future work. However, since the relevant graviton modes simply become ordinary light scalar fields in a $(1+1)$-dimensional reduction of the evaporation, as do all the other bosonic modes that contribute to the evaporation, it seems reasonable to expect that their contributions to the bulk entropy will be qualitatively the same as any other mode.}

We first recall our argument, from the very beginning of this section, that the maximin prescription implies that the Ryu-Takayanagi surface must become non-empty at the Page time, even when the greybody factors are non-trivial. After this time, assuming that the quantum maximin prescription is valid, there must exist a non-empty quantum extremal surface.\footnote{In fact, since we continue to assume rotational symmetry, this argument would not actually require the full power of the assumption of quantum maximin. Instead, we can restrict our maximisation and minimisation to rotational symmetric Cauchy slices and surfaces $\chi$ respectively, thereby avoiding most of the subtleties that would be involved in defining the quantum maximin prescription and showing its equivalence to the quantum extremal surface prescription.} We shall show that this is indeed the case

What can we say about the location of this quantum extremal surface? We know that the entanglement wedge must contain the causal wedge \cite{engelhardt2015quantum}, so the extremal surface must lie in the interior of the black hole.\footnote{We shall prove explicitly, later in this section, that the quantum extremal surface that we find is indeed inside the event horizon of the black hole.} However we also know that there does not exist a classical extremal surface anywhere in this spacetime. This means that, at the non-empty quantum extremal surface, the gradient of the bulk entropy term must be $O(1/G_N)$ (at least in Eddington-Finkelstein coordinates where the gradient of the area is $O(1)$ everywhere). 

Since the bulk entropy itself is $O(1)$, the only way that this can happen is if the extremal surface is very close in Eddington-Finkelstein coordinates to a point where the bulk entropy diverges. The extremal surface must therefore approach the past lightcone of the boundary, which is always outside the event horizon and only approaches the event horizon at infalling times that are far into the past.\footnote{Technically, the bulk entropy will also diverge near the future lightcone. However there cannot be a quantum extremal surface near the future lightcone, because, at the future lightcone, the bulk entropy will only diverge as function of the infalling time $v$, while $dA = (d-1)\, r_s^{d-2} \Omega_{d-1} dr$.} In Eddington-Finkelstein coordinates, the quantum extremal surface must therefore both approach the event horizon, with respect to the radius $r$, and diverge into the infinite past, with respect to the infalling time $v$, in the limit $G_N \to 0$. Again, we shall explicitly verify that this is the case. 

It is helpful at this point to switch from Eddington-Finkelstein coordinates to lightlike, Kruskal-Szekeres-like coordinates. At radii close to the event horizon, and over infalling timescales that are small compared to the evaporation time, the metric of the evaporating black hole in Eddington-Finkelstein coordinates is given by
\begin{align}
ds^2 =  - \frac{4 \pi} {\beta}(r - r_s(v)) dv^2 + 2 dv\, dr + r^2 d\Omega_{d-1}^2,
\end{align}
where we can assume that the inverse temperature $\beta$ and the evaporation rate $d r_s/ dv$ are constant at leading order. Substituting the Kruskal-like coordinates,
\begin{align} \label{eq:V}
V = \frac{\beta}{2 \pi} \exp(2 \pi v /\beta),
\end{align}
and
\begin{align} \label{eq:U}
U = (r_\text{hor}(v) - r) \exp(- 2 \pi v / \beta),
\end{align}
where $r_\text{hor} = r_s + \beta (d r_s / d v) / 2 \pi$ as in \eqref{eq:rhor}, we find
\begin{align} \label{eq:kruskal}
ds^2  = -2 dU dV + r^2 (U, V) d\Omega_{d-1}^2.
\end{align}
Note that the definition \eqref{eq:U} is only intended to be valid in the near horizon region where outgoing lightrays escape exponentially in Eddington-Finkelstein coordinates. More generally, the coordinate $U$ should be defined in the exterior region by 
\begin{align}
U \propto - \exp(- 2 \pi u / \beta),
\end{align}
where $u$ is the boundary time at which an outgoing lightray reaches the boundary, and then the metric should be analytically extended to the interior where $U > 0$. This will ensure that the coordinates $U, V$ are exactly lightlike everywhere. However, we are only interested in the near horizon region where the definition \eqref{eq:U} and the metric \eqref{eq:kruskal} are valid.\footnote{Even within the near horizon region, our conventions for $U$ and $V$ differ from the more standard conventions for Kruskal coordinates in AdS space by constant factors \cite{harlow2016jerusalem}. However, within the near horizon region, our convention will be somewhat more convenient.} Note that $V>0$ everywhere; the infinite past with respect to infalling time $v$ corresponds to the limit $V \to 0^+$. Also note that the past lightcone of the current boundary time (i.e. $v = 0$) is at $U_{l.c.} = - O(r_s)$.\footnote{As discussed in Section \ref{sec:classicalmaximin}, for near-extremal black holes, we actually have $U_{l.c} = - O(r_s^2/ \beta)$. Also, recall that we are assuming for convenience that the past lightcone escapes the \emph{near-horizon} region at $v=0$. Hence for parametrically small AdS-schwarzschild black holes, the current boundary time is really $v = \pi l_{AdS} + O(\beta)$. These subtleties will be unimportant for our purposes; the key point is that $U_{l.c.}$ is independent of $G_N$.}

Our basic strategy will be to show that $\partial S_\text{bulk}/ \partial U$ should approach a well-defined $O(1)$ limit as $V \to 0$, whereas $$
\frac{1}{4G_N} \frac{\partial A}{ \partial U} = O\left(\frac{V}{ G_N}\right).
$$
We will also find that, in the same limit, $\partial S_\text{bulk}/ \partial V = O(1/V)$, while $$\frac{1}{4G_N}  \frac{\partial A}{ \partial V} = O\left(\frac{1}{G_N}\right) + O\left(\frac{1}{V}\right).$$ We will therefore be able to argue that the quantum extremal surface must be at some fixed $U$ that is independent of $G_N$ and at $V = O(G_N)$, which corresponds to an infalling time exactly one scrambling time (plus subleading corrections) into the past.

As with the Eddington-Finkelstein coordinates $r, v$, the metric \eqref{eq:kruskal} is approximately constant in the near horizon region in terms of the coordinates $U, V$. We can therefore consistently cut-off ingoing modes at the quantum extremal surface with a constant cut-off in units of $V$, and outgoing modes with a constant cut-off in units of $U$. Since
\begin{align} \label{eq:dVdv}
\frac{\partial V}{\partial v} =  \exp(2 \pi v /\beta),
\end{align}
and
\begin{align}
\frac{\partial U}{\partial r} = \exp(-2 \pi v /\beta),
\end{align}
have non-trivial infalling-time dependence, the gradient of the entropy of either the ingoing or outgoing modes \emph{alone} will depend on the set of cut-offs that we use. In particular, in Section \ref{sec:extremal}, when the cut-offs were constant in Eddington-Finkelstein coordinates, there was an increase in bulk entropy, when moving the RT surface forwards in infalling time along an outgoing lightcone, that came from outgoing bulk modes . With constant cut-offs in Kruskal-like coordinates, the same increase in bulk entropy exists, but it comes from the infalling modes. The gradient of the total bulk entropy is the same in both cases.

An advantage of using constant cut-offs in units of $U$ and $V$ is that outgoing modes are not blueshifted, with respect to $U$, as we evolve them backwards in time. The outgoing modes that are contained in the entanglement wedge of $\mathcal{H}_\text{CFT}$ will be determined only by $U$, and we don't have to worry about blueshifting the cut-off at the outgoing lightcone. If the cut-off at the quantum extremal surface is also constant in units of $U$, the entropy of the outgoing modes in the entanglement wedge of the CFT will be independent of $V$. To calculate $\partial S_\text{bulk}/\partial V$ we therefore only need to worry about the ingoing modes near the quantum extremal surface. 

In Section \ref{sec:extremal}, the infalling modes were in the vacuum state with respect to the infalling time $v$. We therefore argued that, so long as the cut-off was constant in units of $v$, the gradient of the entropy of the infalling modes would tend to zero as the quantum extremal surface diverged into the infinite past in the semiclassical limit.

\begin{figure} [t]
\centering
\vspace{-1cm}
\begin{subfigure}{.48\textwidth}
  \centering
 \includegraphics[width = 0.8\linewidth]{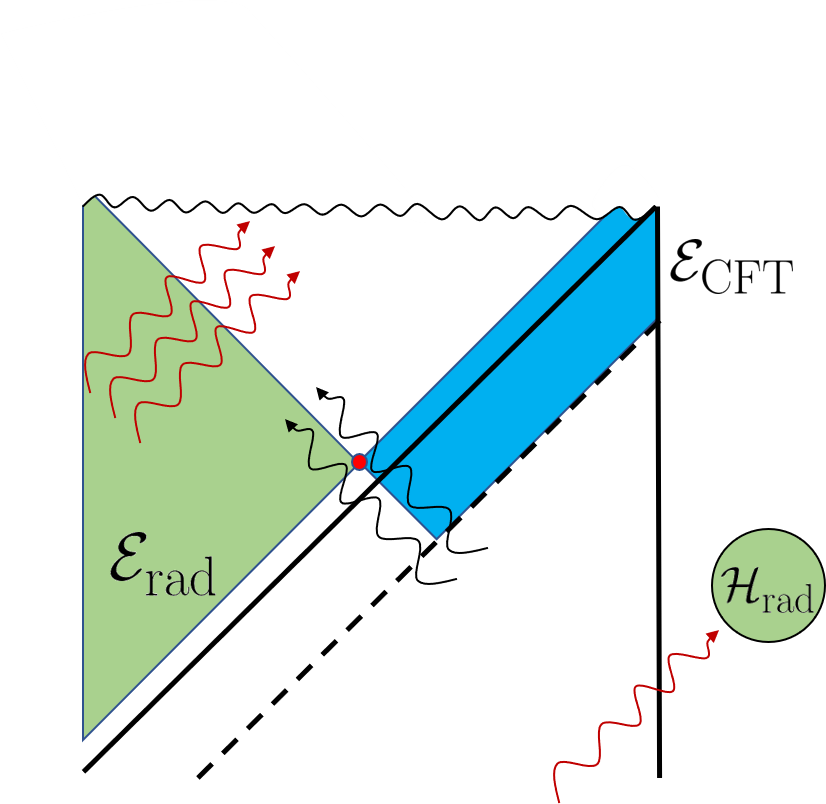}
\end{subfigure}
\begin{subfigure}{.48\textwidth}
 \includegraphics[width = 0.87\linewidth]{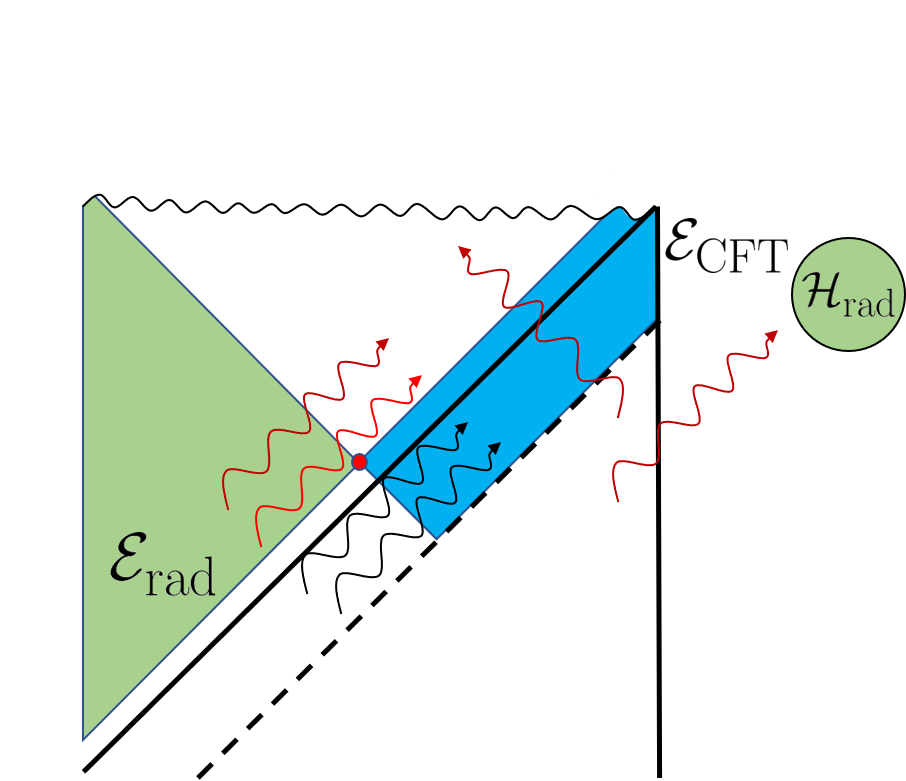}
 \centering

\end{subfigure}
\caption{In the left figure, infalling modes near the quantum extremal surface are in a infalling-time-translation invariant mixed state. These mixed state modes (black) are purified by modes deep in the interior, together with modes that escaped into the reservoir (both red). Both of these are in the entanglement wedge of the reservoir $\mathcal{H}_\text{rad}$. In the right figure, outgoing modes near the quantum extremal surface (black) are entangled with outgoing modes in the interior, but also with outgoing modes that recently escaped into the reservoir and late-time infalling modes that were reflected back into the black hole (all in red). The first two are in the entanglement wedge of $\mathcal{H}_\text{rad}$, while the last is in the entanglement wedge of the CFT. The bulk entropy of the modes in the entanglement wedge of the CFT depends on $U$ no longer depends on $U$ in the same simple way that it did when no greybody factors were present in Section \ref{sec:extremal}.}
\label{fig:reflected}
\end{figure}

When there are non-trivial greybody factors, with part of the Hawking radiation being reflected back into the black hole, the infalling modes will instead be in a \emph{mixed} state that is invariant with respect to translations in infalling time. The mixed-state infalling modes will be purified by outgoing modes that escaped the black hole into the reservoir, as well as modes deep in the interior of the black hole. As shown in Figure \ref{fig:reflected}, these modes are all in the entanglement wedge of $\mathcal{H}_\text{rad}$. 

In the semiclassical limit, when the extremal surface diverges into the infinite past, there will be no entanglement between ingoing modes near the quantum extremal surface and outgoing modes in the entanglement wedge of the CFT. We therefore find
\begin{align} \label{eq:greybodydSdv}
\frac{2 \pi V}{\beta}\left.\frac{\partial S_\text{bulk}}{\partial V}\right|_U = \left.\frac{\partial S_\text{bulk}}{\partial v}\right|_U = \frac{c_\text{evap} \pi}{6 \beta} - \frac{d S_\text{in}}{d v},
\end{align}
where $d S_\text{in} / d v \geq 0$ is the constant entropy per unit infalling time of the infalling modes assuming a constant cut-off in units of the infalling time $v$.  $d S_\text{in} / d v$ can in principle be calculated from a numerical approximation of the reflection coefficients for modes escaping the near-horizon zone. (See, for example, similar calculations in \cite{page1976particle}.) The additional term $c_\text{evap} \pi / 6 \beta$ shows up because the cut-off at the extremal surface should be constant in units of $V$, rather than constant in units of $v$, if we want to ignore the outgoing modes. Since
\begin{align}
\left.\frac{\partial}{\partial V}\right|_U = e^{-2 \pi v / \beta} \left.\frac{\partial}{\partial v}\right|_U,
\end{align}
and the logarithmic divergence $(c_\text{evap} / 12) \log(1/\varepsilon)$ with respect to the cut-off length $\varepsilon$ is universal,\footnote{Recall that $\varepsilon$ is the cut-off on only the ingoing modes, at only one end of an interval.} we can immediately obtain \eqref{eq:greybodydSdv}.

Formally, there are infinitely many angular momentum modes and so $c_\text{evap}$ is infinite. However, modes with large angular momentum are almost entirely reflected back into the black hole. The ingoing modes are in a thermal state at the same inverse temperature $\beta$ as the black hole. They therefore satisfy \cite{calabrese2004entanglement, calabrese2009entanglement}
\begin{align}
\frac{d S_\text{in}}{d v} = \frac{c_\text{evap} \pi}{6 \beta}.
\end{align}
It follows that only the finite number of low angular momentum modes, which actually partially escape the black hole, contribute to \eqref{eq:greybodydSdv}. So long as we include the same modes in calculating $dS_{\text{in}}/dv$ that we use in calculating $c_\text{evap}$, \eqref{eq:greybodydSdv} should be independent of the choice of any sufficient large angular momentum cut-off on the modes we consider.

From \eqref{eq:V} and \eqref{eq:U}, we have
\begin{align} \label{eq:rUV}
r = r_\text{hor}(v) - \frac{2 \pi}{\beta} UV,
\end{align}
and hence
\begin{align} \label{eq:drdV}
\left. \frac{\partial r}{\partial V} \right|_U = \frac{\beta}{2 \pi V} \frac{d r_s}{d v} - \frac{2 \pi U}{\beta}.
\end{align}
It follows that
\begin{align}
\frac{1}{4 G_N} \left. \frac{\partial A}{\partial V} \right|_U &= \frac{(d-1) r_s^{d-2} \Omega_{d-1}}{4 G_N} \left. \frac{\partial r}{\partial V} \right|_U,
\\ & = \frac{\beta^2}{2 \pi V} \frac{d M}{d v} - \frac{\pi (d-1)\, r_s^{d-2} \,\Omega_{d-1}\, U}{2 \beta G_N}. \label{eq:dAdV}
\end{align}
In the second equality we used \eqref{eq:drdV} and the first law of black hole thermodynamics $\beta dM = dA_\text{hor}/4G_N$. The quantum extremal surface must therefore satisfy
\begin{align}
0 &= \frac{1}{4 G_N} \left. \frac{\partial A}{\partial V} \right|_U + \left.\frac{\partial S_\text{bulk}}{\partial V}\right|_U,
\\  r_{\text{hor}} - r = \frac{2 \pi}{\beta} U V &=  \frac{2 G_N\,\beta}{\pi\,(d-1)\, r_s^{d-2} \,\Omega_{d-1}}\left[\beta \frac{d M}{d v} + \frac{c_\text{evap} \pi}{6 \beta} - \frac{d S_\text{in}}{d v}\right], \label{eq:marginalgreybody}
\end{align}
where we used \eqref{eq:greybodydSdv} and \eqref{eq:dAdV}.

It can be shown, as follows, that the right hand side of \eqref{eq:marginalgreybody} must be non-negative, and so the extremal surface is inside the event horizon of the black hole. Suppose that the infalling modes were in a thermal state at inverse temperature $\beta'$. We would then have
\begin{align}
\frac{dM}{dv} = \frac{d M}{d v} =\frac{c_\text{evap} \pi}{12} (\frac{1}{\beta'^2} - \frac{1}{\beta^2}),
\end{align}
and \cite{calabrese2004entanglement, calabrese2009entanglement}
\begin{align}
\frac{d S_\text{in}}{d v} = \frac{c_\text{evap} \pi}{6 \beta'}.
\end{align}
The right hand side of \eqref{eq:marginalgreybody} would then be given by 
\begin{align} \label{eq:positiverhs}
\frac{G_N \,\beta^2\,c_\text{evap}}{6\,(d-1)\, r_s^{d-2} \,\Omega_{d-1}}\left(\frac{1}{\beta} - \frac{1}{\beta'}\right)^2 \geq 0,
\end{align}
which is non-negative at any inverse temperature $\beta'$.\footnote{For detailed calculations of quantum extremal surfaces for finite temperature infalling modes, see Appendix \ref{app:finitetemp}.} Since thermal states have maximal entropy for any fixed energy flux, the right hand side of \eqref{eq:marginalgreybody} must therefore be non-negative for any state of the infalling modes, thermal or otherwise. The quantum extremal surface is always inside the event horizon.

So far we have only demanded that the surface be extremal if we vary $V$ at constant $U$. The quantum extremal surface should also be extremal when we vary $U$ at constant $V$. From \eqref{eq:rUV}, we have
\begin{align} \label{eq:dA/dU}
\frac{1}{4 G_N} \left. \frac{\partial A}{\partial U} \right|_V =  - \frac{(d-1)\, r_s^{d-2} \,\Omega_{d-1}\, V}{4 G_N}.
\end{align}
What about the variation of the bulk entropy term? By varying $U$, we change the outgoing modes that are included in the entanglement wedge of the CFT. These outgoing modes are entangled with the other outgoing modes, both inside the quantum extremal surface and outside the past lightcone of the boundary. 

In Section \ref{sec:extremal}, all the outgoing modes that were not in the entanglement wedge of the CFT were in the entanglement wedge of the reservoir. We therefore found
\begin{align} \label{eq:nogreybody}
\left. \frac{\partial S_\text{bulk}}{\partial U} \right|_V = \frac{c_\text{evap}}{6(U - U_{l.c})},
\end{align}
where $U_{l.c.} < 0$ labels the past lightcone of the boundary.\footnote{Of course, our actual calculations in Section \ref{sec:extremal} were in Eddington-Finkelstein coordinates, but were equivalent to \eqref{eq:nogreybody}. With constant cut-offs in Kruskal-like coordinates, the outgoing bulk entropy $S_\text{out} = (c_\text{evap}/ 6)\log((U - U_{l.c})/ \sqrt{\varepsilon_1 \varepsilon_2})$ where the cut-offs $\varepsilon_1$ and $\varepsilon_2$ are both constant in units of $U$.} 

When greybody factors are present, this will no longer be the case, as shown in Figure \ref{fig:reflected}. Outgoing modes outside the past lightcone will be partially reflected back into the black hole, and end up as ingoing modes, which \emph{are} in the entanglement wedge of the CFT. The functional form of $\partial S_\text{bulk}/\partial U$ will therefore be much more complicated. 

However, in the semiclassical limit where the extremal surface diverges into the infinite past, there will still be no entanglement between outgoing modes in the entanglement wedge of the CFT and ingoing modes \emph{near the extremal surface}. The gradient $\partial S_\text{bulk}/\partial U$ will therefore be independent of $V$, so long as $V$ is sufficiently small. Indeed, this follows directly from the fact that $\partial S_\text{bulk}/\partial V$ is independent of $U$, by the symmetry of mixed partial derivatives. We conclude that $\partial S_\text{bulk}/\partial U$ has a well-defined limit as $V \to 0$, which is some (presumably complicated) function of $U$.

If the surface is extremal, we must have
\begin{align}
0 &=\frac{1}{4 G_N} \left. \frac{\partial A}{\partial U} \right|_V + \left. \frac{\partial S_\text{bulk}}{\partial U} \right|_V,
\\  \left. \frac{\partial S_\text{bulk}}{\partial U} \right|_V & = \frac{(d-1)\, r_s^{d-2} \,\Omega_{d-1}\, V}{4 G_N},
\\ U \left. \frac{\partial S_\text{bulk}}{\partial U} \right|_V &= \frac{\beta}{2\pi}\left[\beta \frac{d M}{d v} + \frac{c_\text{evap} \pi}{6 \beta} - \frac{d S_\text{in}}{d v}\right], \label{eq:greybodyextremal}
\end{align}
where in the first equality we used \eqref{eq:dA/dU} and in the last equality we used \eqref{eq:marginalgreybody}. The right hand side of \eqref{eq:greybodyextremal} is constant over timescales that are small compared to the evaporation time, while the left hand side is a function of $U$. 

If a non-empty quantum extremal surface is to exist in the near horizon region for any sufficiently small $G_N$, as we expect from the maximin prescription, there must exist a solution $U_0$ to \eqref{eq:greybodyextremal}. Importantly, since \eqref{eq:greybodyextremal} does not depend on $G_N$, this solution must be independent of $G_N$.
 
 In the simple example from Section \ref{sec:extremal} where there are no greybody factors, $dM/dv$ is given in \eqref{eq:energyflux}, $dS_\text{in}/dv = 0$ and  $\partial S_\text{bulk}/\partial U$ is given in \eqref{eq:nogreybody}. Hence
 \begin{align}
 U_0 = -U_{l.c.}/3,
 \end{align}
in agreement with our calculations in Eddington-Finkelstein coordinates.

What about when greybody factors are present? In this case, it turns out that we can still calculate $\partial S_\text{bulk}/\partial U$ in the limit $U \to \infty$, and this is sufficient to argue that a solution $U_0$ must exist. For this argument it is simplest to use the Rindler-like coordinates, $u_L =- (\beta / 2 \pi) \log (U/r_s)$ for $U > 0$ and $u_R =- (\beta / 2 \pi) \log (|U|/ r_s)$ for $U< 0$.\footnote{The factors of $r_s$ are included only to make the logarithms dimensionless.} Up to a $G_N$-independent shift, $u_R$ is the time at which an outgoing lightray would reach the boundary, while $u_L$ is its `mirror' coordinate inside the event horizon.

The outgoing modes are near the horizon are in the thermofield double state with respect to these Rindler coordinates. A key point will be that the thermofield double state only has significant correlation between the left and right region when $u_R = u_L \pm O(\beta)$. If the RT surface is at $(U^{RT}, V^{RT})$, the interior outgoing modes are in the entanglement wedge for $u_L > u_L^{RT} =- (\beta / 2 \pi) \log (U^{RT}/r_s)$. 

What about the exterior outgoing modes? For $u_R/\beta \gg 0$, they are completely in the entanglement wedge, since they haven't had time to escape. For $u_R/\beta \ll 0$, the situation is more complicated. In this case, the outgoing near-horizon modes are encoded in a combination of the Hawking radiation that has escaped into $\mathcal{H}_\text{rad}$, and the reflected ingoing modes at an infalling time $v = u_R + O(\beta)$. For $u_R \gg v^{RT}  =  (\beta / 2 \pi) \log (V^{RT}/r_s)$, the reflected modes are in the entanglement wedge, but the escaped Hawking radiation is not. For $u_R \ll v^{RT}$, neither is in the entanglement wedge. Finally, since we are assuming that \eqref{eq:marginalgreybody} holds, we note that $u_L^{RT} \gg v^{RT}$.

In summary, for $u_L,u_R \ll v^{RT}$, none of the interior or exterior outgoing modes are in the entanglement wedge. For $u_L^{RT} \gg u_L,u_R \gg v^{RT}$, the interior modes are not in the entanglement wedge and only the reflected part of the exterior modes is in the entanglement wedge. For $0 \gg u_L,u_R \gg u_L^{RT}$, the interior modes and the reflected part of the exterior modes are in the entanglement wedge. Finally for $u_L,u_R \gg 0$, all the modes are in the entanglement wedge.

Since the thermofield double state has a local entanglement structure, as discussed above, then we can ignore any effects that come from the finite size of each of these regions, in the semiclassical limit $G_N \to 0$. Instead there are only three contributions to the gradient of the bulk entropy as a function of $u_L^{RT}$. The first is that increasing $u_L^{RT}$ increases the range of $u_L, u_R$ for which only the reflected part of the interior modes is in the entanglement wedge. This gives a contribution equal to the entropy density $dS_\text{in}/dv$ of the reflected modes. The second is that increasing $u_L^{RT}$ \emph{decreases} the range of $u_L, u_R$ for which both the interior modes and the reflected part of the exterior modes are in the wedge. Since the tripartite state of interior modes, reflected exterior modes and escaped exterior modes is pure, this gives a contribution equal to $-d S_\text{rad}/dv$. Finally, we need the cut-off to be constant in units of $U$, we means the cut-off length must exponentially grow as a function of $u_L$. This gives a contribution to the gradient of $-\pi c_\text{evap}/6\beta$. We therefore find that
\begin{align} \label{eq:Utoinfty}
U \frac{\partial S_\text{bulk}}{\partial U} = - \frac{\beta}{2 \pi} \frac{\partial S_\text{bulk}}{\partial u_L} =  \frac{\beta}{2 \pi}\left[\frac{d \Srad}{dv} - \frac{d S_\text{in}}{d v} + \frac{\pi c_\text{evap}}{6 \beta}\right].
\end{align}
To build some intuition for this formula, note that, in the limit $U \to \infty$ and radius $r$ given in \eqref{eq:marginalgreybody},
\begin{align}
\left.\frac{\partial S_\text{bulk}}{\partial v}\right|_r = - \frac{2 \pi U}{\beta} \left.\frac{\partial S_\text{bulk}}{\partial U}\right|_V + \frac{2 \pi V}{\beta} \left.\frac{\partial S_\text{bulk}}{\partial V}\right|_U =  - \frac{d \Srad}{dv}.
\end{align}
This is in exact agreement with our heuristic picture from way back in Figure \ref{fig:schematic} of Bell pairs, entangled between the interior and the Hawking radiation, that are evenly distributed along the wormhole.

So long as the black hole is evaporating and hence $dM/dv < 0$, it follows from \eqref{eq:Utoinfty} that the left hand side of \eqref{eq:greybodyextremal} should be larger at large $U$ than the right hand side, which is independent of $U$.\footnote{If $dM/dv > 0$, then this will stop being true, and the quantum extremal surface will stop existing, at exactly the moment when the rate of change of the Bekenstein-Hawking entropy $1/ 4G_N dA_\text{hor}/dv = \beta dM/dv$ becomes greater than the rate of increase in the entropy of the Hawking radiation. This is exactly the moment when boundary unitarity becomes consistent with no information ever escaping the black hole. See Appendix \ref{app:finitetemp} for a more detailed discussion of this.} Meanwhile, at the horizon, the left hand side is zero, while the right hand side is positive, as shown in \eqref{eq:positiverhs}. Hence, assuming that the derivative $\partial S_\text{bulk}/ \partial U$ is a continuous function of the location of the rotationally symmetric surface, then by the intermediate value theorem there must indeed exist a solution $U_0$ to \eqref{eq:greybodyextremal}, as we expected from the maximin arguments.\footnote{One might wonder whether there could exist multiple solutions and hence multiple non-empty quantum extremal surfaces. We first note that, even if there did exist multiple solutions, the solution that minimised the generalised entropy would be independent of $G_N$, for sufficiently small $G_N$, and so we could simple ignore the other solutions. However, in practice, it seems that the left hand side of \eqref{eq:greybodyextremal} should be a monotonically increasing function of $U$ and so only one solution will exist. If no exterior modes were in the entanglement wedge of the CFT, we would have $\partial S_\text{bulk}/\partial U \propto 1/U$ and the left hand side of \eqref{eq:greybodyextremal} would be constant. The existence of exterior outgoing modes in the entanglement wedge of the CFT should only slow the rate of decay of $\partial S_\text{bulk}/\partial U$ as a function of increasing $U$.}
 
Since we know a solution $U_0$ must exist, we can substitute it into \eqref{eq:marginalgreybody} and find that
\begin{align}
V &=  \frac{G_N \,\beta^2}{\pi^2\,U_0\,(d-1)\, r_s^{d-2} \,\Omega_{d-1}}\left[\beta \frac{d M}{d v} + \frac{c_\text{evap} \pi}{6 \beta} - \frac{d S_\text{in}}{d v}\right].
\end{align}
Hence
\begin{align}
v = \frac{\beta}{2 \pi} \log \frac{2 \pi V}{\beta} = \frac{\beta}{2 \pi} \log S_{BH} + O(\beta).
\end{align}
In the last equality, we assumed for simplicity that all other scales are held fixed in the semiclassical limit $G_N \to 0$. We can therefore derive the Hayden-Preskill decoding criterion even when the greybody factors are non-trivial. 

We can also find the Page curve (up to subleading corrections) using the Ryu-Takayanagi formula. Furthermore, our argument from Section \ref{sec:haydenpage}, showing that the outgoing modes automactically have exactly the right entanglement to reproduce the Page curve, without any firewall paradox, only used entanglement wedge reconstruction and the fact that the Ryu-Takayanagi surface is an extremum of the generalised entropy. All our main results can therefore be derived, even in the presence of greybody factors. 

\section{State Dependence} \label{sec:statedependence}
In deriving the results of Section \ref{sec:evaporation}, we were careful to focus on a single spacetime where a black hole was formed by collapse in some particular initial state of the matter fields. Although the details of how the black hole was formed did not affect any of our calculations, we always implicitly assumed that those details were known by the observer reconstructing interior operators. We never considered the task of reconstructing a diary from the Hawking radiation with the initial state of the black hole partially, or completely, unknown. 

We therefore avoided the issue of whether, and to what extent, reconstructions of bulk interior operators necessarily depend on the initial state of the black hole. Such questions will be the focus of this section. In particular, we will find that this state dependence is crucial in resolving the information paradox. If the interior partners of the late-time Hawking modes were encoded in the early radiation in a state-independent way, there would be no way for the final state of the Hawking radiation to depend on the initial state of the matter falling into the black hole, even if the final state was pure rather than thermal. The state dependence allows the entanglement of early and late radiation to depend on the state, despite the entanglement of late radiation and interior modes being fixed. This allows information to escape.

We begin the section by briefly reviewing results from \cite{hayden2018learning} that show how state dependence can arise in entanglement wedge reconstruction.

\subsection{State Dependence in Entanglement Wedge Reconstruction} \label{sec:abits}
Entanglement wedge reconstruction is best understood in the language of holographic quantum error correction \cite{almheiri2015bulk}. Bulk operators in AdS/CFT are only well defined within the ``code subspace'' $\mathcal{H}_{\text{code}} \subseteq \mathcal{H}_\text{CFT}$ of boundary states with the correct smooth bulk geometry. The claim of entanglement wedge reconstruction can then be phrased as follows:  the noisy quantum channel, mapping states $\rho$ in the code space to their restriction $\rho_B$ to some boundary region $B$, forms an approximate operator algebra quantum error correcting code for the bulk operators in the entanglement wedge $b$ of $B$. This means that there exists a decoding channel $\mathcal{D}$ such that, for all states $\rho$ in the code subspace,
\begin{align}
\mathcal{D} (\rho_B) \approx \rho_b,
\end{align}
where $\rho_b$ is the restriction of the bulk state $\rho$ to the algebra of operators associated with the entanglement wedge $b$ of $B$.\footnote{Here, restriction can be thought of as a partial trace, although it is really the projection of the state onto the von Neumann subalgebra associated with the bulk region $b$.} We can then use the adjoint channel $\mathcal{D}^\dagger$ to map bulk operators $\phi_b$, acting within the entanglement wedge, to boundary `reconstructions' $\phi_B = \mathcal{D}^\dagger (\phi_b)$, acting only on region $B$, whose action is (approximately) the same as the bulk operator $\phi_b$ on states in the codespace. It is important to note that the decoding channel $\mathcal{D}$ is in general highly non-unique; two very different boundary reconstructions of the same bulk operator may be both be valid, so long as their action on states in the code subspace is (approximately) the same.

Because the spacetime geometry is dynamical, the entanglement wedge of a given boundary region depends on the bulk geometry of the particular CFT state that we are interested in. Even if we only consider a code space of bulk states with a fixed spacetime geometry, the quantum extremal surface may be state-dependent because of the bulk entropy term. When we say that operators in the entanglement wedge can be reconstructed in the boundary region, we should be careful to specify \emph{which} states need to have their entanglement wedge contain the bulk operator.

The initial derivation of entanglement wedge reconstruction in \cite{dong2016reconstruction} suggested that one could always reconstruct a bulk operator so long as it was contained within the entanglement wedge of the boundary region for all pure states in the code space of states for which the reconstruction was meant to work. However, this derivation ignored the fact that entanglement wedge reconstruction is only approximate at finite $G_N$. (More specifically it ignored the fact that the equality between bulk and boundary relative entropies \cite{jafferis2016relative} is only approximate at finite $G_N$.) It should therefore only be trusted for code spaces whose dimension is relatively small (and, in particular, independent of $G_N$).

A more rigorous derivation of entanglement wedge reconstruction, using the tools of approximate quantum error correction, makes it clear that the bulk operator needs to be contained in the entanglement wedge even for \emph{mixed states} with support only in the code space \cite{cotler2017entanglement}.\footnote{For an alternative derivation that is more directly equivalent to the derivation in \cite{dong2016reconstruction}, but which reaches the same conclusion as \cite{cotler2017entanglement}, see \cite{hayden2018learning}.} Rather than directly considering mixed states, it is often more natural, and is mathematically equivalent, to assume that the mixed states are purified by an arbitrary `reference system' $\mathcal{H}_R$, whose dimension is equal to the code space dimension.\footnote{The reference system $\mathcal{H}_R$ should not be confused with the Hawking radiation reservoir $\mathcal{H}_\text{rad}$ that we use when studying evaporating black holes. The second is an actual physical system, whereas the first is purely a mathematical `accounting trick'.} In this picture, the bulk operator must be contained in the entanglement wedge of $B$ for all pure states, including entangled states, in $\mathcal{H}_\text{code} \otimes \mathcal{H}_R$.

If, instead, the bulk operator is \emph{only} contained in the entanglement wedge of $B$ for all pure states in the code space (with no reference system), entanglement wedge reconstruction will still be possible, but only if we allow the reconstruction to be state-dependent~\cite{hayden2018learning}. 

It is a general fact about quantum error correction that, if exact state-dependent reconstruction of operators is possible all states in a finite-dimensional code space, then exact state-independent reconstruction is also possible for that code space \cite{alphabits,hayden2018learning}.\footnote{In the Schr\"{o}dinger picture, which is usually used to describe quantum error correction, this corresponds to the fact that exact universal subspace quantum error correction implies full quantum error correction \cite{alphabits}.} This is why it was sufficient to only consider pure states in \cite{dong2016reconstruction}, where the reconstruction errors were ignored. 

Even if the state-dependent reconstruction is only approximate, state-independent reconstruction will necessarily also be possible (with a somewhat larger error), so long as the code space is not too large. In holography, this corresponds to the fact that entanglement with a reference system cannot affect the location of the Ryu-Takayanagi surface in the semiclassical limit, so long as the dimension of the code space is $O(1)$, because any entanglement with the reference system will give a subleading correction to the generalised entropy.

In contrast, if the dimension of the code space is very large, for example when one considers a large number of black hole microstates, approximate state-dependent reconstruction may be possible, even when state-independent reconstruction is not.

\begin{figure}[t]
\begin{subfigure}{.48\textwidth}
\includegraphics[width = 0.8\linewidth]{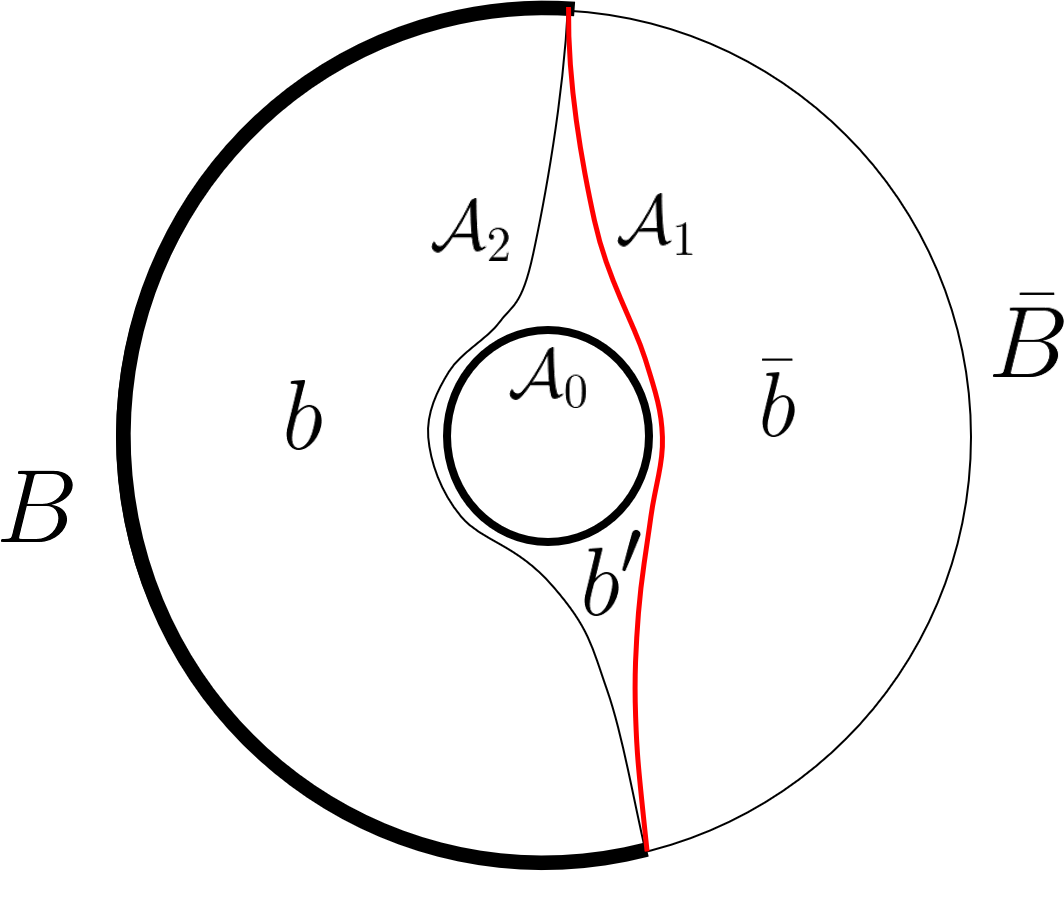}
\centering
\caption{}
\label{fig:BHent}
\end{subfigure}
\begin{subfigure}{.48\textwidth}
\includegraphics[width = 0.95\linewidth]{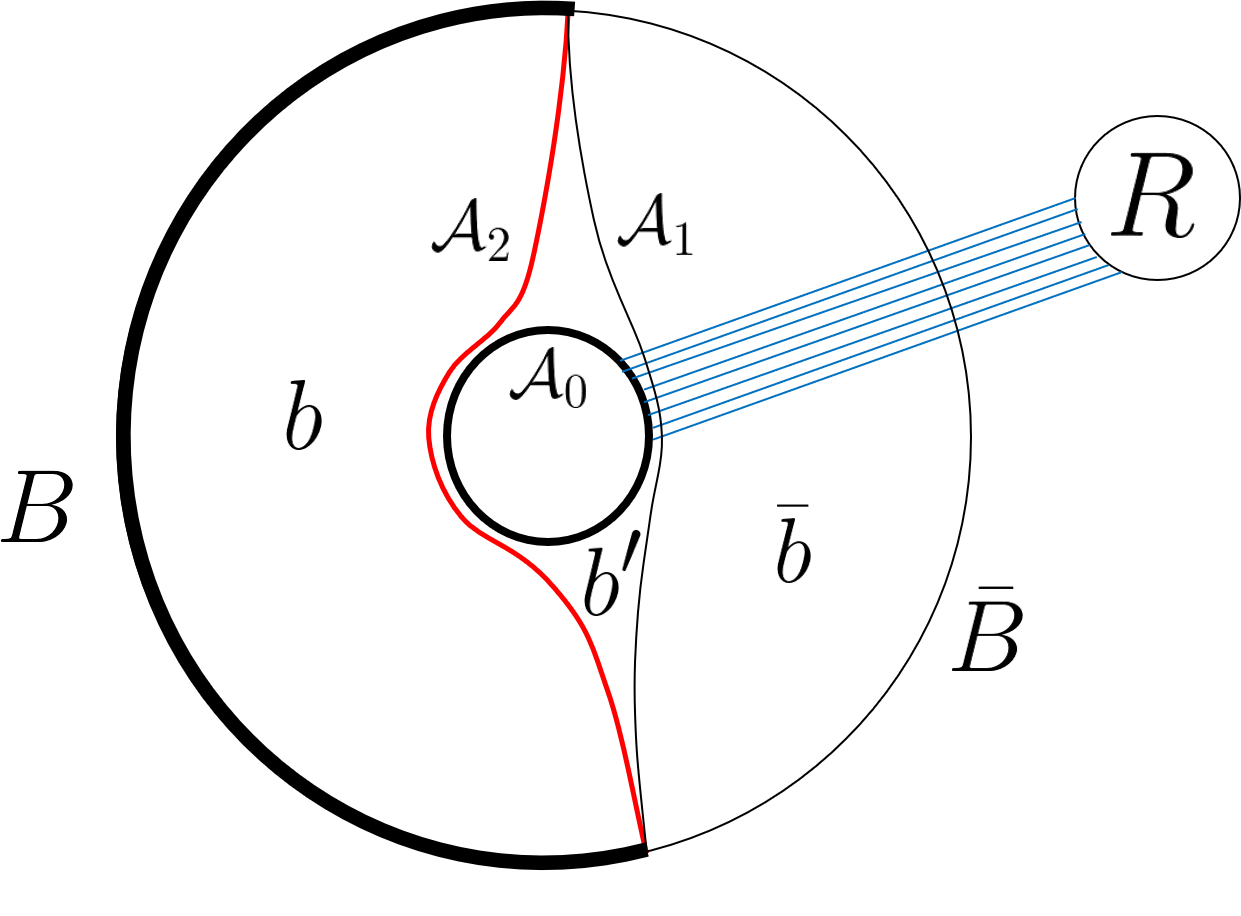}
\centering
\caption{}
\label{fig:BHRent}
\end{subfigure}
\caption{The exterior geometry of a black hole, with horizon area $\mathcal{A}_0$, in AdS/CFT. The boundary is divided into two regions $B$ and $\bar B$; there exists a locally minimal surface separating $B$ from $\bar{B}$ on either side of the black hole, with areas $\mathcal{A}_2$ and $\mathcal{A}_1$. These divide the bulk into three regions $b$, $b'$ and $\bar{b}$, where region $b'$ lies between the two minimal surfaces and contains the black hole. If $\mathcal{A}_2 > \mathcal{A}_1$, region $b'$ is contained in the entanglement wedge of region $B$ for all pure microstates, shown in Figure \ref{fig:BHent}. However, if the black hole is entangled with a reference system with $S_\text{bulk} / 4G_N> \mathcal{A}_2 - \mathcal{A}_1$, as shown in Figure \ref{fig:BHRent}, the Ryu-Takayanagi surface will jump to $\mathcal{A}_2$ and the entanglement wedge of $B$ will no longer contain region $b'$. As a result, a state-independent reconstruction of region $b'$ on region $B$ only exists for code spaces of microstates with dimension $|\mathcal{H}_\text{code}| \leq e^{(\mathcal{A}_2 -\mathcal{A}_1)/4G_N}$. }
\label{fig:bhentwedge}

\end{figure}

A simple example of this was studied in \cite{hayden2018learning}. It shows up when one considers code spaces consisting of a large number of black hole microstates, plus bulk degrees of freedom outside the horizon, as illustrated in Figure \ref{fig:bhentwedge}. We emphasize that, unlike in the rest of this paper, we will not be interested in the details of the interior of these black hole microstates.

Suppose that we try to reconstruct bulk operators in a simply connected region $B$ consisting of slightly more than half of the boundary. For any pure black hole microstate, the Ryu-Takayanagi surface, with area $\mathcal{A}_1$, lies between the black hole and the complementary region $\bar{B}$. The entanglement wedge of region $B$ contains the black hole.

However, for a two-side black hole, such as the thermofield double state, the homology constraint means that the Ryu-Takayanagi surface lies between region $B$ and the black hole, so long as the area $\mathcal{A}_2$ of this surface is less than the area $\mathcal{A}_1$ plus the horizon area $\mathcal{A}_0$. The entanglement wedge will no longer contain the bulk region $b'$ that lies between the two extremal surfaces.

Replacing the second CFT by an arbitrary reference system cannot affect whether bulk operators in region $b'$ can be reconstructed in region $B$. Region $b'$ therefore cannot be encoded in region $B$ for any purification of the thermal density matrix. So long as we correctly use the quantum extremal surface prescription to define the Ryu-Takayanagi surface, we do indeed find that this is the case. Instead of classical area and a homology constraint, we now have a large amount of bulk entanglement between the black hole and the reference system. However, the quantum Ryu-Takayanagi surface, and its generalised entropy, are unchanged.

The black hole does not need to be maximally entangled with the reference system in order to exclude region $b'$ from the entanglement wedge. We only need the entanglement entropy $S$ to satisfy
\begin{align}\label{eq:diff7}
S > \frac{\mathcal{A}_2 - \mathcal{A}_1}{4G_N}.
\end{align}

At face value, we now have something of a paradox. The bulk region $b'$ is encoded in region $B$ for \emph{any} pure microstate, but for sufficiently entangled state it can be reconstructed on the combination of region $\bar B$ and the reference system. By linearity, a reconstruction that works for any pure state will also work for entangled states.\footnote{For approximate reconstructions, this fact is somewhat non-trivial to prove. However, it is indeed true, up to a dimension-independent increase in the error size \cite{kretschmann2004tema}.} But the no cloning theorem says that quantum information can't be simultaneously encoded in region $B$, and in the combination of region $\bar B$ and the reference system.

The resolution, of course, is the fact that entanglement wedge reconstruction can only be made state independent, if the bulk operator is contained in the entanglement wedge even for mixed states with support only in the code space. In this case, the reconstruction will have to be state dependent, precisely when the entropy $S$ of the code space satisfies \eqref{eq:diff7}. There is therefore no single reconstruction that could be used for the entangled state.

Again, we emphasize that this resolution is only consistent because of the approximate nature of entanglement wedge reconstruction. Otherwise it would be impossible for a state-dependent reconstruction to exist for every state in the code space, without state-independent reconstruction also being possible. There would be no way to evade the paradox described above.

If we make the code space $\mathcal{H}_\text{code}$ of microstates as large as possible, so that 
\begin{align}
\log |\mathcal{H}_\text{code}| \approx S_{BH} = \frac{\mathcal{A}_0}{4 G_N},
\end{align}
at leading order,  a single reconstruction will only exist for any \emph{subspace} of the code space with dimension less than $|\mathcal{H}_\text{code}|^\alpha$, where
\begin{align}
\alpha = \frac{\mathcal{A}_2 - \mathcal{A}_1}{\mathcal{A}_0}.
\end{align}
This is an example of something known as an `$\alpha$-bit code' \cite{alphabits}. Indeed, many of the results about state-dependence in evaporating black holes that we will derive in this section can be rephrased in the language of \cite{alphabits} as statements about the existence (or non-existence) of $\alpha$-bit codes for various values of $\alpha$. However, since the terminology of `$\alpha$-bits' was developed for asymptotic quantum resource equalities and is, perhaps, more misleading than clarifying in the present context, we will not use it any further in this paper.

Using the results of \cite{alphabits}, it is possible to put strict lower bounds on the size of the non-perturbative error that must exist to avoid a cloning paradox. Specifically we find that the error in the reconstruction of region $b'$ on region $B$ for a code space $\mathcal{H}_\text{code}$ must be at least $e^{-O(\Delta S)}$, where 
\begin{align}
\Delta S = \mathcal{A}_2/4G_N - \mathcal{A}_1/4G_N - \log |\mathcal{H}_\text{code}|,
\end{align}
is the minimum difference, for any state in the code space, between the generalised entropies of the two extremal surfaces\cite{hayden2018learning}.

\subsection{State Dependence in Evaporating Black Holes} \label{sec:statedepend}
Exactly the same effects that were described in \cite{hayden2018learning} also happen in evaporating black holes. The boundary $\mathcal{H}_\text{CFT}$ and the reservoir $\mathcal{H}_\text{rad}$ play the roles of the two boundary regions $\mathcal{H}_B$ and $\mathcal{H}_{\bar B}$. 

Suppose, as in Section \ref{sec:haydenpage}, that we throw a small diary into a black hole. This time, however, rather than knowing the initial state of the black hole exactly, we only know that the initial state was in some large code space of possible states. 

For example, we can imagine starting with a small black hole, with Bekenstein-Hawking entropy $S_\text{code}$, in a \emph{completely unknown} state. If we then throw a large amount of additional energy (this time in a known state) into this black hole, we end up with a larger black hole, whose state is partially unknown. It lies in a code space of possible states, which has entropy $\log |\mathcal{H}_\text{code}| = S_\text{code}$.

When can we reconstruct the diary from the Hawking radiation? As discussed in Section \ref{sec:evaporation}, for any pure initial black hole state, there is a phase transition in the Ryu-Takayanagi surface at the Page time. After this point, the diary will be in the entanglement wedge of the Hawking reservoir $\mathcal{H}_\text{rad}$ (assuming it was thrown into the black hole at least one scrambling time into the past).

Unfortunately, our lack of knowledge about the state of the black hole prevents us from taking advantage of this fact. Instead, we can only successfully decode the state of the diary once a state-independent reconstruction becomes possible that works for the entire code space of possible initial states.
\begin{figure}[t]
\includegraphics[width = 0.4\linewidth]{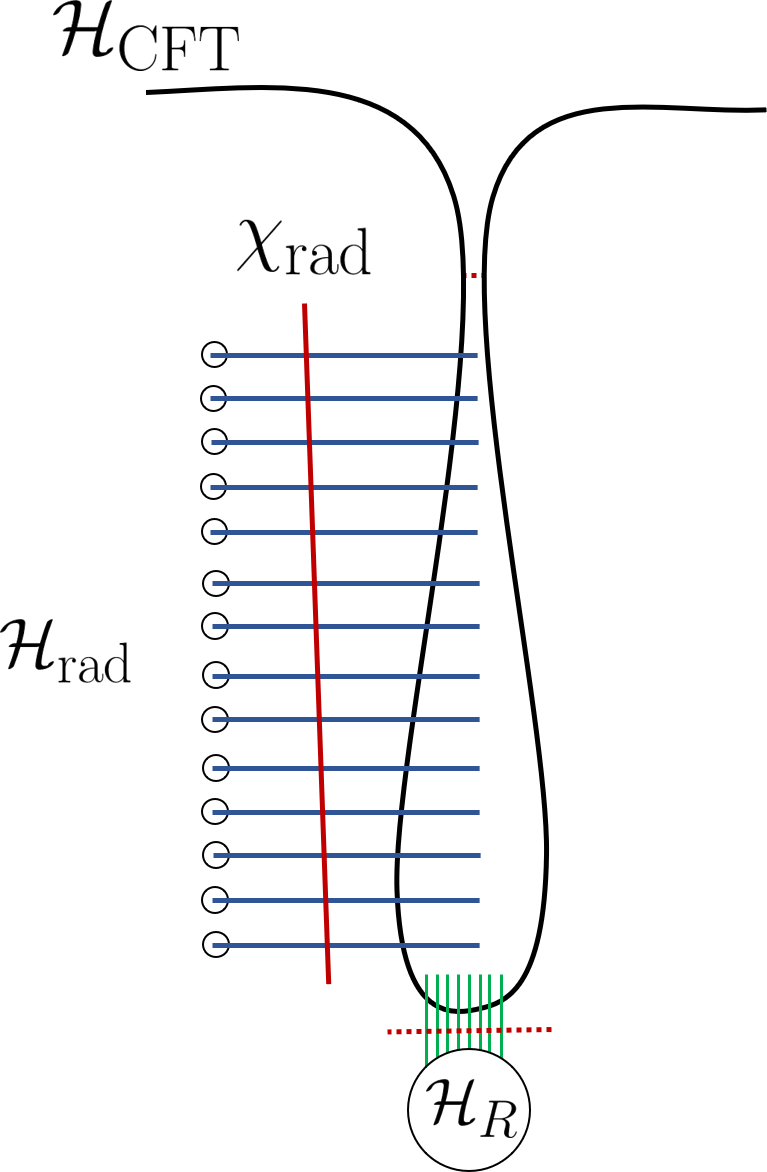}
\centering
\caption{After the Page time, the entanglement wedge of the reservoir contains most of the interior for any pure initial black hole microstate. However, if we only know that the initial state was in some large class of possible microstates, we cannot take advantage of this fact to do a Hayden-Preskill recovery, unless the interior is also contained in the entanglement wedge for states in the code space of possible microstates that are highly entangled with a reference system. This entanglement entropy increases the generalised entropy of the non-empty quantum extremal surface (dotted lines) and can make the Ryu-Takayanagi surface $\chi_\text{rad}$ become empty (solid line), preventing us for doing a Hayden-Preskill reconstruction until the black hole has evaporated further.}
\label{fig:afterpage_reference}
\end{figure}

Such a reconstruction will only be possible once the diary is contained in the entanglement wedge of the reservoir $\mathcal{H}_\text{rad}$ even for highly mixed states in the code space of initial microstates (or equivalently code space states that are highly entangled with a reference system $\mathcal{H}_R$). These states will have a large bulk entropy in the interior, which will increase the generalised entropy of the non-empty quantum extremal surface for $\mathcal{H}_\text{rad}$, as shown in Figure \ref{fig:afterpage_reference}. Note that, because we are no longer considering bipartite pure states, the Ryu-Takayanagi surfaces for $\mathcal{H}_\text{rad}$ and $\mathcal{H}_\text{CFT}$ will no longer necessarily be the same. The Ryu-Takayanagi surface of $\mathcal{H}_\text{rad}$ will therefore remain empty for highly mixed states until
\begin{align} \label{eq:entropydifference}
\Srad - S_{BH} > S_\text{code}.
\end{align}
Immediately after the Page time, we need to know the exact initial state of the black hole in order to reconstruct the diary. However, as the black hole continues to evaporate, the amount of state-dependence required in the reconstruction decreases; a single reconstruction can work for an increasingly large class of microstates. Happily, \eqref{eq:entropydifference} agrees exactly with the amount of state-dependence required for the Hayden-Preskill decoding criterion in simple random unitary toy models \cite{hayden2018learning}.

This state dependence does not just make the entanglement wedge version of Hayden-Preskill compatible with toy models. It also provides the \emph{mechanism} by which information about the initial state of the black hole is able to escape out into the Hawking radiation. The Hawking radiation that escapes the boundary is always entangled with interior modes in the same way, regardless of the initial state of the black hole. However, the way that the interior modes are encoded in $\mathcal{H}_\text{rad}$ depends on the initial state of the black hole. The Hawking radiation will be purified by a different subsystem of $\mathcal{H}_\text{rad}$, depending on the initial state of the black hole.\footnote{We emphasize that, for any single state, the subsystem that purifies some Hawking quanta is not uniquely defined because the decoding channel used to reconstruct the interior mode on $\mathcal{H}_\text{rad}$ is not unique. The point here is that there is \emph{no} subsystem that purifies the Hawking quanta for all the possible initial states.} As a result, the reservoir $\mathcal{H}_\text{rad}$, plus the additional Hawking radiation, contains more information about the initial state of the black hole than the reservoir alone. The bulk evaporation is consistent with information escaping, even though the state of the Hawking radiation and the interior mode does not care about the initial microstate of the black hole.

Interestingly, even if the initial microstate is completely unknown, and so the code space entropy $S_\text{code}$ is equal to the initial Bekenstein-Hawking entropy of the black hole, \eqref{eq:entropydifference} will be satisfied long before the black hole has completely evaporated, because of the thermodynamic entropy increase from the evaporation. The information in the diary, as well as all the information about the initial state of the black hole, will be revealed, in a completely state-independent way, even while the black hole is still an $O(1)$ fraction of its original size. This is closely related to the fact that, even for a completely thermal initial black hole state, the von Neumann entropy of the reservoir will peak and begin decreasing, even while the black hole is still an $O(1)$ fraction of its initial size, as discussed by Page in \cite{page2013time}. For black holes in our universe, both events occur when the black hole has approximately $90\%$ evaporated  \cite{page2013time}.

\begin{figure}[t]
\includegraphics[width = 0.5\linewidth]{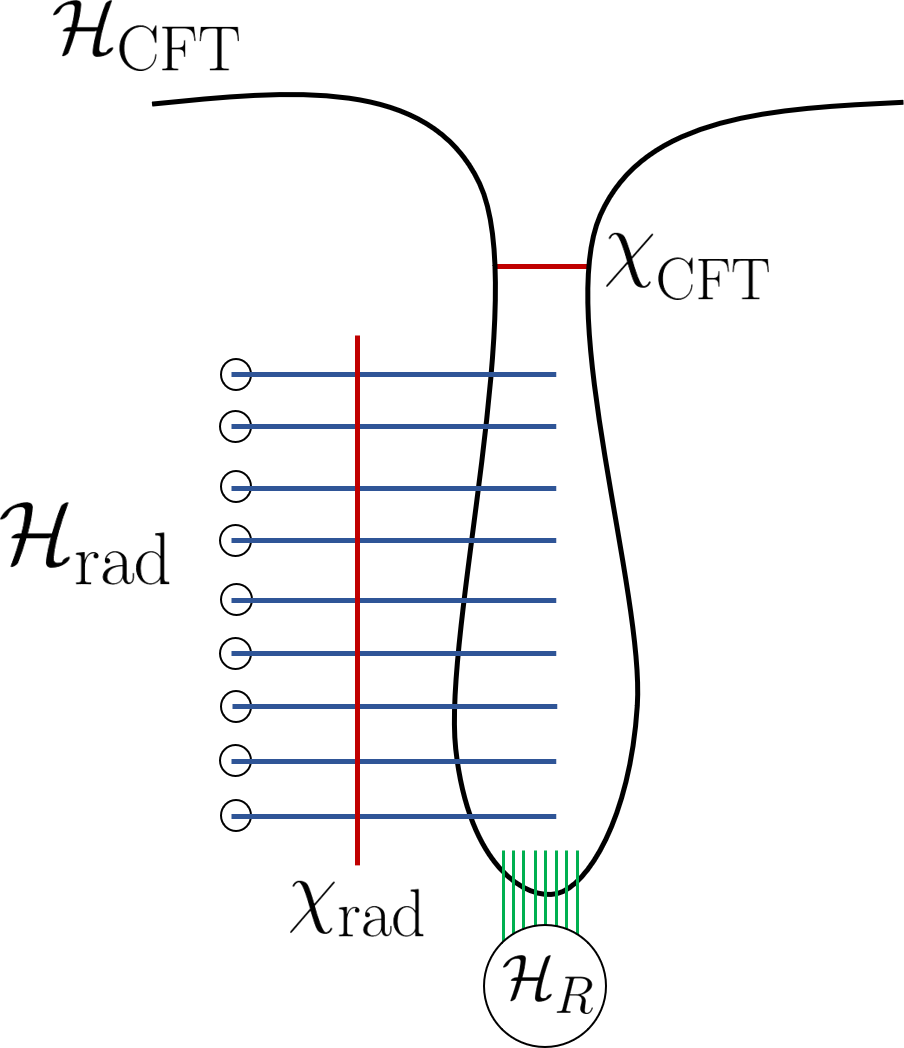}
\centering
\caption{Before the Page time, the interior is in the entanglement wedge of the CFT for any pure state. However, if the initial state is allowed to be highly entangled with a reference system, the Ryu-Takayanagi surface $\chi_{{}_\text{CFT}}$ for the CFT may jump to the non-empty quantum extremal surface. (In contrast, the RT surface $\chi_\text{rad}$ of the reservoir will remain empty.) This means that a reconstruction of the interior that acts only on the CFT, and not on the reservoir, will have to be at least somewhat state dependent, if the code space is too large.}
\label{fig:beforepage_reference}
\end{figure}

The same effect happens with reconstructions of the interior on the boundary CFT, before the Page time. In this case, a large amount of bulk entropy in the interior will increase the generalised entropy of the empty surface for $\mathcal{H}_\text{CFT}$, and so can make the Ryu-Takayanagi surface for $\mathcal{H}_\text{CFT}$ become non-empty. As shown in Figure \ref{fig:beforepage_reference}, the Ryu-Takayanagi surface of $\mathcal{H}_\text{CFT}$ will only be empty for highly mixed/entangled states if
\begin{align} \label{eq:statedependencebeforepage}
 S_\text{code} < S_{BH} - \Srad.
\end{align}
The interior can therefore only be reconstructed on $\mathcal{H}_\text{CFT}$ in a state-independent way for code spaces with entropy $S_\text{code}$ satisfying \eqref{eq:statedependencebeforepage}. Initially, we don't need to know very much about the state of the black hole to reconstruct the entire interior on the boundary CFT. However, as the black hole evaporates, the reconstruction becomes more and more state-dependent. Eventually, just before the Page time, one needs to know the exact initial state of the black hole. Again, \eqref{eq:statedependencebeforepage} agrees with toy models.

The part of the interior that lies between the non-empty quantum extremal surface and the boundary is always encoded in the CFT in a completely state independent way, both before and after the Page time. No amount of bulk entanglement with a reference system can stop the entanglement wedge of the CFT from including this region.

The non-empty extremal surface lies at a radius $O(G_N)$ inside the event horizon of the black hole. An outgoing lightcone starting from this extremal surface will therefore hit the singularity after an infalling time equal to the scrambling time plus an $O(\beta)$ correction after the infalling time of the extremal surface. This infalling time is within $O(\beta)$ of the `current' time, when the last radiation was extracted into $\mathcal{H}_\text{rad}$. The worldline of an observer, who jumps into the black hole at an $O(\beta)$ time into the future,  will remain entirely within the entanglement wedge of the CFT. The entire experience of the observer in the interior, until the curvature becomes large close to the singularity, will be encoded in the CFT in a completely state-independent way.

This is a somewhat comforting fact. It means that the accessible part of the black hole interior is always encoded in the state of the black hole itself, plus $O(1)$ number of recent quanta of outgoing Hawking radiation.\footnote{This is somewhat similar to recent work by Yoshida \cite{yoshida2019firewalls}, where it was shown,  in a qubit toy model of a black hole, that swapping an $O(1)$ number of degrees of freedom and then applying a scrambling unitary was sufficient to make new Hawking radiation be unentangled with the early Hawking radiation, even long after the Page time. Yoshida therefore argued that swapping in the new degrees of freedom had made the interior be encoded in the black hole degrees of freedom, plus the purification of these new degrees of freedom, with no dependence on the initial black hole state. In our case, it is simply the continuous increase with time of the combined thermodynamic entropy of the black hole and Hawking radiation that makes part of the interior be encoded in the CFT. See Appendix \ref{app:finitetemp} for an example of the extremal surface that lies exactly on the event horizon because there is no net increase in thermodynamic entropy.} The initial state of the black hole does not matter, nor do manipulations, even arbitrarily complicated ones, of Hawking radiation that escaped from the black hole a long time in the past.

In particular, so long as $r_s \ll l_{AdS}$, an observer who jumps into the black hole from the boundary will never leave the entanglement wedge of the CFT at the time that they left the boundary. This seems to still be true, even for large AdS black holes, at least when the Hawking radiation is extracted from inside the zone, as in Section \ref{sec:extremal}. Potentially, this is important for precomputation versions of the firewall paradox \cite{almheiri2013apologia}, where an observer attempts to extract a mode from $\mathcal{H}_\text{rad}$, which is expected to take exponential time \cite{harlow2013quantum}, before jumping into the black hole. 

However, as discussed in Appendix \ref{app:finitetemp}, by making the infalling modes be at a temperature very close to the black hole temperature, we can make the extremal surface be arbitrarily close to the event horizon. By tuning the state of the infalling modes, we can therefore always ensure that the observer is able to escape the entanglement wedge of the CFT and encounter the infalling mode that they extracted into $\mathcal{H}_\text{rad}$. The fundamental answer to the precomputation version of the firewall paradox seems to simply be that we can indeed manipulate the interior using $\mathcal{H}_\text{rad}$, so long as we are able to do very complicated, non-semiclassical manipulations. After all, it is well known that it is possible to manipulate the interior of a two-sided black hole in the thermofield double state, just by acting on the left CFT.

\subsection{Approximation to the Rescue} \label{sec:approx}

Just like the $\alpha$-bit codes found in \cite{hayden2018learning} and summarised in Section \ref{sec:abits}, the results that we found in Section \ref{sec:statedepend} only make sense because entanglement wedge reconstruction is approximate. This fact should be somewhat apparent from our discussion in Section \ref{sec:abits} and \ref{sec:statedepend}. However, in the interests of clarity, we now give a simple, explicit example of a paradox that would otherwise occur.

Suppose that we allow a black hole to evaporate until slightly before the Page time, storing the Hawking radiation in a reservoir $\mathcal{H}_{1}$. We then allow it to continue to evaporate until slightly after the Page time, storing the Hawking radiation in a different reservoir $\mathcal{H}_{2}$. Let the entropy of the Hawking radiation in $\mathcal{H}_{1}$ be $(1-\delta)S_{BH}$ and the entropy of the Hawking radiation in $\mathcal{H}_{2}$ be $2 \delta S_{BH}$. where $\delta>0$ is small and $S_{BH}$ is the Bekenstein-Hawking entropy of the black hole after all the evaporation has taken place. For reasons that will become clear, we shall refer to the combined state of the evaporating black hole and the Hawking radiation as the `entangled state'.

The black hole has evaporated beyond the Page time. The Ryu-Takayanagi surface of $\mathcal{H}_\text{rad}$ is therefore non-empty and lies just inside the horizon of the black hole. Most of the interior of the black hole is encoded in $\mathcal{H}_{1} \otimes \mathcal{H}_{2}$.

Now suppose that we do a complete measurement of $\mathcal{H}_{2}$ in some arbitrary basis. Regardless of the outcome of such a measurement, and regardless of the basis that we measure in, the Ryu-Takayanagi surface for the CFT will now be empty and so the interior will be encoded in $\mathcal{H}_\text{CFT}$. We shall refer to the states that result from such a measurement as `pure states', in contrast with the original `entangled state', because they are pure states in $\mathcal{H}_\text{CFT} \otimes \mathcal{H}_{1}$.

We now have exactly the same apparent paradox that was found in \cite{hayden2018learning} and discussed in Section \ref{sec:abits}. For \emph{any} pure state, the interior is encoded in the CFT. However, in the entangled state, it is encoded in $\mathcal{H}_{1} \otimes \mathcal{H}_{2}$ and so, by the no-cloning theorem, it \emph{cannot} be encoded in the CFT. The boundary Hilbert space $\mathcal{H}_\text{CFT}$ plays the role of the boundary subregion Hilbert space $\mathcal{H}_B$ that we had access to in Section \ref{sec:abits}; the early Hawking radiation reservoir $\mathcal{H}_{1}$ plays the role of the complementary boundary subregion Hilbert space $\mathcal{H}_{\bar B}$ and the later Hawking radiation reservoir $\mathcal{H}_{2}$ plays the role of the reference system $\mathcal{H}_R$.

As before, the resolution of this paradox is two-fold. Firstly, the reconstruction of the interior on $\mathcal{H}_\text{CFT}$ necessarily depends on the state of the interior modes that were previously entangled with $\mathcal{H}_{2}$.  Hence, there is no single reconstruction that will work for all possible states of those interior modes. This would be necessary to reconstruct the interior of the entangled state in $\mathcal{H}_\text{CFT}$. 

However, this resolution only works because the reconstruction is approximate. Since the reduced density matrix of the entangled state on $\mathcal{H}_2$ consists of a large number of approximately independently and identically distributed thermal modes, the smooth max entropy of $\mathcal{H}_{2}$ in the entangled state is approximately equal to its von Neumann entropy $2\delta S_{BH}$, up to subleading corrections of order $O(\sqrt{S})$ \cite{tomamichel2009fully}. By the definition of the smooth max entropy, this means that we can construct a code space $\mathcal{H}_\text{code} \subseteq \mathcal{H}_\text{CFT} \otimes \mathcal{H}_{1}$ satisfying
\begin{align}
\log |\mathcal{H}_\text{code}| = 2 \delta S_{BH} + O(\sqrt{S_{BH}}),
\end{align}
such that the entangled state can be approximated, with very high fidelity, by a state in $\mathcal{H}_\text{code} \otimes \mathcal{H}_{2}$.

Any interior operator $\mathcal{O}$ should have a state-independent global reconstruction $\mathcal{O}^\text{code}$ on $\mathcal{H}_\text{CFT} \otimes \mathcal{H}_{1}$ that works for the entire code space $\mathcal{H}_\text{code}$. If entanglement wedge reconstruction were exact, then, for any state $\ket{\psi}$, the state-dependent reconstruction $\mathcal{O}_\text{CFT}^{\psi}$ that acts only on the CFT would satisfy
\begin{align} \label{eq:exactequal}
\mathcal{O}_\text{CFT}^{\psi} \ket{\psi} = \mathcal{O}^\text{code} \ket{\psi}.
\end{align}
However, as shown in \cite{hayden2012weak, alphabits}, this would imply that there must also exist a state-independent reconstruction $\mathcal{O}_\text{CFT}^\text{code}$ that works for the entire code space and acts only on the CFT. The cloning paradox can only be resolved if there is an error term in \eqref{eq:exactequal} with size at least $\exp(-O(\delta \,S_{BH}))$ \cite{alphabits}.

This error is tiny; it is non-perturbatively small in $G_N$ for any fixed $\delta > 0$. It is expected that non-perturbative $\exp(-O(S))$ corrections to the bulk physics of black holes in AdS/CFT must exist in order for the decay of correlators to be consistent with boundary unitarity \cite{maldacena2003eternal}. However, there has been debate about whether such tiny errors can explain the large-scale ``$O(1)$'' paradoxes that show up in evaporating black holes. The answer is that they can and do. If the code space of allowed microstates has exponentially large dimension, exponentially small errors can be amplified in very entangled states and become $O(1)$ in size. 

Once we have measured $\mathcal{H}_2$, regardless of the measurement outcome we obtain, a diary in the interior of the black hole will be encoded in $\mathcal{H}_\text{CFT}$ and only has a non-perturbatively small effect on the state of $\mathcal{H}_1$. However, the entanglement with $\mathcal{H}_2$ amplifies these tiny differences, so that orthogonal diary states are almost exactly orthogonal on $\mathcal{H}_1 \otimes \mathcal{H}_2$. No magic is required; just the same mechanism that occurs in random unitary toy models \cite{alphabits}.

\subsection{Large Diaries} \label{sec:largediaries}
So far we have assumed that any diaries thrown into the black hole are small, both in energy and entropy. We have therefore been able to ignore both their backreaction on the geometry and their contribution to the bulk entropy. In this section, we remove those assumptions. 

If a heavy diary is thrown into a black hole before the Page time, it can still be reconstructed immediately after the Page time (so long as the entropy of the diary is small). The only change is that the Page time will be delayed by the increase in the horizon area of the black hole caused by the diary. By almost identical arguments to those in Section \ref{sec:statedepend}, if the diary also has a large entropy $S_\text{diary}$, we have to wait until
\begin{align}
\Srad - S_{BH} \geq S_\text{diary},
\end{align}
 so that the diary is contained in the entanglement wedge of $\mathcal{H}_\text{rad}$ even for highly mixed diary states. 

A more interesting situation occurs when a large diary is thrown into the black hole after the Page time. Let us first consider the case where the entropy of the diary is small, but the energy is large. The diary now causes a large backreaction on the geometry that significantly increases the horizon area. We assume, for simplicity, that the diary is encoded only in s wave modes and so the rotational symmetry of the spacetime is preserved. After we throw the diary into the black hole, the Bekenstein-Hawking entropy of the black hole will again be significantly larger than its entanglement entropy; heuristically, we will have made the black hole young again.

What happens to the quantum extremal surface in such a spacetime? Before the diary is thrown into the black hole, the quantum extremal surface will continuously move forwards in infalling time, at a radius just inside the event horizon, as radiation escapes the black hole into $\mathcal{H}_\text{rad}$. However, radiation that would have escaped shortly after the diary was thrown into the black hole will instead fall into the larger black hole created by the backreaction of the diary.

The actual radiation that escapes instead comes from close to the new event horizon, which, being teleological, already began moving out from the apparent horizon $r_s$ along an outgoing lightcone in anticipation of the diary falling in. Even at very late times, the Hawking radiation comes from outside the past lightcone of the boundary, which in turn will always lie outside the event horizon, although it will approach the horizon exponentially at any fixed time as we evolve the boundary forwards in time. 

Using \eqref{eq:dSdv} and \eqref{eq:dSdr}, it is therefore easy to see that the quantum extremal surface stops tracking along the horizon after the diary is thrown into the black hole, and instead asymptotes to a radius
\begin{align} \label{eq:prediaryr}
r = r_s - \frac{G_N c_\text{evap}}{3 (d-1) \Omega_{d-1} r_s^{d-2}},
\end{align}
at the infalling time $v$ when
\begin{align}
r_\text{hor}(v) = r_s + \frac{G_N c_\text{evap}}{3 (d-1) \Omega_{d-1} r_s^{d-2}}.
\end{align}
For simplicity, we are assuming here, as in Sections \ref{sec:extremal} and \ref{sec:haydenpage}, that the Hawking radiation is extracted from close to the horizon and so there are no greybody factors. 

If the change in horizon area $\delta A_\text{diary}$ caused by the diary is small compared to the original horizon area $A_\text{hor}$, we find that
\begin{align} \label{eq:prediaryv}
v = v_\text{diary} - \frac{\beta}{2\pi} \log \frac{\delta A_\text{diary}}{c_\text{evap} G_N} + O(\beta),
\end{align}
where $v_\text{diary}$ is the infalling time at which the diary is thrown into the black hole. In deriving \eqref{eq:prediaryv}, we have used the fact that the event horizon is an outgoing lightcone, obeying \eqref{eq:arbitraryoutgoinglc}, and that, at $v_\text{diary}$, the radius of the event horizon should be approximately equal to the new Schwarzschild radius of the black hole.

\begin{figure}[t]
\begin{subfigure}{.48\textwidth}
  \centering
 \includegraphics[width = 0.65\linewidth]{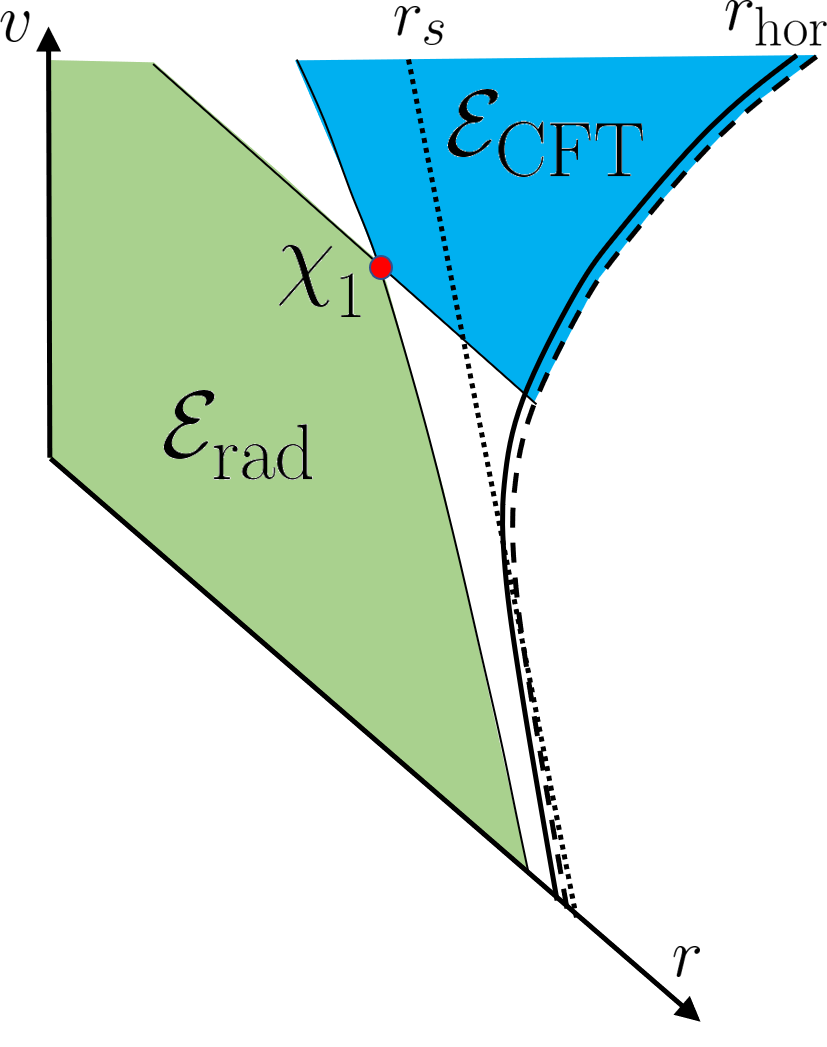}
\end{subfigure}
\begin{subfigure}{.48\textwidth}
  \centering
  \vspace{-2cm}
 \includegraphics[width = 0.9\linewidth]{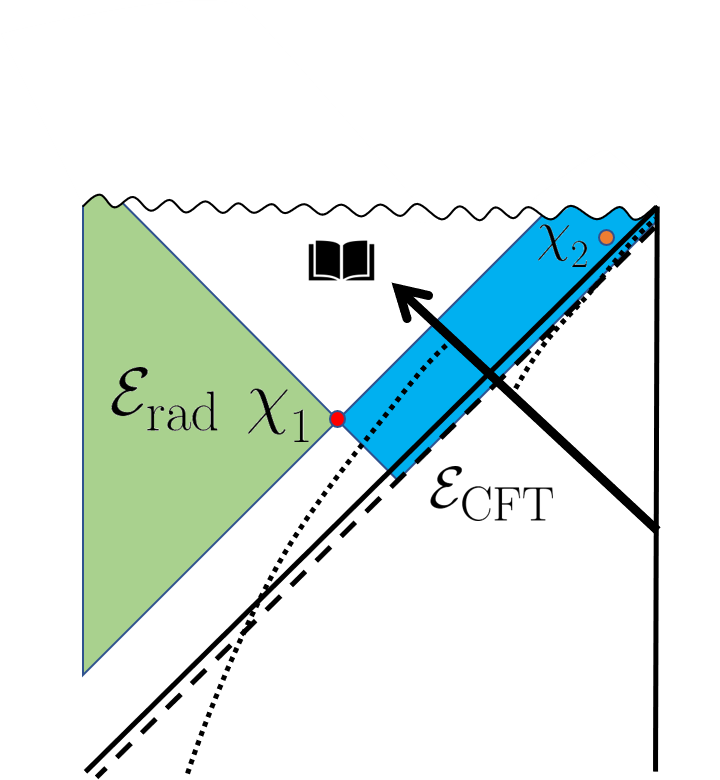}
\end{subfigure}
\centering
\caption{When a large diary is thrown into a black hole, the radius $r_\text{hor}$ of the event horizon (solid line) begins increasing in anticipation of the diary falling in, while the radius $r_s$ of the apparent horizon (dotted line) continues to slowly decrease until the diary is actually thrown into the black hole. As the black hole continues to evaporate, the past lightcone (dashed line) of the current boundary remains outside the event horizon. The Ryu-Takayanagi surface $\chi_1$ asymptotes to a point approximately the scrambling time before the diary was thrown in. There is a second quantum extremal surface $\chi_2$ at an infalling time after the diary is thrown in. Initially, this extremal surface is not the RT surface because its area is larger than the area of $\chi_1$. However, eventually, at the new Page time, there will be a phase transition, with $\chi_2$ becoming the new RT surface, and the diary can finally be reconstructed from the Hawking radiation, so long as its entropy is small. (Left: Eddington-Finkelstein coordinates, right: a Penrose diagram.)}
\label{fig:largediary}
\end{figure}

The entanglement wedge of $\mathcal{H}_\text{rad}$ will not contain the diary. The large amount of energy thrown into the black hole has stopped any information from escaping. The location of the extremal surface and the entanglement wedges in shown in Figure \ref{fig:largediary}.

However, this quantum extremal surface will not remain the Ryu-Takayanagi surface forever. As new radiation escapes from the black hole into $\mathcal{H}_\text{rad}$, the bulk entropy, calculated using this quantum extremal surface, will increase. Heuristically, the new Hawking radiation is entangled with interior modes that lie close to the new, larger black hole horizon and hence lie in the entanglement wedge of $\mathcal{H}_\text{CFT}$. More formally, as in \eqref{eq:changecutoff}, the cut-off $\varepsilon_2$ in \eqref{eq:bulkentropyoutgoing} shrinks exponentially as
\begin{align}
\varepsilon_2 \propto \exp(-2 \pi v_\text{rad} / \beta)
\end{align}
where $v_\text{rad}$ is the `current' infalling time  (i.e. the point in time where we last extracted Hawking radiation into $\mathcal{H}_\text{rad}$). In contrast, the radial distance
\begin{align}
r_{l.c.} - r \approx r_{hor} - r,
\end{align}
between the past lightcone and the RT surface is approximately constant. As a result, we find using \eqref{eq:bulkentropyoutgoing} that
\begin{align}
\frac{\partial S_\text{bulk}}{\partial v_\text{rad}} = \frac{c_\text{evap} \pi}{6 \beta}.
\end{align}
This is the increase in entropy that one finds with thermal outgoing modes that are purified by degrees of freedom in $\mathcal{H}_\text{CFT}$ \cite{calabrese2004entanglement, calabrese2009entanglement}. By adding energy, we have stopped information escaping the black hole and made the Hawking radiation be purely entangled with the CFT. The black hole did indeed become young again.

The surface $\chi_1$ at \eqref{eq:prediaryr}, \eqref{eq:prediaryv} remains a quantum extremal surface even at boundary times long after the diary was thrown into the black hole. However, it is not the only non-empty quantum extremal surface at such late times. There will also be a second non-empty quantum extremal surface $\chi_2$ that lies, as usual, approximately the scrambling time before the current boundary time. Because of the increase in horizon area created by the diary, the generalised entropy of the extremal surface $\chi_2$ will initially be significantly larger than the generalised entropy of the surface $\chi_1$.

However the generalised entropy of $\chi_1$ steadily increases over time because of the increase in bulk entropy discussed above. Meanwhile, the generalised entropy of $\chi_2$ decreases over time as the black hole evaporates. Eventually, after the new Page time of the black hole, the generalised entropy of $\chi_2$ will become smaller than the generalised entropy of $\chi_1$. The later extremal surface $\chi_2$ will be the Ryu-Takayanagi surface, and the diary will finally be in the entanglement wedge of $\mathcal{H}_\text{rad}$ and can be decoded.

\begin{figure}[t]
\begin{subfigure}{.48\textwidth}
  \centering
 \includegraphics[width = 0.8\linewidth]{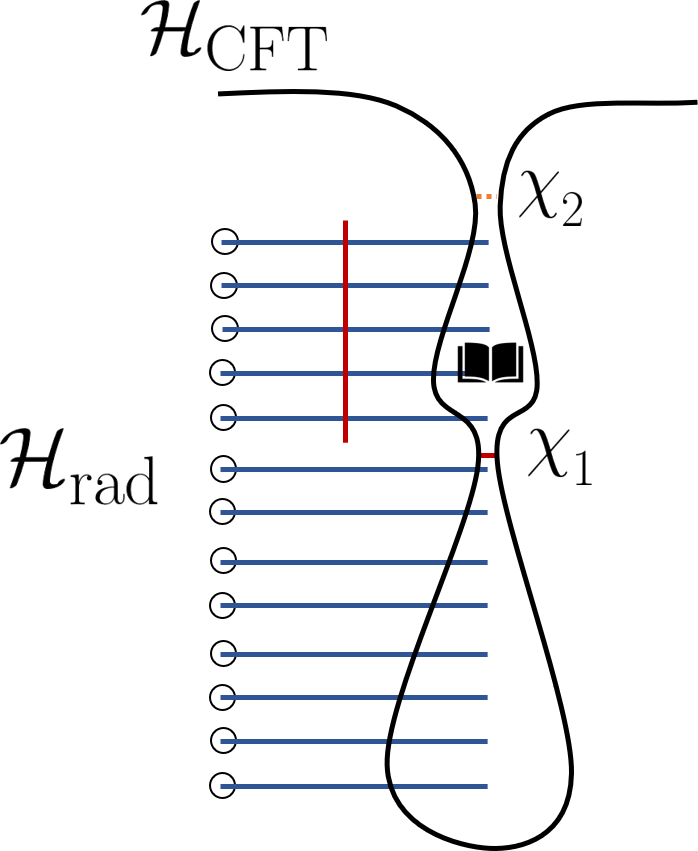}
\end{subfigure}
\begin{subfigure}{.48\textwidth}
  \centering
 \includegraphics[width = 0.8\linewidth]{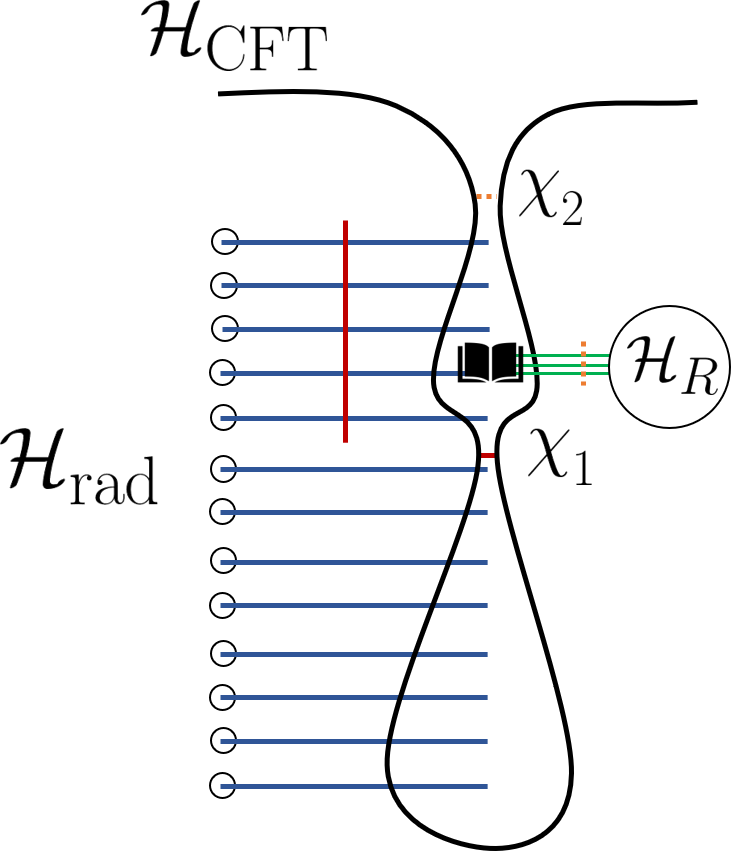}
\end{subfigure}
\centering
\caption{If a heavy diary, left, is thrown into the black hole, it can only be reconstructed from $\mathcal{H}_\text{rad}$ once the generalised entropy (solid lines) of the surface $\chi_1$ for $\mathcal{H}_\text{rad}$ is greater than the generalised entropy (dotted lines) of the surface $\chi_2$. The generalised entropy of $\chi_1$ has contributions from both the area term and the entropy of Hawking radiation emitted after the diary was thrown into the black hole. In contrast, so long as the diary has small entropy, the surface $\chi_2$ only has a contribution from its area, plus an $O(1)$ bulk entropy correction. However, if the diary also has a large entropy (right), the diary also needs to be in the entanglement wedge, even when it is in a highly mixed state, or is entangled with a reference system. This increases the generalised entropy of $\chi_2$.}
\label{fig:largediaryevap}
\end{figure}

If the diary had a large amount of entropy, as well as a large amount of energy, we would have to wait even longer in order to recover the diary. Even after the new Page time, when the Ryu-Takayanagi surface has a phase transition for pure diary states, the Ryu-Takayanagi surface of $\mathcal{H}_\text{rad}$ will still not contain the diary for highly mixed diary states. To actually recover the diary from the Hawking radiation, we need to wait until the diary is contained in the entanglement wedge of $\mathcal{H}_\text{rad}$, even for such highly mixed states, as shown in Figure \ref{fig:largediaryevap}. This requires
\begin{align} \label{eq:largediarycondition}
S_\text{rad}^\text{new} - \frac{\delta A_\text{diary} + \delta A_\text{evap}}{4G_N} > \log |\mathcal{H}_\text{diary}|
\end{align}
where $S_\text{rad}^\text{new}$ is the bulk entropy of the new radiation emitted after the diary was thrown into the black hole, $\delta A_\text{diary} > 0$ is the change in horizon area from throwing the diary into the black hole, $\delta A_\text{evap} < 0$ is the change in horizon area from the black hole evaporation after the diary is thrown into the black hole  and $\mathcal{H}_\text{diary}$ is the Hilbert space of the diary. 

Note that the generalised second law implies that 
\begin{align} \label{eq:thermo1}
\delta A_\text{diary}/ 4G_N \geq \log |\mathcal{H}_\text{diary}|
\end{align}
 and 
\begin{align} \label{eq:thermo2}
 S_\text{rad}^\text{new} \geq -\delta A_\text{evap}.
 \end{align}
Hence, whenever  \eqref{eq:largediarycondition} is satisfied, we will also have
\begin{align}
S_\text{rad}^\text{new} > \log |\mathcal{H}_\text{diary}|.
\end{align}
It follows that there is always sufficient entropy in the new Hawking radiation to encode the diary. Entanglement wedge reconstruction is consistent with quantum capacity bounds, so long as we consider mixed (or entangled) states in the code space.

It can be verified \eqref{eq:largediarycondition} is consistent with random unitary toy models \cite{hayden2018learning}, although, as discussed in Section \ref{sec:haydenpage}, most of the focus in random unitary models has been on evaporation that is either perfectly or approximately thermodynamically reversible, where \eqref{eq:thermo1} and \eqref{eq:thermo2} are equalities.

As in Section \ref{sec:statedepend}, the period when the interior reconstruction depends on the state of the diary does not merely make the amount of information encoded in the Hawking radiation be compatible with quantum capacity bounds; it also provides the mechanism by which information about the state of the diary escapes the black hole. The new Hawking radiation is entangled with the same interior modes, regardless of the state of the diary. However, because the encoding of those interior modes in $\mathcal{H}_\text{rad}$ depends on the state of the diary, an observer with access to the reservoir $\mathcal{H}_\text{rad}$ can learn information about the diary from the new Hawking radiation.

So far, both in this section and in Section \ref{sec:haydenpage}, we have not worried too much about the errors that exist in entanglement wedge reconstruction, even though we showed in Section \ref{sec:approx} that their existence was crucial to the consistency of our results. Reconstruction errors are also present in the Hayden-Preskill protocol in random unitary toy models of black holes. Since all our other results have been consistent with random unitary toy models, we might hope that the error in the Hayden-Preskill entanglement wedge reconstruction will also be consistent with random unitary toy models.

Unfortunately, the actual size of the errors in entanglement wedge reconstruction remains unknown. However, the \emph{lower bound} on their size that was derived in \cite{hayden2018learning} and discussed briefly in Section \ref{sec:abits} suggests that, for reconstruction with error $\varepsilon$ to be possible, the difference $\Delta S$ between the generalised entropy of an extremal surface for which the bulk operator would not be in the entanglement wedge and the generalised entropy of the Ryu-Takayanagi surface, where the bulk operator is in the entanglement wedge, must satisfy
\begin{align} \label{eq:deltaSSSSS}
\Delta S \geq O(\log \frac{1}{\varepsilon}).
\end{align}
Note that \eqref{eq:deltaSSSSS} needs to be satisfied for all states, both pure and mixed, in the code space. The exact coefficient in \eqref{eq:deltaSSSSS} depends on how the error is measured, and we will not worry about it here. 

The generalised entropy of the extremal surface $\chi_1$, where the diary is not in the entanglement wedge of $\mathcal{H}_\text{rad}$, increases by an $O(1)$ amount in $O(\beta)$ time. Similarly, the generalised entropy of the extremal surface $\chi_2$, where the diary is in the entanglement wedge of $\mathcal{H}_\text{rad}$, decreases by an $O(1)$ amount in $O(\beta)$ time.

If we make the strong assumption that the lower bound on the error $\varepsilon$ derived in \cite{hayden2018learning} is approximately saturated, we find that to reconstruct the diary with error $\varepsilon$, we need to wait for an additional time
$$
O(\beta \log(\frac{1}{\varepsilon}),
$$
even after the condition \eqref{eq:largediarycondition} is satisfied. Up to the (unstated) linear coefficient, this agrees, yet again, with random unitary toy models \cite{hayden2007black, alphabits}.

\subsection{Minimal State Dependence} \label{sec:minimal}

So far, we have avoided talking about state dependence for interior operators in black holes that have not evaporated at all, where there is no auxiliary reservoir $\mathcal{H}_\text{rad}$. Yet this is the situation in which state dependence is most commonly discussed \cite{papadodimas2013infalling, papadodimas2014state, kourkoulou2017pure}.

There is a very good reason for our reticence. Every proof of entanglement wedge reconstruction assumes that there is a global isometry from the bulk code space to the larger boundary Hilbert space. It is this isometry, combined with a partial trace over some of the boundary degrees of freedom, that creates a noisy quantum channel and, potentially, a quantum error correcting code. Yet, so long as such an isometry exists, there must always exist state-independent global boundary reconstructions. 

If the CFT is the only Hilbert space and pure bulk states correspond to pure boundary states, interior operators cannot be encoded in the CFT in a state-dependent way, within the framework of quantum error correction.

Nonetheless, suppose we take as an axiom the idea that boundary reconstructions are state independent if, and only if, the bulk operator is contained in entanglement wedge even for states in a `code space' of the bulk effective field theory that are entangled with a reference system.\footnote{We emphasize that this may no longer be a code space of a quantum error correcting code in the traditional sense.} If a code space of interior states $\mathcal{H}_\text{code}$ satisfies
\begin{align}\label{eq:smallenoughspace}
\log |\mathcal{H}_\text{code}| < S_{BH},
\end{align}
the entanglement wedge of the boundary should always include the interior, even if the state is entangled with a reference system.\footnote{The dimension of the code space here includes both allowed interior degrees of freedom, as well as any degrees of freedom describing an end-of-the-world brane that are also included in the code space, if our geometry ends in such a brane. See, for example, \cite{kourkoulou2017pure, cooper2018black} for discussion of interior geometries ending on end-of-the-world branes.}. A single reconstruction of interior operators should therefore exist that works for the entire code space.

In contrast, if
\begin{align} \label{eq:extrabigspace}
\log |\mathcal{H}_\text{code}| \geq S_{BH},
\end{align}
then we can make states in $\mathcal{H}_\text{code} \otimes \mathcal{H}_R$ where the RT surface is non-empty, and part of the interior is no longer contained in the entanglement wedge of the boundary.

To distinguish this idea from most of the literature on interior state dependence, which has focussed on constructing interior operators for a single microstate, or at most a small code space of microstates with $O(1)$ dimension, we shall refer to it as `minimal state dependence'. We emphasize, however, that this paper is far from the first to suggest it, see, for example, the discussion near the end of \cite{papadodimas2013infalling}.

As discussed, given that we are taking results derived using quantum error correction and applying them outside of that framework, we should be somewhat cautious about this idea. In particular, rather than making too many claims about code spaces for which \eqref{eq:extrabigspace} is true, it seems better to focus on the fact that, so long as a code space satisfies \eqref{eq:smallenoughspace}, we should be relatively confident that everything is well behaved and that all bulk operators have global, state-independent boundary reconstructions that work for the entire code space. 

An important question is how small
\begin{align}
\Delta S = S_{BH} - \log |\mathcal{H}_\text{code}|
\end{align}
can be without causing any problems. However, since we don't have the tools to answer such a question with any confidence, we shall be maximally cautious and assume that $\Delta S$ needs to be non-zero at leading order. In other words, we shall require
\begin{align}
\log |\mathcal{H}_\text{code}| \leq (1-\delta) S_{BH},
\end{align}
for some $\delta > 0$, which should be fixed in the semiclassical limit.

We can now show how minimal state dependence neatly avoids the so-called AMPSS argument \cite{almheiri2013apologia} that generic black hole microstates must have firewalls. 
 
The AMPSS argument goes as follows. Consider the subspace of CFT states within some narrow, but $O(1)$ width, energy band $M \leq E \leq M + \delta M$. If $M$ is sufficiently large and $\delta M$ is sufficiently small then all such states have a bulk description as a black hole. We then assume that there exists some state-independent operator $b_\omega^\dagger$ that acts as a raising operator for an interior Hawking mode, as well as an inverse operator $(1 + b_\omega^\dagger b_\omega)^{-1}\, b_\omega$. 

Acting with the operator $b_\omega^\dagger$ decreases the Schwarzschild energy by $\omega$ and so maps our subspace of CFT states into the energy band $M - \omega \leq E \leq M + \delta M - \omega$. But the number of states in this energy band is smaller than the number in our original energy band by a factor of approximately $e^{-\beta \omega}$. So the map cannot be invertible, contradicting our original assumption. The authors of AMPSS conclude that $b_\omega^\dagger$, and hence the interior, cannot exist.

It is, of course, well known that the argument above breaks down if the operator $b_\omega^\dagger$ is state dependent. Our point here is simply to emphasize that this is still true even if the operator $b_\omega^\dagger$ is only \emph{minimally} state dependent. 

Up to logarithmic corrections, the entropy of the maximally mixed state in the energy band $M \leq E \leq M + \delta M$ is equal to the Bekenstein-Hawking entropy. Assuming minimal state-dependence, we won't be able to find a single state-independent operator $b_\omega^\dagger$. Our code space is too large. If instead we restrict to a random code subspace, within the energy band and with entropy at most $(1-\delta)S_{BH}$, there will be plenty of space for the image of the code subspace in the energy band $M - \omega \leq E \leq M + \delta M - \omega$.

Of course, instead of looking at a random subspace of states within the energy window, we can alternatively make $\delta M$ so small that the entire code space of states in the energy band has a single state-independent reconstruction. To do so, it would be necessary to have 
\begin{align}
\delta M = O(e^{-\delta S_{BH}}).
\end{align}
If we insisted that the energy band $M \leq E \leq M + \delta M$ be mapped invertibly into the energy band $M - \omega \leq E \leq M + \delta M - \omega$, we would still have a paradox, because the latter band is still strictly smaller than the former. 

This would be far too strong a requirement however. More reasonably, we should only expect the energy of a state to decrease by $\omega$ plus non-perturbatively small corrections. However, since $\delta M$ is itself non-perturbatively small, this non-perturbatively small uncertainty in $\omega$ can significantly increase the size of the energy window. We can therefore avoid the paradox.

\section{Discussion} \label{sec:discuss}

\subsection{Summary of Results}
In this paper, we have argued that the key expected features of unitary black hole evaporation in AdS/CFT can be derived from the bulk semiclassical description of an evaporating black hole, so long as we assume entanglement wedge reconstruction. We now review those arguments.
\begin{itemize}

\item We studied entanglement wedge reconstruction in an evaporating black hole, formed from collapse, where the Hawking radiation was extracted out of the AdS space containing the black hole, with boundary Hilbert space $\mathcal{H}_\text{CFT}$, and into an auxiliary Markovian reservoir $\mathcal{H}_\text{rad}$. Importantly, we assumed that the bulk description of the evaporation was semiclassical, and so the bulk entanglement between the Hawking radiation and the interior of the black hole continued to grow, even after the Page time.

\item Using the maximin prescription, we saw that the quantum Ryu-Takayanagi surface, associated to the entire boundary Hilbert space $\mathcal{H}_\text{CFT}$, must become non-empty at the Page time. Since the overall state $\ket{\psi} \in \mathcal{H}_\text{CFT} \otimes \mathcal{H}_\text{rad}$ is pure, this non-empty RT surface will also be the RT surface for $\mathcal{H}_\text{rad}$.

\item If the Hawking radiation is extracted into $\mathcal{H}_\text{rad}$ from deep inside the zone, near the horizon, the greybody factors will all be either zero or one, depending on whether a given angular momentum mode is extracted or not. The location of the non-empty quantum Ryu-Takayanagi surface can then be calculated explicitly. It lies at a radius
\begin{align}
r = r_s -  \frac{c_\text{evap} G_N}{3 (d-1) r_s^{d-2} \Omega_{d-1}} = r_\text{hor} -  \frac{c_\text{evap} G_N}{6 (d-1) r_s^{d-2} \Omega_{d-1}},
\end{align}
where $r_s$ is the radius of the classical apparent horizon of the black hole, $r_\text{hor}$ is the radius of the event horizon of the black hole and $c_\text{evap}$ is the number of modes that are extracted into the reservoir $\mathcal{H}_\text{rad}$. The infalling time of the quantum extremal surface is given by
\begin{align}
v = - \frac{\beta}{2 \pi} \log\frac{S_{BH}}{c_\text{evap}} + O(\beta),
\end{align}
where $v=0$ is the current boundary time. A large part of the black hole interior lies in the entanglement wedge of the Hawking radiation reservoir $\mathcal{H}_\text{rad}$, rather than the boundary Hilbert space $\mathcal{H}_\text{CFT}$.

\item A small diary thrown into the black holes early in the evaporation can therefore be reconstructed from the Hawking radiation immediately after the Page time. A small diary thrown into the black hole after the Page time can be reconstructed after waiting for the scrambling time. These two results constitute the Hayden-Preskill decoding criterion \cite{hayden2007black}.

\item If the number of angular momentum modes $c_\text{evap}$ extracted into the Hawking radiation is large, there is a small, logarithmic decrease in the delay before the diary can be decoded from the radiation. This decrease is consistent with a heuristic picture of fast scrambling where a perturbation spreads exponentially through the degrees of freedom.

\item More generally, we showed that, in \emph{any} evaporating black hole after the Page time, the RT surface lies at an infalling time
\begin{align}
v = - \frac{\beta}{2 \pi} \log\frac{1}{G_N} + O(\beta),
\end{align}
where the subleading corrections depend on the details of the evaporation and, in general, cannot be analytically calculated, because of greybody factors. We can therefore derive the Hayden-Preskill decoding criterion, up to unknown, but subleading, corrections, even when non-trivial greybody factors are present.

\item Given the location of the Ryu-Takayanagi surface, it is an immediate consequence of the Ryu-Takayanagi formula that the entanglement between the CFT and the reservoir is given by the Page curve. Moreover, entanglement wedge reconstruction explains \emph{how} the entanglement entropy ends up decreasing (and how the AMPS firewall paradox is avoided). It decreases because the outgoing radiation is entangled with interior modes that are in the entanglement wedge of, and so are encoded in, $\mathcal{H}_\text{rad}$. 

\item If we consider the change in bulk entanglement entropy from transferring a small amount of Hawking radiation from $\mathcal{H}_\text{CFT}$ to $\mathcal{H}_\text{rad}$, we find exact quantitative agreement with the change in entropy given by the Ryu-Takayanagi formula, i.e. the Page curve.

\item This quantitative agreement does not only exist in the simple cases where we can calculate the Ryu-Takayanagi surface explicitly. It is a general consequence of the fact that the quantum Ryu-Takayanagi surface is an extremum of the generalised entropy. As with the Hayden-Preskill decoding criterion, we are therefore able to derive the Page curve, and avoid the firewall paradox, even when there are greybody factors.

\item As argued in \cite{hayden2018learning}, based on results about approximate operator algebra quantum error correcting codes derived in \cite{beny2007generalization, beny2009conditions, beny2010general}, \emph{state-independent} entanglement wedge reconstruction is only possible for a given code space if the bulk operator is contained in the entanglement wedge of the boundary region for all states, \emph{both pure and mixed} that have support only within the code space.

\item \emph{State-dependent} entanglement wedge reconstruction is possible so long as the bulk operator is contained in the entanglement wedge of the boundary region for all \emph{pure states} in the code space.

\item Using these results, we were able to derive the correct state dependence for Hayden-Preskill reconstructions. Immediately after the Page time, the diary can only be reconstructed if the exact initial black hole microstate is known. As the black hole continues to evaporate, less state dependence is required, in exact agreement with toy models. Specifically, a single reconstruction will work for a large code space of possible initial black hole states, with entropy $S_\text{code}$, so long as
\begin{align}
S_\text{code} < \Srad - S_{BH},
\end{align}
where $\Srad$ is the bulk entanglement entropy between the Hawking radiation and the interior and $S_{BH}$ is the Bekenstein-Hawking entropy of the black hole.

\item Similarly, before the Page time, reconstructions of the interior on the boundary CFT become increasingly state dependent as the black hole evaporates. The entropy $S_\text{code}$ of the code space of allowed initial states must satisfy
\begin{align}
S_\text{code} < S_{BH} - \Srad.
\end{align}
Eventually, at the Page time, the reconstruction is only possible if the exact initial black hole state is known.

\item The state dependence in the encoding of interior partners of Hawking modes in the early radiation (after the Page time) explains how the final (microscopic) state of the combined early and late radiation can depend on the initial state of the black hole (i.e. how we can avoid information loss) despite the state of the late Hawking mode and its interior partner having no dependence on the initial state.

\item These results are only consistent because entanglement wedge reconstruction is only approximate. Tiny, non-perturbatively small errors build up and have $O(1)$ effects.

\item When a heavy diary is thrown into the black hole, the Ryu-Takayanagi surface stops tracking along the horizon and instead asymptotes to a location approximately one scrambling time before the diary was thrown in. The Hawking radiation will contain no further information until the new Page time is reached, at which point the Ryu-Takayanagi surface will jump forwards in time and the diary can be decoded from the Hawking radiation. 

\item If the \emph{entropy} of the diary is large, as well as its energy, we have to wait even longer before it can be decoded. Specifically, we have to wait until the generalised entropy of the earlier quantum extremal surface, where the diary is not in the entanglement wedge of $\mathcal{H}_\text{rad}$, is greater than the generalised entropy of the later quantum extremal surface plus the entropy of the diary code space. 

\item All our results about Hayden-Preskill reconstructions of large diaries are consistent with random unitary toy models \cite{hayden2018learning}.

\item If we assume that the lower bound, derived in \cite{hayden2018learning}, on errors in entanglement wedge reconstruction is saturated up to a linear coefficient, we find that the errors in Hayden-Preskill reconstructions are consistent with toy models.

\item Finally, our arguments suggest that, even when a black hole has not evaporated at all, its interior can only be reconstructed with `minimal' state dependence. Such state dependence is beyond the framework of quantum error correction, but it provides a natural resolution to the AMPSS typical state firewall paradox.
\end{itemize}
In appendices,
\begin{itemize}
\item We give a simple pedagogical example of the importance of the coordinate dependence of cut-offs in entanglement entropy calculations.

\item We calculate the location of the Ryu-Takayanagi surface explicitly when the infalling modes are replaced by thermal modes or by pure modes of constant energy density. In each case we find that information stops escaping in the Hawking radiation (from the perspective of an observer with access either only to the reservoir, or to the reservoir and a purification of the thermal infalling modes) exactly at the moment that no information escaping becomes consistent with boundary unitarity.

\item We show that the Kourkoulou-Maldacena state-dependent interior reconstruction in the SYK model can be trivially extended to give minimally state-dependent reconstructions.
\end{itemize}

\subsection{Entanglement Wedge Reconstruction in Toy Models}
Throughout this paper, we have found that bulk calculations, using entanglement wedge reconstruction and the Ryu-Takayanagi formula, agree perfectly with random unitary and fast scrambling toy models of the boundary dynamics. 

Not only do our results agree with the Page curve and the Hayden-Preskill decoding criterion, but we also found exactly the right state dependence, for both reconstructions acting on the CFT before the Page time and reconstructions acting on the reservoir after the Page time. Our results for large diaries were consistent with toy models as a function of the both the energy and the entropy of the diary, as were the reconstruction errors so long as we assumed that the lower bound from \cite{hayden2018learning} was approximately saturated.

This seems either to be a somewhat remarkable coincidence, or to involve some deep magic of quantum gravity. In fact, it is neither of these things. We simply have the direction of causation in reverse. Rather than random unitary models determining the consequences of entanglement wedge reconstruction, entanglement wedge reconstruction determines the behaviour of random unitary toy models.

A random unitary toy model of black hole evaporation is an exceptionally trivial example of a random tensor network \cite{hayden2016holographic}. An iterated random isometry model like that in Figure \ref{fig:irreversible} is another, more complicated, example.

However, random tensor networks are well known to obey the Ryu-Takayanagi formula and hence have entanglement wedge reconstruction \cite{hayden2016holographic}. It is therefore inevitable that random unitary models agree with results derived \emph{from} Ryu-Takayanagi and entanglement wedge reconstruction. Indeed, one of the most popular methods to prove results about error correction in random unitaries, the so-called decoupling approach \cite{hayden2008decoupling, dupuis2010decoupling}, essentially involves deriving entanglement wedge reconstruction from the Ryu-Takayanagi formula.

Of course, random tensor networks do not have all the properties of holographic spacetimes. In particular, they are not covariant. A tensor network corresponds to a single timeslice of a bulk spacetime; RT surfaces, in a tensor network, are minimal, not extremal, surfaces. 

For many of the calculations in this paper, for example the scrambling time delay in the Hayden-Preskill criterion, the extremality of the surface, or equivalently the maximisation over Cauchy slices in the maximin prescription, was crucial in deriving the correct results. In particular, the covariant Ryu-Takayanagi surface somehow knows about the fast scrambling dynamics of the boundary theory. If there is any magic going on, it seems to be here.

\subsection{The Post Evaporation State and the Bulk-to-Boundary Map} \label{sec:postevaporation}
While we have studied evaporating black holes both before and after the Page time in this paper, we have not discussed the final state of the system, after the evaporation is complete. We make some comments about this state now.

When the black hole has nearly completely evaporated, the horizon curvature will become large and so stringy and Planckian effects will become important. We can no longer trust semi-classical calculations.
\begin{figure}[t]
\includegraphics[width = 0.37\linewidth]{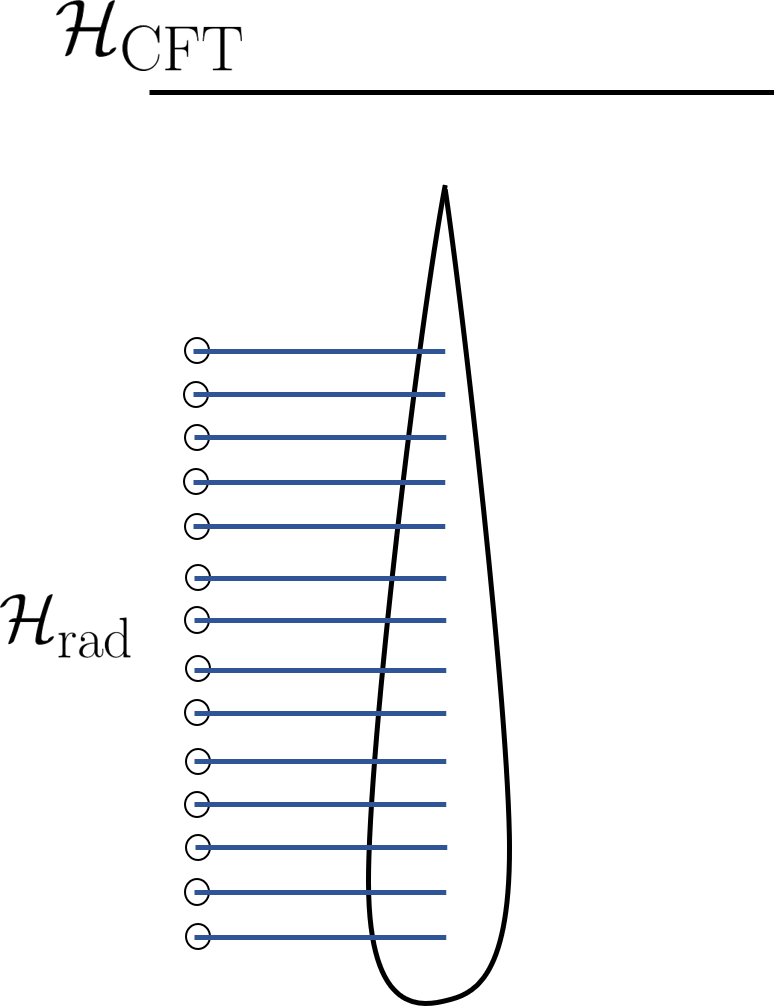}
\centering
\caption{After the black hole has completely evaporated, the bulk encoded in the CFT will be in the vacuum state. However, we can choose a (disconnected) bulk Cauchy slice where the radiation in the reservoir $\mathcal{H}_\text{rad}$ is still be entangled with the pinched-off interior wormhole, which lies in its entanglement wedge. The Ryu-Takayanagi surface is empty and has zero generalised entropy; the two systems $\mathcal{H}_\text{CFT}$ and $\mathcal{H}_\text{rad}$ are therefore unentangled.}
\label{fig:finished}
\end{figure}

However it is reasonable to expect that, after an indeterminate, but short, time, there will no longer be any sort of smooth connected geometry between the wormhole and the AdS space that previously contained the black hole. The mouth of the wormhole will have closed; the black hole will have completely evaporated. Let us assume we have extracted all the remaining energy out of the AdS space and into the reservoir $\mathcal{H}_\text{rad}$, so that the original bulk AdS space lies in the vacuum state. This is shown schematically in Figure \ref{fig:finished}.

Because the closed wormhole geometry has no boundary, on its own, the Ryu-Takayanagi \emph{surface} is no longer sufficient to define an entanglement wedge, and hence a generalised entropy. For example, if the RT surface is empty, we need to further specify whether the wormhole is in the entanglement wedge of the reservoir or of the CFT, before the generalised entropy is well-defined. In this case, it is obvious that the RT surface should be empty, with the wormhole in the entanglement wedge of the reservoir, since this gives zero generalised entropy.\footnote{Recall that the empty surface is contained in every Cauchy slice, and so will be the RT surface if there is any Cauchy slice for which it has minimal generalised entropy.}

There is no entanglement between $\mathcal{H}_\text{CFT}$ and $\mathcal{H}_\text{rad}$; both states are pure. Entanglement wedge reconstruction tells us that the state of the wormhole is encoded in the Hawking radiation. Any information thrown into the black hole during the evaporation will be contained in the entanglement wedge of $\mathcal{H}_\text{rad}$, no matter how large the entropy of the initial state. The information thrown into the black hole is therefore encoded in $\mathcal{H}_\text{rad}$ in a completely state-independent way.\footnote{In contrast, the interior of the black hole, from before it began to evaporate, appears to be encoded in the reservoir $\mathcal{H}_\text{rad}$ in exactly the same minimally state dependent way that it was originally encoded in the CFT.} All the information has been preserved.

Even though we know from the Ryu-Takayanagi formula that the state of the Hawking radiation in $\mathcal{H}_\text{rad}$ is pure, it appears from a bulk perspective that it is still entangled with the closed-off wormhole. How should we understand this seeming contradiction? 

An entangled state of the Hawking radiation and wormhole can be written as
\begin{align} \label{eq:hawkpostevapentangled}
\ket{\psi} = \sum_i \sqrt{p_i} \ket{\phi_i} \ket{\chi_i},
\end{align}
where the states $\ket{\phi_i}$ describe the radiation and the states $\ket{\chi_i}$ describes the wormhole. The Hilbert space of a holographic theory is associated with its boundary. However, the wormhole has no boundary. We therefore conclude (perhaps somewhat controversially) that its Hilbert space is trivial; it is isomorphic to the complex numbers $\mathbb{C}$. The states $\ket{\chi_i}$ are therefore simply complex coefficients $c_i$. The state
\begin{align} \label{eq:hawkpostevap}
\ket{\psi} = \sum_i \sqrt{p_i} c_i \ket{\phi_i}
\end{align}
is therefore simply some complicated, pure state in $\mathcal{H}_\text{rad}$.

Of course, there exists a perfectly valid bulk description of the same bulk state $\ket{\psi}$ from \eqref{eq:hawkpostevap} where the Hawking radiation is simply in a bulk state that has a complicated entanglement structure, but no wormhole.\footnote{If it is not clear that the Markovian reservoir $\mathcal{H}_\text{rad}$ has a bulk description at all, recall that we can imagine throwing each small chunk of Hawking radiation into its own copy of anti-de Sitter space.} Some version of black hole complementarity, or the $ER=EPR$ duality \cite{maldacena2013cool}, makes it equally valid to suppose that the wormhole still exists, or that it has vanished leaving some complicated pure state of the Hawking radiation.

If we apply some complicated unitary operator to $\mathcal{H}_\text{rad}$, we can transform the complicated state $\ket{\psi}$ into a simple state, say $\ket{\psi_0}$. From the perspective where the wormhole continues to exist, we will still be in an `entangled state'
\begin{align}
\sum_i \sqrt{p_i'} \ket{\psi_i} \ket{\chi_i'},
\end{align}
where there are many non-zero $p_i'$. However, the states $\ket{\chi_i'}$ are now very complicated superpositions of the original `simple' wormhole states $\ket{\chi_i}$. Each term in each superposition is simply a complex number. For $\ket{\chi'_0}$ the coefficients in the superposition must add constructively, so that $\ket{\chi'_0}$ is some large complex number $c_0'$. For $\ket{\chi'_i}$ with $i \neq 0$, they must interfere destructively so that $\ket{\chi'_i} = 0$. The `entangled state' really is just $\ket{\psi_0}$.

Just like in ordinary AdS/CFT, we have a linear `bulk to boundary' map from the state of bulk fields $\mathcal{H}_\text{bulk}$ on a fixed background spacetime (in this case the closed wormhole) to a `boundary' state of the spacetime itself. However, since the `boundary' Hilbert space is trivial, up to normalisation, this map simply projects the bulk fields on the wormhole into a particular, very complicated state. 

The coefficients $c_i$ which describe the `boundary' state associated to a particular wormhole state are simply the coefficient of that state in the projected wormhole state.
This story is, in effect, simply the Horowitz-Maldacena final state postselection proposal \cite{horowitz2004black}. From a bulk perspective, Horowitz and Maldacena suggested that the projection happens at the singularity. Our arguments suggest that, in the microscopic boundary description of the theory, it happens the moment that the wormhole closes off, even if we choose a bulk Cauchy slice that includes the wormhole. 

In fact, the process that ends in this final state projection begins immediately after the Page time, long before the black hole fully evaporates. At this point, there are, for the first time, more `orthogonal' states of the interior fields, that are needed to describe the entanglement with the Hawking radiation, than there are microstates of the black hole according to the Bekenstein-Hawking entropy. 

As a result, even though each `orthogonal' pair of bulk states will be almost orthogonal on the boundary, there must exist very complicated superpositions of states of the interior fields that are annihilated by the map to the boundary. Indeed, this is what allows the entanglement entropy between the reservoir and the CFT to be much less than the bulk entanglement entropy between the radiation and the interior, in accordance with the Ryu-Takayanagi formula.

As with the post-evaporation state, the combined state of the black hole and radiation, after the Page time can be written as
\begin{align}
\ket{\psi} = \sum_i \sqrt{p_i} \ket{\phi_i} \ket{\chi_i},
\end{align}
where the probabilities $p_i$ are determined by the semiclassical bulk evaporation, the states $\ket{\phi_i}$ describe the Hawking radiation in the reservoir and the states $\ket{\chi_i}$ now describe black hole microstates, which are encoded as CFT states. If there was an isometry, or even an approximate isometry, mapping each apparently distinct microstate $\ket{\chi_i}$ to a different CFT state, the entanglement entropy between the reservoir and the CFT would be equal to $\Srad$. However, minimal state dependence says that the bulk `code space' of microstates $\ket{\chi_i}$ is too large for such an isometry to exist. Since the map from bulk states to boundary states is not an isometry, there is no inconsistency with the entanglement entropy between $\mathcal{H}_\text{CFT}$ and $\mathcal{H}_\text{rad}$ actually being the Bekenstein-Hawking entropy $S_{BH}$.

It is often suggested that state dependence makes quantum mechanics nonlinear. However the map from bulk states to boundary states is perfectly linear in this proposal; it just isn't an isometry. In effect, the na\"{i}ve inner product on the bulk effective field theory is very different from the pull back of the boundary inner product to bulk states, which defines the actual microscopic inner product of the quantum theory. What then is the bulk inner product? The most natural answer, which is consistent with recent work on JT gravity \cite{saad2019jt}, is that the bulk inner product is a statistical average of the boundary inner product over an ensemble of microscopic Hamiltonians, such as a small range of couplings.

It is well known that final state projection models can lead to issues with describing measurements done by an observer falling into the black hole \cite{lloyd2014unitarity, bousso2014measurements, marolf2017black}. While we leave a detailed accounting of these issues to future work, the solution to at least some of these problems appears to be quantum error correction. 

So long as we only consider a sufficiently small code spaces of allowed states, for example the code space of states describing the state of the observer jumping into the black hole, along with her experimental apparatus, there is always an isometry from the bulk to the global boundary (including the reservoir). 

On the boundary, the evolution is described by standard quantum mechanics with no postselection. In the bulk, the observer is free to manipulate the unitary evolution of the experiment. The state of the experiment can also decohere and become entangled with the state of the observer; in more conventional phrasing, the observer can measure the experiment.\footnote{It is important to note that causality prevents this decoherence from escaping the black hole. No measurement by an interior observer will ever `have happened' from the perspective of an observer that remains outside the black hole.} The isometry from the code space to global boundary states maps all these events to ordinary unitary quantum mechanics with no postselection on the global boundary. The only change that happens as the black hole evaporates is that the observer and the experiment end up being encoded in the reservoir $\mathcal{H}_\text{rad}$, rather than the CFT.

\subsection{The Peak of the Page Curve}
The other part of the evaporation that we skipped, in the interests of avoiding speculative discussion, was the period of time, very close to the Page time, when the Page curve peaks and begins to decline. 

In a simple random unitary toy model of black hole evaporation, the entanglement entropy is almost exactly equal to the number of qubits in the radiation, until the black hole is within an $O(1)$ number of qubits of half its original size. There is then an $O(1)$ correction to the entropy at the peak of the curve, before the entropy becomes approximately equal to the number of qubits describing the black hole an $O(1)$ number of qubits later \cite{page1993average}.

For an actual black hole, at times within $O(t_\text{page}/\sqrt{S_{BH}})$ of the Page time, $O(\sqrt{G_N})$ fluctuations in the horizon area of the black hole and $O(\sqrt{S})$ fluctuations in the total energy of the Hawking radiation mean that there will not be a single well-defined Ryu-Takayanagi surface. At such times, we should therefore expect that the entanglement entropy will neither increase as fast as the bulk entropy of the radiation, nor decrease as fast as the Bekenstein-Hawking entropy of the black hole. 

However, we can write the total state of the black hole and Hawking radiation as a superposition of $O(\sqrt{S})$ states, each of which has only $O(G_N)$ fluctuation in the horizon area and where the entanglement spectrum of the Hawking radiation, in each state, has $O(1)$ width.\footnote{We cannot reduce the area fluctuations by more than this without substantially altering the bulk geometry, see \cite{bao2018beyond, bao2019holographic}, for example, for details.}  

For almost all of the states in such a superposition, there is still a well defined Ryu-Takayanagi surface. Indeed, if we believe that the lower bound on the reconstruction error from \cite{hayden2018learning} is approximately saturated, then, at any given time, the quantum extremal surface prescription will be valid with very small error for all but an $O(1)$ number of the states in this superposition. 

For some fraction $f$ of the states in the superposition, the interior will be encoded in the reservoir $\mathcal{H}_\text{rad}$, while for almost all the rest, it will be encoded in the CFT. As the black hole evaporates, the fraction $f$ will increase, until eventually the fraction $f$ approaches one and the interior can be reconstructed using only the reservoir, with only a very small error. From the fraction $f$, as a function of time, it should, in principle, be possible to calculate the shape of the peak of the Page curve.

Because the fluctuations in the evaporation rate smear out the Page time over an $O(\beta \sqrt{S_BH})$ time window, it seems plausible that, with sufficient work, one could calculate the entropy, up to an $O(1/\sqrt{S_{BH}})$ error, at all times. In other words, the fluctuations in the area of the black hole and the energy of the Hawking radiation may make it feasible to calculate the Page curve much more precisely than would otherwise be possible.

\subsection{Explicit Interior Reconstruction} \label{sec:explicit}
While we were able to make very precise statements in this paper about when and where information was encoded in an evaporating black hole, we said comparatively little about \emph{how} the information was encoded. In particular, one would ideally want to have explicit, even if not necessarily practical, reconstructions of the interior operators. 

There has been considerable work done in recent years on understanding how entanglement wedge reconstruction can be done explicitly \cite{cotler2017entanglement, faulkner2017bulk, chen2019entanglement}. However, these approaches often assume knowledge of some global reconstruction, such as HKLL \cite{hamilton2006local}. For interior operators, it is not clear even what global reconstruction to use as a starting point. Infalling modes can, of course, simply be evolved back in time to give exterior operators, but this is not possible for outgoing interior modes. 

This is not a new problem; it has been a major focus of research in AdS/CFT for many years. In particular, there are some credible suggestions of ways in which one can construct a state-dependent interior operator, given a particular choice of microstate \cite{papadodimas2013infalling, papadodimas2014state, kourkoulou2017pure}. If we believe the arguments from Section \ref{sec:minimal}, however, reconstructions should in principle be possible for much larger code spaces.

In Appendix \ref{app:syk}, we give a simple generalisation of the Kourkoulou-Maldacena state-dependent interior reconstruction for the SYK model that works for code spaces with entropy almost as large as the Bekenstein-Hawking entropy.

We can also make the following more general argument for extending state-dependent reconstructions that work for a single microstate to minimally state-dependent reconstructions. Suppose we assume the existence of a single \emph{unknown} reconstruction $\phi_\text{code}$ of an interior operator $\phi$ for a large code space $\mathcal{H}_\text{code}$ with orthonormal basis $\ket{i}$. Moreover, suppose we also assume that for any state $\ket{i}$ in this basis, we know an explicit reconstruction $\phi_i$, which is consistent with the unknown reconstruction $\phi_\text{code}$. In other words, for all $\ket{i}$,
\begin{align}
\phi_\text{code} \ket{i} \approx \phi_i \ket{i}.
\end{align}
But then for any state 
$$\ket{\psi} = \sum_i c_i \ket{i}$$
we have
\begin{align}
\sum_i \phi_i \ket{i}\braket{i|\psi} = \sum_i c_i \phi_i \ket{i} \approx \phi_\text{code} \ket{\psi}.
\end{align}
Hence
\begin{align}
\tilde{\phi}_\text{code} = \sum_i \phi_i \ket{i}\bra{i}
\end{align}
is an explicit reconstruction that approximates $\phi_\text{code}$ when acting on any state in the code space.

\subsection{The Information Paradox Beyond AdS/CFT}
This paper is entirely about AdS/CFT. However, by understanding the information paradox in AdS/CFT, we hope to eventually learn something about the information paradox in more general quantum gravity. Does information escape black holes in our universe (in the absence of the cosmic microwave background etc.) and if so how does it do so?

So far, entanglement wedge reconstruction, and the Ryu-Takayanagi formula, are only understood in the context of AdS/CFT. However, there is no obvious reason to think that they are specific to spacetimes with a negative cosmological constant. In particular, for asymptotically flat spacetimes, one can anchor a `Cauchy' slice at some `boundary' surface in asymptotic future infinity, and then calculate quantum extremal surfaces based on this. Indeed, in this case, one does not even need to do anything special to get absorbing boundary conditions. There is no timelike boundary for modes to reflect from. Instead, early modes will simply automatically not be included if they reach the asymptotic infinity at an earlier outgoing time than the time at which we anchored our `Cauchy' slice.

For de Sitter spacetimes, which most resemble our universe, there is no timelike or lightlike asymptotic region that we can use to anchor spacelike slices. However, one would still hope that the basic conceptual ideas of this paper -- essentially the fact that there is a state-dependent encoding of the black hole interior in the early Hawking radiation after the Page time -- might be relevant.

As an intermediate step, consider the case of black holes in AdS/CFT that are small enough to be microcanonically unstable. These black holes are so small that we do not need to extract Hawking radiation into an auxiliary system for the black hole to evaporate; the black hole will have already evaporated by the time the Hawking radiation can reach the boundary and come back.

If we don't extract the Hawking radiation, there is no entanglement wedge that can show us that the interior is encoded in the Hawking radiation after the Page time. The Hawking radiation entirely surrounds the black hole horizon; there is no boundary region whose entanglement wedge contains the radiation, but not the black hole.

However, if we do extract the Hawking radiation, which we can do using non-local boundary dynamics, it is clear from entanglement wedge reconstruction that the interior is indeed encoded in the Hawking radiation, just like for larger AdS black holes. The interior must still have been encoded in the Hawking radiation before we extracted the radiation; we just had no way to learn this using only entanglement wedge reconstruction.

To directly see that the interior was encoded in the radiation, even before we extracted the radiation, we would need a way to distinguish the microscopic degrees of freedom encoding a small neighbourhood of the black hole from the microscopic degrees of freedom describing the Hawking radiation further out. This would require understanding holography beyond asymptotic boundaries. There has been considerable recent progress in that direction using $T \bar T$ deformations of conformal field theories \cite{cavaglia2016mathrm, mcgough2018moving, kraus2018cutoff, donnelly2018entanglement, gorbenko2018ds}.

\section{Acknowledgements}
I would like to thank Raphael Bousso, Netta Engelhardt, Daniel Harlow, Frances Kirwan, Lampros Lamprou, Juan Maldacena, Don Page, Daniel Ranard, Phil Saad, Eva Silverstein, Jon Sorce, Steve Shenker, Douglas Stanford, Alex Streicher, Lenny Susskind, Mae Teo and Aron Wall for valuable discussions. In particular, I would like to thank my advisor, Patrick Hayden, for his invaluable support throughout, Edward Witten, for first suggesting that arguments similar to those in \cite{hayden2018learning} could potentially explain the black hole information paradox and for detailed feedback on this manuscript, and Ahmed Almheiri, for explaining his paper \cite{almheiri2018holographic} to me and for other very valuable discussions. This work was supported in part by AFOSR award FA9550-16-1- 0082 and DOE award {DE-SC0019380}.

\appendix

\section{Cut-offs in Rindler Space} \label{app:rindler}
In this appendix, we study a simple pedagogical example of a situation where understanding the coordinate dependence of cut-offs is vital, if we want to calculate entanglement entropies correctly. The example is closely related to, but distinct from, the black hole extremal surface calculations in Sections \ref{sec:extremal} and \ref{sec:greybody}.

Consider the interval $[0,r]$ of the vacuum state in some $(1+1)$-dimensional conformal field theory. The entanglement entropy of this interval is given by
\begin{align} \label{eq:cutoffsrindler}
S = \frac{c}{3} \log \frac{r}{\sqrt{\varepsilon_1, \varepsilon_2}}
\end{align}
where $\varepsilon_1$ and $\varepsilon_2$ are the cut-offs at each end of the interval \cite{calabrese2004entanglement, calabrese2009entanglement}. We therefore find that the derivative of the entanglement entropy
\begin{align} \label{eq:entropymink}
\frac{d S}{d r} = \frac{c}{3 r}.
\end{align}
We can also do this calculation in Rindler space, where it corresponds to finding the derivative of the entropy of an infinite half-line of a thermal state at inverse temperature $\beta = 2 \pi$. The entropy of a long interval of a thermal state of a CFT is equal to
\begin{align}
S = \frac{\pi c\, l}{3 \beta}
\end{align}
where $l$ is the length of the interval \cite{calabrese2004entanglement, calabrese2009entanglement}. We therefore find that
\begin{align}
\frac{d S}{d r^{*}} = \frac{c}{6}
\end{align}
where the Rindler position $r^* = \log r$. Hence
\begin{align} \label{eq:entropyrind}
\frac{d S}{d r} = \frac{1}{r} \frac{d S}{d r^*} = \frac{c}{6 r}.
\end{align}
However, this differs from \eqref{eq:entropymink} by a factor of two. We apparently have a contradiction. 

We can also look at the entanglement entropy of the interval $[r,\infty]$. In the Minkowski space calculation, the gradient of the entropy of an interval tends to zero as the length of the interval tends to infinity. However, for the thermal state in Rindler space, the gradient is simply the negative of \eqref{eq:entropyrind}.

The resolution of the paradox is, of course, the cut-offs. Implicitly, we assumed in the Minkowski space calculation that the cut-off was constant in units of $r$, while in the Rindler space calculation we assumed that it was constant in terms of the Rindler position $r^*$. However, $r$ and $r^*$ are nonlinearly related. So the cut-off cannot be constant in both units.

Let us assume that we actually wanted the cut-off to be constant in terms of the Rindler position $r^*$. Recall that a constant cut-off in units of $r^*$ really means that the cut-off is equal to
$$
\varepsilon_0 \frac{\partial}{\partial r^*},
$$
for some constant $\varepsilon_0$. Since
\begin{align}
\frac{\partial}{\partial r^*} = r \frac{\partial}{\partial r}
\end{align}
 the cut-off $\varepsilon_2$ in \eqref{eq:cutoffsrindler}, which is in units of $r$, is $r\, \varepsilon_0$. We therefore find that
 \begin{align}
 \frac{d S}{d r} = \frac{c}{6 r},
 \end{align}
 while the derivative of the entropy for the interval $[r,\infty]$ is $-c / 6 r$. The results now agree with the Rindler space calculation.

\section{Finite Temperature Infalling Modes} \label{app:finitetemp}

In this appendix, we generalise the explicit calculation of the location of the non-empty quantum extremal surface from Section \ref{sec:extremal} to thermal infalling modes at finite temperature, and to pure infalling modes with constant energy density and without long range entanglement. As in Section \ref{sec:extremal}, we assume that the outgoing modes are extracted from close to the horizon and so there are no greybody factors. We assume throughout the section that we are after the Page time, and so the non-empty quantum extremal surface is the Ryu-Takayanagi surface.\footnote{The exception is when the temperature, or energy density of the infalling modes is sufficient to prevent the black hole ever reaching the Page time. As we shall see, in those cases, there does not exist a non-empty quantum extremal surface at all.} Throughout this section, we shall work in Eddington-Finkelstein coordinates, as in Section \ref{sec:extremal}. However, the calculations can also easily be done in Kruskal-like coordinates, as in Section \ref{sec:greybody}.

We begin by studying the case of thermal infalling modes at a temperature $T'$ that may be different from the black hole temperature $T$. We assume that the infalling modes are purified by an auxiliary Hilbert space $\mathcal{H}_\text{pur}$ and hence, importantly, are unentangled with the outgoing Hawking radiation. We consider both the information learned by an observer with access to $\mathcal{H}_\text{rad} \otimes \mathcal{H}_\text{pur}$, and an observer with access only to $\mathcal{H}_\text{rad}$.

Taking into account the new infalling thermal flux, we find that \eqref{eq:energyflux} becomes
\begin{align}
\frac{d M}{d v} =\frac{c_\text{evap} \pi}{12} (T'^2 - T^2).
\end{align}
Hence we have
\begin{align} \label{eq:drsT'}
\frac{d r_s}{d v} = \frac{c_\text{evap} \pi G_N}{3 T(d-1) r_s^{d-2} \Omega_{d-1}}(T'^2 - T^2),
\end{align}
and, using \eqref{eq:rhor}, the event horizon is at
\begin{align}
r_\text{hor} = r_s + \frac{c_\text{evap} G_N}{6 (d-1) r_s^{d-2} \Omega_{d-1}}\frac{T'^2 - T^2}{T^2}.
\end{align}
Using \eqref{eq:vclassical}, the classical maximin surface lies on the classical apparent horizon $r_s$ at
\begin{align}
v =  -\frac{\beta}{2 \pi} \log \frac{S_{BH}T^2}{c_\text{evap} \,(T^2 - T'^2)} + O(\beta).
\end{align}
As the infalling radiation temperature $T'$ approaches the black hole temperature $T$, the location of the classical maximin surface diverges into the infinite past because $d r_s/d v \to 0$. In contrast, we shall see that the non-empty quantum extremal surface for the CFT remains well-behaved at this temperature.

We first calculate the Ryu-Takayanagi surface associated to $\mathcal{H}_\text{CFT}$. (Since the overall tripartite state is pure, this is also the Ryu-Takayanagi surface for $\mathcal{H}_\text{rad} \otimes \mathcal{H}_\text{pur}$.) The entropy of the outgoing modes is the same as \eqref{eq:outgoingentropy}, but, because the ingoing modes are now at finite temperature, the entropy of the infalling modes in the entanglement wedge of the CFT is now
\begin{align} \label{eq:dSindv}
S_\text{in} = - \frac{c_\text{evap} \pi  T' v}{6} + \dots,
\end{align}
where, as usual, we have dropped constant terms and we have assumed that the cut-off is independent of position in units of $v$.\footnote{We are also, as usual, ignoring the corrections associated with the finite infalling time range of the infalling modes because these corrections should vanish in the semiclassical limit.} The total bulk entropy is therefore
\begin{align} \label{eq:outgoingentropyT'}
S_\text{bulk} = \frac{c_\text{evap}}{6}\log\left(r_{lc}(v) - r\right) - \frac{c\pi v}{6}(T+T') + \dots
\end{align}

Note that for $T' = T$, which corresponds to the Hartle-Hawking state, we expect that the total entanglement entropy of ingoing and outgoing modes will agree with the Minkowski vacuum formula for the entanglement entropy based on the proper distance between the quantum extremal surface and the point where the outgoing modes are extracted. Using the Schwarzschild time translation symmetry, or boost symmetry, of the Hartle-Hawking state, we can map this interval to a small interval close to the bifurcation surface, where the Hartle-Hawking state locally looks like the Minkowski vacuum.\footnote{In contrast, the case where the infalling modes have zero temperature corresponds to the Unruh state, which is singular at the white hole horizon and so does not locally look like the Minkowski vacuum near the bifurcation surface.} It can easily be verified that this is indeed the case.

Using \eqref{eq:drsT'} and \eqref{eq:outgoingentropyT'}, it is easy to calculate the location of the Ryu-Takayanagi surface of $\mathcal{H}_\text{CFT}$. We find that
\begin{align}
0 &= \frac{\partial S_\text{bulk}}{\partial v} + \frac{1}{4G_N} \frac{\partial A}{\partial v},
\\& = \frac{\partial S_\text{bulk}}{\partial v},
\\& = \frac{\partial r_{lc}/\partial v}{6(r_{lc} - r)} - \frac{\pi}{6} (T+T'),
\\(T+T') (r_{lc}(v) - r)& = 2 T (r_{lc} - r_s(v)),
\end{align}
while \eqref{eq:dSdr} continues to be valid. Hence the quantum extremal surface lies at
\begin{align} \label{eq:rqT'}
r = r_s + \frac{T' - T}{T} \frac{G_N c_\text{evap}}{3(d-1) \Omega_{d-1} r_s^{d-2}} = r_\text{hor} - (T'-T)^2 \frac{ G_N c_\text{evap}}{6 (d-1) r_s^{d-2} \Omega_{d-1} T^2}, 
\end{align}
and satisfies
\begin{align} \label{eq:rlcT'}
r_{lc}(v) = r_s + \frac{T' + T}{T} \frac{G_N c}{3(d-1) \Omega_{d-1} r_s^{d-2}} = r_\text{hor} +\left[4- \frac{(T'-T)^2}{T^2}\right] \frac{ G_N c_\text{evap}}{6 (d-1) r_s^{d-2} \Omega_{d-1}}.
\end{align}
Thus
\begin{align} \label{eq:vT'}
v = - \frac{\beta}{2 \pi} \log\frac{S_{BH}}{c_\text{evap} (4- \frac{(T'-T)^2}{T^2})} + O(\beta).
\end{align}

When the infalling modes are at the same temperature as the black hole, the Ryu-Takayanagi surface lies exactly on the event horizon.\footnote{Since we only calculated the radius to $O(G_N)$, higher order corrections can potentially move the quantum extremal surface outside the horizon. When calculating the entanglement wedge of the CFT \emph{plus} all the future ingoing thermal modes that will be thrown into the black hole, the RT surface cannot end up outside the event horizon without creating a paradox. However this is not true for the entanglement wedge of the CFT alone. In that case, corrections from only having a finite interval of infalling thermal modes pushes the RT surface an $O(G_N^2)$ radial distance outside the horizon, as was shown in \cite{almheiri2019islands} after this paper first appeared on arXiv. (The result follows most obviously as a consequence of the time-reversal symmetry of the state forcing the RT surface to lie in the static slice.)} At all other temperatures, it lies strictly inside the event horizon. When $T'=T$, and only when $T' = T$, the total thermodynamic entropy of the black hole and exterior radiation does not increase with time. The entropy of the new Hawking radiation is cancelled by the loss of entropy from radiation falling into the black hole, while the entropy of the black hole itself stays constant. Because all outgoing modes in the interior are in the entanglement wedge of $\mathcal{H}_\text{rad} \otimes \mathcal{H}_\text{pur}$, the Hawking radiation, even Hawking radiation far in the future, is perfectly thermally entangled with $\mathcal{H}_\text{rad} \otimes \mathcal{H}_\text{pur}$. Since the ingoing modes are also thermally entangled with $\mathcal{H}_\text{pur}$ and the horizon area is constant, the black hole entanglement entropy stays constant. 

This is consistent with the Page curve, which can also be derived using Ryu-Takayanagi formula. Indeed, at any temperature $T'$, the entanglement structure of the Hawking radiation will be exactly consistent with the Page curve, because of our general argument from Section \ref{sec:haydenpage}. It can easily be verified that this is indeed the case.

The Ryu-Takayanagi surface remains approximately one scrambling time in the past so long as the temperature of the infalling modes is relatively low (the latest infalling time is obtained at $T'=T$), but diverges into the past at $T' = 3T$. At higher temperatures, the radial distance between the quantum extremal surface and the event horizon, required by \eqref{eq:rqT'}, will be greater than the radial distance between the quantum extremal surface and the outgoing lightcone, required by \eqref{eq:rlcT'}. Since the outgoing lightcone can never be inside the event horizon, no quantum extremal surface can exist.

An observer with access to $\mathcal{H}_\text{rad} \otimes \mathcal{H}_\text{pur}$ will only ever learn the state of a diary thrown into the black hole if the temperature $T' < 3T$. Interestingly, $T'  = 3 T$ is exactly the temperature at which thermal Hawking radiation, unentangled with $\mathcal{H}_\text{rad} \otimes \mathcal{H}_\text{pur}$, becomes consistent with unitarity. At this temperature,
\begin{align} \label{eq:etaentropyincrease}
\frac{1}{4 G_N} \frac{d A_\text{hor}}{d v} = \frac{2 c_\text{evap} \pi}{3 \beta} =   \frac{d S_\text{in}}{d v} +\frac{d \Srad}{d v},
\end{align}
where $d S_\text{in}/d v \geq 0$ is the entropy of the ingoing modes per unit infalling time and $d \Srad/d v \geq 0$ is the rate that entropy is produced in the Hawking radiation. The increase in the Bekenstein-Hawking entropy is therefore just sufficient to purify both the outgoing Hawking radiation, and the purification $\mathcal{H}_\text{pur}$ of the infalling modes.

If the observer only has access to $\mathcal{H}_\text{rad}$, but not to $\mathcal{H}_\text{pur}$, it will affect the information that they learn about the black hole. To understand this, we need to calculate the location of the Ryu-Takayanagi surface for $\mathcal{H}_\text{rad}$.  (Because the system is no longer in a bipartite pure state, this is not the same as the Ryu-Takayanagi surface of $\mathcal{H}_\text{CFT}$.)

The entanglement wedge of $\mathcal{H}_\text{rad}$ contains the part of the interior inside the Ryu-Takayanagi surface, as well as the outgoing modes that were extracted into the reservoir. Since the overall state of the outgoing modes is pure, the outgoing entropy in the entanglement wedge of $\mathcal{H}_\text{rad}$ for a given candidate RT surface is equal to the outgoing entropy in the entanglement wedge of $\mathcal{H}_\text{CFT}$ for the same candidate surface. (Since the Ryu-Takayanagi surfaces for $\mathcal{H}_\text{rad}$ and $\mathcal{H}_\text{CFT}$ will end up being different, they will have different entropies for the outgoing modes. However as a \emph{function} of the location of the surface, they are the same.)

This is not true for the ingoing modes, which are in a mixed state that is purified by $\mathcal{H}_\text{pur}$. The entanglement wedge of $\mathcal{H}_\text{CFT}$ contains ingoing modes at infalling times that are later than the Ryu-Takayanagi surface, while the entanglement wedge of $\mathcal{H}_\text{rad}$ contains at infalling times that are earlier than the RT surface. Hence, instead of \eqref{eq:outgoingentropyT'}, we have
\begin{align}
S_\text{bulk} =  \frac{c_\text{evap}}{6}\log\left(r_{lc}(v) - r\right) + \frac{c\pi v}{6}(T' - T) + \dots
\end{align}
We therefore find that the quantum extremal surface for $\mathcal{H}_\text{rad}$ lies at
\begin{align}
r = r_s - \frac{T' + T}{T} \frac{G_N c_\text{evap}}{3(d-1) \Omega_{d-1} r_s^{d-2}} = r_\text{hor} - (T'+T)^2 \frac{ G_N c}{6 (d-1) r_s^{d-2} \Omega_{d-1} T^2}, 
\end{align}
and satisfies
\begin{align}
r_{lc}(v) = r_s + \frac{T - T'}{T} \frac{G_N c}{3(d-1) \Omega_{d-1} r_s^{d-2}} = r_\text{hor} +\left[4- \frac{(T'+T)^2}{T^2}\right] \frac{ G_N c_\text{evap}}{6 (d-1) r_s^{d-2} \Omega_{d-1}}.
\end{align}
Thus
\begin{align}
v = - \frac{\beta}{2 \pi} \log\frac{S_{BH}}{c (4- \frac{(T'+T)^2}{T^2})} + O(\beta).
\end{align}
For $T' < T$, information thrown into the black hole will still reappear in the Hawking radiation. However the location of the extremal surface diverges as $T' \to T$; for $T'\geq T$ no information will ever escape in the Hawking radiation for an observer with access only to $\mathcal{H}_\text{rad}$. 

As before, this is exactly the temperature at which the Hawking radiation can look thermal, to an observer with access only to $\mathcal{H}_\text{rad}$ without violating unitarity. The total entropy of $\mathcal{H}_\text{CFT} \otimes \mathcal{H}_\text{rad}$ increases because thermal modes are being thrown into the black hole. The entropy of $\mathcal{H}_\text{rad}$ can therefore increase at the same rate, and the new Hawking radiation can be unentangled with $\mathcal{H}_\text{rad}$, even while the horizon area, and hence the entropy, of the black hole remains constant.

Finally, the Ryu-Takayanagi surface of $\mathcal{H}_\text{pur}$ will always be empty -- it will always be in a thermal state. This is a necessary consequence of the boundary dynamics being unitary; the state of $\mathcal{H}_\text{pur}$ is initially thermal, and this is unchanged when we throw its purification into a black hole.

We can also calculate the location of the quantum extremal surface when the infalling modes are in a pure state, with constant energy density $\eta$, but without long range entanglement, for example, if there is a constant infalling particle flux. Again, it is important that the ingoing modes are unentangled with the outgoing Hawking radiation. In this case, both \eqref{eq:dSdv} and \eqref{eq:dSdr} will continue to be valid as in the vacuum case. However, instead of \eqref{eq:energyflux}, we now have
\begin{align}
\frac{d M}{d v} = - \frac{c_\text{evap} \pi}{12 \beta^2} + \eta,
\end{align}
and
\begin{align}
\frac{d r_s}{d v} = -\frac{c_\text{evap} \pi G_N}{3 \beta (d-1) r_s^{d-2} \Omega_{d-1}} +\frac{4 G_N \beta \eta}{(d-1) r_s^{d-2} \Omega_{d-1}}.
\end{align}
This affects the radius $r_{l.c.}(v)$ of the past lightcone as a function of the infalling time $v$, which is given in \eqref{eq:outgoinglc2}. Hence, while the Ryu-Takayanagi surface still lies at
\begin{align}
r = r_s - \frac{G_N c_\text{evap}}{3(d-1)\Omega_{d-1}r_s^{d-2}},
\end{align}
its infalling time will now be given by
\begin{align}
v = -\frac{\beta}{2 \pi} \log \frac{S_{BH}}{1 - 4 \beta^2 \eta/c_\text{evap} \pi } + O(\beta).
\end{align}
As before, when sufficient energy, in this case $\eta > c_\text{evap} \pi /4 \beta^2$ are thrown into the black hole, no quantum extremal surface exists. The event horizon, and thus the outgoing light cone, are at too large a radius for \eqref{eq:dSdv} and \eqref{eq:dSdr}  to be simultaneously satisfied.

Yet again, this is exactly the point at which it stops being necessary for information to escape the black hole in order to preserve unitarity. At this energy density the rate of increase of the Bekenstein Hawking entropy is equal to the rate of increase in the entropy of the radiation
\begin{align} \label{eq:T'entropyincrease}
\frac{1}{4 G_N} \frac{d A_\text{hor}}{d v} = \frac{c_\text{evap} \pi}{6 \beta} = \frac{d \Srad}{d v}.
\end{align}
Hence the Hawking radiation can remain thermal, and unentangled with $\mathcal{H}_\text{rad}$, forever, without exceeding the entanglement entropy exceeding the Bekenstein-Hawking entropy of the black hole. 

The ingoing energy flux at which information stops escaping the black hole is highest for thermal infalling modes and an observer who has access to both the reservoir $\mathcal{H}_\text{rad}$ and the purification $\mathcal{H}_\text{pur}$ of the ingoing modes. This is because there needs to be sufficient Bekenstein-Hawking entropy in the black hole both to purify outgoing thermal Hawking radiation \emph{and} to purify the degrees of freedom in $\mathcal{H}_\text{pur}$.

In contrast, when the observer only has acccess to $\mathcal{H}_\text{rad}$, the required ingoing energy flux for thermal infalling modes is much smaller. The ingoing entropy makes it \emph{easier} for the Hawking radiation to be unentangled with $\mathcal{H}_\text{rad}$, because $\mathcal{H}_\text{rad}$ can be purified by $\mathcal{H}_\text{pur}$ as well as the black hole.

Finally, when the ingoing modes are in a pure state with no long range entanglement, information stops escaping at an intermediate ingoing energy flux. The increase in the Bekenstein-Hawking entropy needs to be sufficient to purify the Hawking radiation in $\mathcal{H}_\text{rad}$; there is no ingoing bulk entropy to make this either harder or easier.

\section{Minimal State Dependence in the SYK Model} \label{app:syk}
In this appendix, we construct minimally state-dependent interior reconstructions in a simple toy model of quantum gravity, known as the SYK model. This model has been studied in great depth in the last few years \cite{sachdev1993gapless, kitaev2015simple, maldacena2016remarks, polchinski2016spectrum, cotler2017black, saad2018semiclassical}; here we provide only the bare minimum of background detail necessary for our purposes. 

The SYK model is a $0+1$-dimensional quantum mechanical model that consists of $N$ Majorana fermions $\psi_i$. These satisfy $$\{\psi_i, \psi_j\} = \delta_{i,j}.$$ Using a Jordan-Wigner transformation, it can be easily seen that there is a single qubit degree of freedom associated with every pair of Majorana fermions. The Hilbert space therefore has dimension $2^{N/2}$. The model has Hamiltonian
\begin{align}
H = \sum_{iklm} j_{iklm} \psi_i \psi_k \psi_l \psi_m,
\end{align}
where $j_{iklm}$ are independent Gaussian random couplings with $\langle j_{iklm}^2\rangle = 6 J^2/N^3$. 

In the limit $N \to \infty$, at fixed $\beta J \gg 1$, the SYK model appears to become holographic; it has many features that resemble nearly-AdS$_2$ gravity. Although the precise dual gravity description, if one exists, remain unknown, both the SYK model and simple nearly-AdS$_2$ gravity theories such as Jackiw-Teitelboim gravity \cite{almheiri2015models, engelsoy2016investigation, jensen2016chaos, maldacena2016conformal, saad2019jt} have an emergent reparameterisation symmetry that is both spontaneously and explicitly broken, with the explicit symmetry-breaking term proportional to the so-called Schwarzian action
\begin{align} \label{eq:schwarz}
S = \frac{\alpha_S N}{J} \int d\tau \left(\frac{f''}{f'}\right)' - \frac{1}{2} \left(\frac{f''}{f'}\right)^2,
\end{align}
where $f(\tau)$ is the reparameterisation and $\alpha_S$ is a numerical constant. From a gravity perspective, this action appears as a boundary term when we cut-off the nearly-AdS$_2$ geometry at some fixed dilaton value $\phi_b$; for Jackiw-Teitelboim gravity, in particular, it describes the entire dynamics of the theory, which can be interpreted as the dynamics of a boundary particle, describing the location of the cut-off, in a rigid AdS$_2$ background.

A complete basis for the entire Hilbert space of the SYK model is given by the states $\ket{B_s}$, satisfying
\begin{align}
\left(\psi_{2k-1} - i s_k \psi_{2k} \right) \ket{B_s} = 0,
\end{align}
where for all $k$, we have $s_k = \pm 1$. If we evolve these states in Euclidean time, we get a set of states
\begin{align}
\ket{B_s (\beta)} = e^{-\beta H / 2} \ket{B_s},
\end{align}
which form an approximate, overcomplete basis for the low energy states of the theory. In fact, if we allow arbitrary superpositions of these states, they still form a complete basis for the entire Hilbert space, because the map $e^{-\beta H}$ is invertible. However, to create a high energy state, we need a very finely tuned superposition
\begin{align}
\ket{\psi} = \sum_s c_s \ket{B_s (\beta)},
\end{align}
where
\begin{align} \label{eq:finetuning}
\braket{\psi|\psi} \ll \sum_s |c_s|^2 \braket{B_s (\beta) |B_s (\beta)}.
\end{align}
However we will only allow `generic' superpositions of the states $\ket{B_s (\beta)}$ that do not satisfy \eqref{eq:finetuning}.

\begin{figure} [t]
\centering
\begin{subfigure}{.48\textwidth}
  \centering
 \includegraphics[width = 0.8\linewidth]{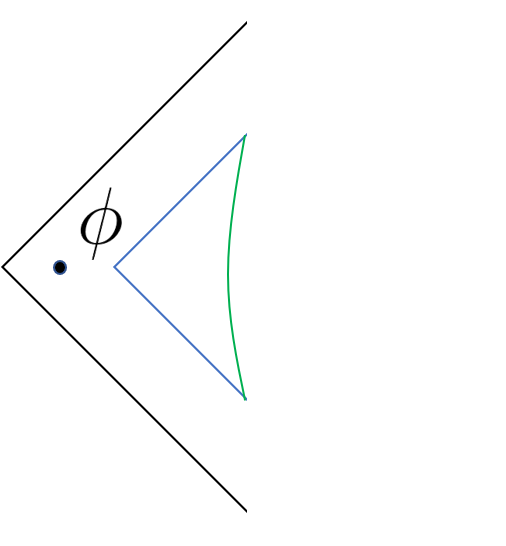}
 \caption{}
 \label{fig:sykuntuned}
\end{subfigure}
\begin{subfigure}{.48\textwidth}
 \includegraphics[width = 0.8\linewidth]{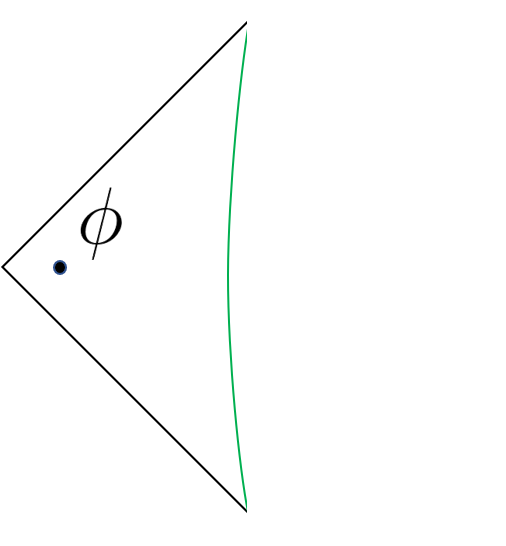}
 \centering
 \caption{}
 \label{fig:syktuned}
\end{subfigure}
\caption{The states $\ket{B_s(\beta)}$ have a gravity description as a one sided black hole, ending on an end-of-the-world brane (black). If the system is evolved using the unperturbed SYK Hamiltonian, shown in Figure \ref{fig:sykuntuned}, the bulk operator $\phi$ lies behind the black hole horizon (blue). However if the Hamiltonian is perturbed in a state-dependent way, shown in Figure \ref{fig:syktuned}, the boundary (green) is pulled inwards and so the operator no longer lies behind a horizon. In the original construction, the Hamiltonian was precisely tuned for a single state  $\ket{B_s(\beta)}$. However one can easily adapt the Hamiltonian to work for a large set of states.}
\label{fig:SYK}
\end{figure}

It was shown in \cite{kourkoulou2017pure} that the states $\ket{B_s (\beta)}$ have a natural gravity interpretation as black hole microstates with a smooth interior ending on an `end-of-the-world brane' (Figure \ref{fig:SYK}). Excitations in the interior can be created by acting with additional boundary operators during the Euclidean time evolution. 

If the system is evolved with the unperturbed Hamiltonian $H$, then such excitations can never reach the boundary. However, if we perturb the Hamiltonian by
\begin{align} \label{eq:origpert}
\delta H = - \varepsilon J \sum_{k=1}^{N/2} s_k i \psi_{2k -1} \psi_{2k},
\end{align}
then, from a gravity perspective, the Schwarzian `boundary particle' is pulled into the bulk of the AdS$_2'$ space, as shown in Figure \ref{fig:SYK}. By evolving the system with the perturbed Hamiltonian $H + \delta H$, we can render the interior of the black hole visible to the boundary. Interior operators can be reconstructed using boundary operators time-evolved using this perturbed Hamiltonian.

The perturbation $\delta H$ was carefully adapted to the state $\ket{B_s (\beta)}$. The reconstruction is therefore highly state-dependent. A natural question is whether we can reduce this state dependence, and find a reconstruction that works for a larger class of microstates.

Instead of using the perturbation \eqref{eq:origpert}, suppose instead that we perturb the Hamiltonian $H$ by
\begin{align}
\delta H_f = - \varepsilon J  \sum_{k=1}^{f N/2} s_k i \psi_{2k-1} \psi_{2k},
\end{align}
where $0 < f < 1$ is a fixed $O(1)$ fraction. Since the number of terms in $\delta H_f$ continues to be $O(N)$,  the perturbation $\delta H_f$ will also make the interior of the black hole microstate $\ket{B_s(\beta)}$ visible to the boundary.\footnote{The argument that the perturbation $\delta H$ can be used to make the interior visible to a boundary observer is given in Section 7 of \cite{kourkoulou2017pure}. For the full details, we simply refer readers to that work. However the basic strategy is to assume that for $\varepsilon \ll 1$ we can treat $\delta H$ as a perturbation to the Schwarzian action \eqref{eq:schwarz} of the reparameterisation modes. So long as we have $O(N)$ terms, the perturbation will appear in the semiclassical action at large $N$. It can then be shown that, at sufficiently large, but fixed, $\beta J$, we can choose $\varepsilon$ so that the semiclassical large $N$ equation of motion for the Schwarzian `boundary particle' makes the entire black hole interior visible. The argument for $\delta H_f$ is identical, except for the addition of the $O(1)$ factor $f$ in the perturbation to the large $N$ semiclassical Schwarzian action.}

However this perturbation only depended on the first $f N/2$ spins $s_k$ of the microstate $\ket{B_s(\beta)}$. We have found a single reconstruction that works for $2^{(1-f)N/2}$ different microstates $\ket{B_s(\beta)}$. By linearity, the same reconstruction should also work for generic superpositions of these microstates.
 
Since the low temperature entropy of the SYK model is approximately
\begin{align}
S_0 \approx 0.23 N, 
\end{align}
we have found a single reconstruction that is valid for more than $e^{S_0}$ microstates. Of course, not all these microstates are independent. Instead, the effective size of the code subspace is determined by the entropy of the mixed state
\begin{align} \label{eq:mixture}
 \rho_{s_1} = 2^{-(1-f)N/2} \sum_{s_2} \ket{B_{s_1 \oplus s_2}(\beta)} \bra{B_{s_1 \oplus s_2}(\beta)}, 
\end{align}
where the sum is over spins $s_k \in s_2$ for $f N/2 \leq k \leq N /2$ while the spins $s_k \in s_1$ for $1 \leq k < f N/2$ are held fixed.

How large is this entropy? Since 
\begin{align}
\frac{e^{-\beta H}}{\Tr(e^{-\beta H})} = 2^{-f N/2}\sum_{s_1} \rho_{s_1},
\end{align}
then the strict concavity of entropy means that 
\begin{align}
\langle S(\rho_{s_1}) \rangle_{s_1} < S_0,
\end{align}
where the expectation is taken over possible states $s_1$ of the fixed spins. If $\rho_{s_1}$ were a uniform mixture of $2^{(1-f)N/2}$ randomly chosen microstates $\ket{B_s(\beta)}$, we would expect that $S(\rho_{s-1})$ would be very close to $S_0$ for $(1-f)N/2 > S_0$. There would be no remaining space to encode the interior degrees of freedom. However this will not be the case for the particular set of microstates in \eqref{eq:mixture}. 

At large $N$, there is an emergent $O(N)$ symmetry of the SYK model. In particular there is a $\mathbb{Z}_2^n$ subgroup of this symmetry group, called the flip subgroup, that acts transitively on the set of states $\ket{B_s(\beta)}$). This means that $S(\rho_{s_1})$ depends, at leading order, only on the number of spins that are held fixed, and not on the signs of those spins.

If no spins are held fixed, the ensemble is simply the canonical ensemble, which has entropy $S_0$ to leading order in $1/N$ for fixed $\beta J \gg 1$. Now suppose that we know the average entropy $S_f$ for a state $\rho_{s_1}$ with a fixed fraction $f$ of the spins held fixed. We then consider the ensembles $\rho_{s_1 \oplus 1}$ and $\rho_{s_1 \oplus -1}$ formed by fixing a single additional spin. These two ensembles therefore have a fraction $f+2/N$ of their spins fixed. Note that
\begin{align}
\rho_{s_1} = \frac{1}{2} \left( \rho_{s_1 \oplus 1} + \rho_{s_1 \oplus -1}\right).
\end{align}
Hence
\begin{align}
2 S(\rho_{s_1}) - S(\rho_{s_1 \oplus 1}) - S(\rho_{s_1 \oplus -1}) &=  S(\rho_{s_1 \oplus 1}|| \rho_{s_1}) + S(\rho_{s_1 \oplus -1}|| \rho_{s_1})
\\&\geq \frac{1}{4} \lVert \rho_{s_1 \oplus 1} - \rho_{s_1 \oplus -1} \rVert_1^2,
\end{align}
where we have used Pinsker's inequality \cite{ohya2004quantum}. However, we can lower bound the trace distance $\lVert \rho_{s_1 \oplus 1} - \rho_{s_1 \oplus -1} \rVert_1$ by
\begin{align}
\lVert \rho_{s_1 \oplus 1} - \rho_{s_1 \oplus -1} \rVert_1 = \max_{\mathcal{O}} \frac{\left|\Tr{\,\left[\mathcal{O} \left(\rho_{s_1 \oplus 1} - \rho_{s_1 \oplus -1}\right)\right]}\right|}{\lVert \mathcal{O}\rVert} \geq \left|\Tr \left[\,\psi_{2k-1} \psi_{2k} \left(\rho_{s_1 \oplus 1} - \rho_{s_1 \oplus -1}\right)\right]\right|,
\end{align}
where $k = Nf/2 + 1$ labels the additional spin fixed in $\rho_{s_1 \oplus \pm 1}$, but not in $\rho_{s_1}$. This last quantity was shown to in \cite{kourkoulou2017pure} to be order one in the limit of large $N$ (although it decays as a function of $\beta J$). Hence
\begin{align}
S_f - S_{f+2/N} = O(1)
\end{align}
and thus
\begin{align}
S_0 - S_f = O(f N),
\end{align}
for any fixed $f > 0$. If the fraction $f$ is small, the entropy is very close to $S_0$ at leading order, but there is still plenty of space left to encode the interior. We have found an explicit, minimally state-dependent reconstruction.

Of course, we have only constructed minimally state-dependent operators for a very particular class of ensembles of microstates. Our arguments in Section \ref{sec:minimal} suggested that there should exist minimally state-dependent reconstructions for \emph{any} sufficiently small code subspace. For the special ensembles that we have considered in this section, the reconstructions only involved a simple perturbation to the SYK Hamiltonian; for arbitrary code subspaces, the reconstructions would presumably be much more complicated. We gave a more general, but much less practical, procedure for constructing minimally state-dependent reconstructions out of reconstructions that only work for individual states in Section \ref{sec:explicit}; it seems reasonable to expect that such a procedure should also work for the SYK model.

\bibliographystyle{unsrt}
\bibliography{biblio}
\end{document}